\documentclass[lettersize,journal]{IEEEtran}
\usepackage{amsmath,amsfonts}
\usepackage{algorithmic}
\usepackage{algorithm}
\usepackage{array}
\usepackage{acro}
\usepackage[caption=false,font=normalsize,labelfont=sf,textfont=sf]{subfig}
\usepackage{textcomp}
\usepackage{stfloats}
\usepackage{url}
\usepackage{verbatim}
\usepackage{graphicx}
\def\BibTeX{{\rm B\kern-.05em{\sc i\kern-.025em b}\kern-.08em
    T\kern-.1667em\lower.7ex\hbox{E}\kern-.125emX}}
\usepackage{balance}
\usepackage{tikz}
\usetikzlibrary{trees}
\usepackage{enumitem}
\usepackage{multirow}
\usepackage[table,xcdraw]{xcolor}

\DeclareAcronym{TVWS}{
  short={TVWS},
  long={TV White Spaces},
}

\DeclareAcronym{ISM}{
  short={ISM},
  long={Industrial, Scientific, and Medical},
}

\DeclareAcronym{CBRS}{
  short={CBRS},
  long={Civil Broadband Radio Service},
}

\DeclareAcronym{UNII}{
  short={U-NII},
  long={Unlicensed National Information Infrastructure},
}

\DeclareAcronym{FCC}{
  short={FCC},
  long={Federal Communications Commission},
}

\DeclareAcronym{WSDB}{
  short={WSDB},
  long={White Space Database},
}

\DeclareAcronym{WSD}{
  short={WSD},
  long={White Space Devices},
}

\DeclareAcronym{ECC}{
  short={ECC},
  long={Electronic Communications Committee},
}

\DeclareAcronym{ETSI}{
  short={ETSI},
  long={European Telecommunications Standards Institute},
}

\DeclareAcronym{BS}{
  short={BS},
  long={Base Station},
}

\DeclareAcronym{UE}{
  short={UE},
  long={User Equipment},
}

\DeclareAcronym{OFDMA}{
  short={OFDMA},
  long={Orthogonal Frequency-Division Multiple Access},
}

\DeclareAcronym{QoS}{
  short={QoS},
  long={Quality of Service},
}

\DeclareAcronym{DIFS}{
  short={DIFS},
  long={DCF Inter-frame Space},
}

\DeclareAcronym{EPC}{
  short={EPC},
  long={Evolved Packet Core},
}

\DeclareAcronym{LPWAN}{
  short={LPWAN},
  long={Low Power Wide Area Networks},
}

\DeclareAcronym{LAN}{
  short={LAN},
  long={Local Area Network},
}

\DeclareAcronym{IoT}{
  short={IoT},
  long={Internet of Things},
}

\DeclareAcronym{FWA}{
  short={FWA},
  long={Fixed Wireless Access},
}

\DeclareAcronym{OAI}{
  short={OAI},
  long={Open Air Interface},
}

\DeclareAcronym{NR}{
  short={NR},
  long={New Radio},
}

\DeclareAcronym{ML}{
  short={ML},
  long={Machine Learning},
}

\DeclareAcronym{IEEE}{
  short={IEEE},
  long={Institute of Electrical and Electronics Engineers},
}

\DeclareAcronym{TWT}{
  short={TWT},
  long={Target Wake Time},
}

\DeclareAcronym{CSMA}{
  short={CSMA},
  long={Carrier Sense Multiple Access},
}

\DeclareAcronym{SRD}{
  short={SRD},
  long={Short Range Devices},
}

\DeclareAcronym{PAL}{
  short={PAL},
  long={Priority Access Licenses},
}

\DeclareAcronym{SAS}{
  short={SAS},
  long={Spectrum Access Systems},
}

\DeclareAcronym{EIRP}{
  short={EIRP},
  long={Effective Isotropic Radiated Power},
}

\DeclareAcronym{LTE}{
  short={LTE},
  long={Long Term Evolution},
}

\DeclareAcronym{LTE-U}{
  short={LTE-U},
  long={LTE Unlicensed},
}

\DeclareAcronym{LAA}{
  short={LAA},
  long={Licensed Assisted Access},
}

\DeclareAcronym{RLAN}{
  short={RLAN},
  long={Radio Local Area Networks},
}

\DeclareAcronym{UL}{
  short={UL},
  long={Uplink},
}

\DeclareAcronym{DL}{
  short={DL},
  long={Downlink},
}

\DeclareAcronym{IU}{
  short={IU},
  long={Incumbent Users},
}

\DeclareAcronym{GAA}{
  short={GAA},
  long={General Authorized Access},
}

\DeclareAcronym{WLAN}{
  short={WLAN},
  long={Wireless Local Area Networks},
}

\DeclareAcronym{WPA}{
  short={WPA},
  long={Wi-Fi Protected Access},
}

\DeclareAcronym{eLAA}{
  short={eLAA},
  long={enhanced LAA},
}

\DeclareAcronym{feLAA}{
  short={feLAA},
  long={further enhanced LAA},
}

\DeclareAcronym{DC}{
  short={DC},
  long={Dual Connectivity},
}

\DeclareAcronym{CA}{
  short={CA},
  long={Carrier Aggregation},
}

\DeclareAcronym{DSS}{
  short={DSS},
  long={Dynamic Spectrum Sharing},
}

\DeclareAcronym{UNB}{
  short={UNB},
  long={Ultra Narrow Band},
}

\DeclareAcronym{SS}{
  short={SS},
  long={Spread Spectrum},
}

\DeclareAcronym{MU}{
  short={MU},
  long={Multiuser},
}

\DeclareAcronym{TDD}{
  short={TDD},
  long={Time Division Duplex},
}

\DeclareAcronym{HARQ}{
  short={HARQ},
  long={Hybrid Automatic Repeat reQuest},
}

\DeclareAcronym{CSMA/CA}{
  short={CSMA/CA},
  long={Carrier Sense Multiple Access with Collision Avoidance},
}

\DeclareAcronym{CLI}{
  short={CLI},
  long={Cross Link Interference},
}

\DeclareAcronym{RIM}{
  short={RIM},
  long={Remote Interference Management},
}

\DeclareAcronym{CAT}{
  short={CAT},
  long={Categoty},
}

\DeclareAcronym{COT}{
  short={COT},
  long={Channel Occupancy Time},
}

\DeclareAcronym{AP}{
  short={AP},
  long={Access Point},
}

\DeclareAcronym{CSAT}{
  short={CSAT},
  long={Carrier Sense Adaptive Transmission},
}

\DeclareAcronym{ITU}{
  short={ITU},
  long={International Telecommunication Union},
}

\DeclareAcronym{PLMN}{
  short={PLMN},
  long={Public Land Mobile Networks},
}

\DeclareAcronym{AIoT}{
  short={AIoT},
  long={Artificial Internet of Things},
}

\DeclareAcronym{TPC}{
  short={TPC},
  long={Transmit Power Control},
}

\DeclareAcronym{DFS}{
  short={DFS},
  long={Dynamic Frequency Selection},
}

\DeclareAcronym{CAC}{
  short={CAC},
  long={Channel Availability Check},
}

\DeclareAcronym{CCA}{
  short={CCA},
  long={Clear Channel Assessment},
}

\DeclareAcronym{AFC}{
  short={AFC},
  long={Automated Frequency Coordination},
}

\DeclareAcronym{TXOP}{
  short={TXOP},
  long={Transmit Opportunity},
}

\DeclareAcronym{PUSCH}{
  short={PUSCH},
  long={Physical Uplink Shared Channel},
}

\DeclareAcronym{LWIP}{
  short={LWIP},
  long={LTE WLAN Radio Level Integration with internet protocol security},
}

\DeclareAcronym{LWA}{
  short={LWA},
  long={LTE-WLAN Aggregation},
}

\DeclareAcronym{URLLC}{
  short={URLLC},
  long={Ultra Reliable Low Latency Communication},
}

\DeclareAcronym{TDMA}{
  short={TDMA},
  long={Time Division Multiple Access},
}

\DeclareAcronym{DCS}{
  short={DCS},
  long={Dynamic Channel Selection},
}

\DeclareAcronym{DCF}{
  short={DCF},
  long={Distributed Coordination Function},
}

\DeclareAcronym{DIFC}{
  short={DIFC},
  long={DCF Inter-Frame Spacing}, 
}

\DeclareAcronym{CS}{
  short={CS},
  long={Carrier Sensing},
}

\DeclareAcronym{ED}{
  short={ED},
  long={Energy Detection},
}

\DeclareAcronym{RTS}{
  short={RTS},
  long={Request-to-Send},
}

\DeclareAcronym{CTS}{
  short={CTS},
  long={Clear-to-Send},
}

\DeclareAcronym{eMBB}{
  short={eMBB},
  long={enhanced Mobile Broadband},
}

\DeclareAcronym{PAWS}{
  short={PAWS},
  long={Protocol to Access White-Space},
}

\DeclareAcronym{mMTC}{
  short={mMTC},
  long={Massive Machine Type Communications},
}

\DeclareAcronym{FR1}{
  short={FR1},
  long={Frequency Range 1},
}

\DeclareAcronym{LBT}{
  short={LBT},
  long={Listen Before Talk},
}
\begin{document}

\title{Communications over Unlicensed sub-8 GHz Spectrum: Opportunities and Challenges}

\author{Karim Saifullin, \IEEEmembership{Student Member,~IEEE},  Hussein Al-Shatri, \IEEEmembership{Senior Member,~IEEE,} and Mohamed-Slim Alouini, \IEEEmembership{Fellow,~IEEE,}
        \thanks{Electrical and Computer Engineering (ECE) {Computer, Electrical and Mathematical Sciences and Engineering (CEMSE)}
        {King Abdullah University of Science and Technology (KAUST)}, 
        {Thuwal, 23955-6900, Kingdom of Saudi Arabia}. 
        karim.saifullin,hussein.shatri,slim.alouini@kaust.edu.sa}
        \thanks{Copyright (c) 20xx IEEE. Personal use of this material is permitted. However, permission to use this material for any other purposes must be obtained from the IEEE by sending a request to pubs-permissions@ieee.org.}
        }

\markboth{INTERNET OF THINGS JOURNAL,~VOL.~XX, NO.~X, XXX~20XX}%
{}

\IEEEpubid{0000--0000/00\$00.00~\copyright~20XX IEEE}

\maketitle

\begin{abstract}
The use of unlicensed spectrum offers a promising solution to spectrum scarcity in densely populated areas and a cost-effective way to connect underserved regions. Recognizing this potential, both academia and industry are actively exploring innovative uses of unlicensed spectrum. This work presents a broad overview of unlicensed spectrum bands below 8 GHz, including TV White Spaces, Civil Broadband Radio Services, Industrial, Scientific, and Medical bands, and the Unlicensed National Information Infrastructure. The paper examines three main aspects: regulations, existing technologies, and applications. Regulations play a central role, as understanding these rules is vital since “unlicensed” does not mean “unregulated”. From a technological perspective, we review current technologies, standards, and products, their features, and related applications. Furthermore, the shared nature of these spectrum bands introduces challenges related to user interference. The communication collisions can be managed through two primary strategies we described: a database-driven approach and coexistence mechanisms at the MAC and PHY layers. This work may serve as a starting point for those interested in the unlicensed spectrum, both in academia and industry. 
\end{abstract}


\section{Introduction}
Global demand for higher wireless data throughput continues to increase annually. A promising strategy to meet this challenge is the use of unlicensed frequency bands, either independently or alongside licensed spectrum. Effective utilization of these unlicensed bands can increase data rates, improve connectivity, and enhance the overall \ac{QoS}. 

Traditionally, unlicensed spectrum has been utilized by indoor, short-range, narrowband, and low-mobility devices, which often exhibit low \ac{QoS} due to factors such as spectrum scarcity, favorable propagation characteristics, and free-of-charge access. However, there is a trend toward utilizing the unlicensed spectrum for high-throughput applications in newly introduced 6 GHz bands. Consequently, academia and industry interest in license-exempt spectrum is rapidly rising.

Within the broad range of unlicensed frequency bands, we focus on the sub-8 GHz range, as it includes many of the most widely adopted technologies worldwide, including cellular networks, Wi-Fi, and other wireless systems. Although “unlicensed” suggests freedom from formal approval, devices operating in these bands must still comply with national and international regulations. In this context, we define the unlicensed spectrum as the portion of the radio frequency spectrum allocated by government authorities for shared use, under specific regulatory conditions and, in some cases, subject to access frameworks or usage fees. Fig. \ref{fig:unlicensed_topics} outlines the main sections addressed in this paper.
\vspace{-1em}
\begin{figure}[ht]
\centering
\tikzstyle{every node}=[draw=black,thick,anchor=west]
\tikzstyle{selected}=[draw=red,fill=red!30]
\tikzstyle{optional}=[dashed,fill=gray!50]
\begin{tikzpicture}[%
  grow via three points={one child at (0.5,-0.7) and
  two children at (0.5,-0.7) and (0.5,-1.4)},
  sibling distance=10mm, 
  edge from parent path={(\tikzparentnode.south) |- (\tikzchildnode.west)}] 
  \node {Unlicensed spectrum}
    child { node {Documents}
        child { node {Standards} }
        child { node {Regulations} }
    }
    child [missing] {}				
    child [missing] {}				
    child { node {Technical Characteristics}
        child { node {Carrier Frequency} }
        child { node {Bandwidth} }
        child { node {Transmit Power} }
        child { node {Communication Range} }
        child { node {Data Rate} }
        child { node {Primary users coexistence} }
        child { node[align=center, yshift=-3mm] {Database and PHY/MAC \\ interference management } }
    }
    child [missing] {}				
    child [missing] {}				
    child [missing] {}
    child [missing] {}				
    child [missing] {}	
    child [missing] {}				
    child [missing] {}	
    child [missing] {}	
    child { node {Applications}
        child { node {High Throughput Systems} }
        child { node {Secure, Low Latency Systems} } 
        child { node {Massive Connectivity} } 
    };
\end{tikzpicture}
\caption{Hierarchical breakdown of the unlicensed spectrum aspects}
\label{fig:unlicensed_topics}
\end{figure}

To clarify the differences, we compare the unlicensed spectrum with the traditional licensed one in Table \ref{tab:licensed_vs_unlicensed}. This paper shows that although the unlicensed spectrum initially developed with its own specific approaches, there is a trend toward convergence with the licensed spectrum in both usage and underlying technologies. However, it is important to recognize that such developments depend on many factors, with policymakers and industry playing a central role.

\IEEEpubidadjcol

\begin{table}[ht] \centering \caption{Comparison between licensed and unlicensed spectrum} \label{tab:licensed_vs_unlicensed} \begin{tabular}{|c|c|c|} \hline Spectrum type & Licensed & Unlicensed \\ \hline \begin{tabular}[c]{@{}c@{}} Device operation\\ must comply with local \\ and global regulations\end{tabular} & Yes & Yes \\ \hline \begin{tabular}[c]{@{}c@{}}Operation incurs \\ a fee\end{tabular} & Yes & Sometimes \\ \hline \begin{tabular}[c]{@{}c@{}}Spectrum access is \\ regulated by\end{tabular} & \begin{tabular}[c]{@{}c@{}}Governmental \\ agencies, typically\\ through auction\end{tabular} & \begin{tabular}[c]{@{}c@{}} Database-driven \\ coordination by \\ private or \\
government companies \\ some bands allow \\ fully open access \end{tabular} \\ \hline \begin{tabular}[c]{@{}c@{}}Spectrum access \\ requires a \\  government license\end{tabular} & Yes & \begin{tabular}[c]{@{}c@{}}No, however, some \\
unlicensed bands allow \\ optional local licensing \end{tabular} \\ \hline \begin{tabular}[c]{@{}c@{}}Exclusive use \\ by single entity\end{tabular} & Yes & \begin{tabular}[c]{@{}c@{}}No. It is shared \\ among anyone, who \\ obeys the rules\end{tabular} \\ \hline \begin{tabular}[c]{@{}c@{}}Well-known standards \\ operating in this band\end{tabular} & 3GPP, LTE, NR & \begin{tabular}[c]{@{}c@{}}LTE Unlicensed, NR-U,\\ Wi-Fi, Bluetooth,\\ LoRa, Sigfox\end{tabular} \\ \hline Specific applications & Cellular networks & \begin{tabular}[c]{@{}c@{}}Local Area Networks,\\ most IoT systems\end{tabular} \\ \hline \end{tabular} \end{table}

{
The contributions of this work lie in providing a comprehensive and structured overview of unlicensed spectrum topics below 8\,GHz. We synthesize information from regulatory documents, existing standards, industry reports, and academic research to offer a coherent narrative that captures both the current state and the evolution of the field. The survey covers legacy technologies alongside emerging and forward-looking solutions, allowing readers to benefit from lessons learned in the past while identifying promising directions for future research. Each subsection concludes with key insights that highlight the main findings and clarify the practical and research implications.
}

\begin{table*}[b]
\caption{Allocation of Sub-8GHz Unlicensed Spectrum}
\label{tab:freq_regions}
\centering
\begin{tabular}{|c|c|c|c|c|c|c|}
\hline
Name  & \begin{tabular}[c]{@{}c@{}}Frequency range \\ MHz\end{tabular}                        & \begin{tabular}[c]{@{}c@{}}Standards and \\ technologies\end{tabular}                              & Applications                                                                                                      & \begin{tabular}[c]{@{}c@{}}{Spectrum}\\ {access}\end{tabular}                & \begin{tabular}[c]{@{}c@{}}{Typical}\\ {Throughput}\end{tabular}                   & Comments                                                                                                                                                                                                                                                                     \\ \hline
TVWS  & \begin{tabular}[c]{@{}c@{}}54 - 72, \\ 76 - 88,\\ 174 - 216,\\ 470 - 694\end{tabular} & \begin{tabular}[c]{@{}c@{}}IEEE 802.11af, \\ IEEE 802.22\end{tabular}                              & \begin{tabular}[c]{@{}c@{}}Fixed Wireless \\ Access (FWA) \\ post-disaster \\ communications, \\ IoT\end{tabular} & \begin{tabular}[c]{@{}c@{}}Database,\\ Sensing\end{tabular}              & 50 Mbps                                                                        & \begin{tabular}[c]{@{}c@{}}This band first introduced database-driven \\ 
spectrum sharing to protect primary users. \\ 
Its limited bandwidth restricts data rates, \\ 
but its long-range propagation makes it \\ 
well suited for rural and underserved areas.
\end{tabular} \\ \hline
ISM   & \begin{tabular}[c]{@{}c@{}}433 - 434.8,\\ 902 - 928,\\ 2400 - 2500\end{tabular}       & \begin{tabular}[c]{@{}c@{}}IEEE 802.11g/n/ax/be,\\ Bluetooth, Zigbee, \\ LoRa, Sigfox\end{tabular} & \begin{tabular}[c]{@{}c@{}}LPWAN, \\ IoT\end{tabular}                                                             & \begin{tabular}[c]{@{}c@{}}Sensing,\\ Arbitrary\end{tabular}             & \begin{tabular}[c]{@{}c@{}}100 kbps, \\ 100-300 \\ Mbps\\ (Wi-Fi)\end{tabular} & \begin{tabular}[c]{@{}c@{}}These bands were the first used for unlicensed \\ 
spectrum. With no primary users to protect, \\ 
devices operate without database regulation, \\ 
relying instead on PHY/MAC coexistence.
\end{tabular}                                             \\ \hline
CBRS  & 3550 - 3700                                                                           & \begin{tabular}[c]{@{}c@{}}LTE, LTE-U, LAA, \\ NR, NR-U\end{tabular}                               & \begin{tabular}[c]{@{}c@{}}Private networks, \\ FWA\end{tabular}                                                  & \begin{tabular}[c]{@{}c@{}}Sensing,\\ Scheduled\end{tabular}             & \begin{tabular}[c]{@{}c@{}}100-300 \\ Mbps\end{tabular}                        & \begin{tabular}[c]{@{}c@{}}Unique to the US, this band employs a novel \\ 
three-tier structure that accommodates both \\ 
unlicensed and primary users. This supports \\ 
the deployment of high-reliability networks.
\end{tabular}                                            \\ \hline
U-NII & 5150 - 7125                                                                           & \begin{tabular}[c]{@{}c@{}}IEEE 802.11ac/ax/be, \\ LTE, LTE-U, LAA, \\ NR, NR-U\end{tabular}       & \begin{tabular}[c]{@{}c@{}}Local, Private, \\ and cellular \\ networks, FWA\end{tabular}                          & \begin{tabular}[c]{@{}c@{}}Database,\\ Sensing,\\ Scheduled\end{tabular} & 3 Gbps                                                                         & \begin{tabular}[c]{@{}c@{}}The newest bands combine database regulation \\
with PHY/MAC coexistence for high-throughput \\
networks. These bands have strong potential for \\
low- to medium-range, high-capacity applications.
\end{tabular}                                       \\ \hline
\end{tabular}
\end{table*}

\section{Unlicensed spectrum overview}
\subsection{Introduction}
The frequency spectrum forms the cornerstone of any wireless technology. Understanding its properties and regulations is crucial for designing effective communication systems. Therefore, prior to delving into any specific technology, it is essential to familiarize ourselves with the nuances of the available spectrum. This work primarily focuses on the spectrum from \ac{FR1}, which extends up to 7125 MHz \cite{FR1REF}, as detailed in Table \ref{tab:freq_regions}. The discussion includes \ac{TVWS}, \ac{ISM} bands, \ac{CBRS}, and finalizes the section with the \ac{UNII} bands.

{Current research trends and industry efforts are particularly focused on the U-NII bands due to their high throughput potential. At the same time, analysing the limitations and lessons learned from TVWS deployments, as well as the advantages of CBRS, provides valuable insights for future system design. The ISM bands are included to provide a more complete picture of the overall spectrum landscape.}

\subsection{TV White Spaces Spectrum}
\subsubsection{Introduction}
\ac{TVWS}, which includes radio spectrum frequencies from 54–72 MHz, 76–88 MHz, 174–216 MHz, and 470–694 MHz \cite{FCCTVWS, SINGAPORETVWS}, was initially allocated for digital broadcasting. Later, several countries recognized its potential for regulated unlicensed applications. Presented in the very high frequency and ultra-high frequency bands, \ac{TVWS} is known for its distinctive qualities, especially its carrier frequency, which offers small attenuation \cite{SIGNALATTENUATION} and extended propagation, resulting in high coverage. These characteristics are particularly important for extending connectivity to sparsely populated rural areas \cite{RURALAFRICA}. In addition, the strong penetration ability of these frequencies \cite{SIGNALATTENUATION2} helps overcome obstacles that block line-of-sight signals. As a result, \ac{TVWS} is well suited for emergency and post-disaster communications \cite{PROFSLIMAPER, INNONET}. With substantial total bandwidth available, \ac{TVWS} can also serve as a core internet backbone in remote regions \cite{WIFROST}. Furthermore, the possibility of unlicensed access to these frequency bands makes them economically attractive \cite{SURVEYTVWS2018}.

The formal development of \ac{TVWS} began in 2002 with the release of a notice of inquiry to assess the feasibility of operations in these bands. Subsequently, the \ac{FCC} initiated investigations into radio frequencies assigned to digital TV broadcasting. These studies revealed that parts of the spectrum remained unused in several regions, presenting opportunities to expand connectivity without interfering with primary broadcasting services. Initial tests in 2007 \cite{TESTSTVWS1} highlighted difficulties in reliably detecting TV signals through spectrum sensing. For example, when moderate TV signals were present on an adjacent channel, some devices failed to consistently detect signals on the primary channel. A second phase of real-world trials in 2008 confirmed that sensing-only devices could not deliver acceptable performance, while devices supported by a \ac{WSDB} demonstrated reliable operation \cite{TESTSTVWS2}. Based on these findings, the Commission concluded that spectrum sensing alone was insufficient to protect licensed users \cite{FCCSECONDMEMOR2008}. Consequently, in 2010, the \ac{WSDB} approach was adopted as the mandatory method for spectrum access, with sensing capabilities retained to enhance \ac{QoS} \cite{FCCSECONDMEMOR2010}.

\subsubsection{Regulations}

Currently, numerous countries adopt the US approach by designating these bands for unlicensed use. Examples include Canada, New Zealand, Singapore, and South Africa \cite{TVWSOVERVIEW2014}. A summary of the relevant regulations is presented in Table \ref{tab:tvws_regulations}.

\begin{table*}[ht]
\caption{TVWS regulations worldwide}
\label{tab:tvws_regulations}
\centering
\begin{tabular}{|c|c|c|c|c|c|ccc|}
\hline

                          &                                                                                                                       &                                                                                    &                                                                                            &                                                                              &                                                                                                 & \multicolumn{3}{c|}{WSDB requirements}                                                                                                                                                                                                                               \\ \cline{7-9} 

\multirow{-2}{*}{Country} & \multirow{-2}{*}{\begin{tabular}[c]{@{}c@{}}TVWS \\ frequencies\\ {[}MHz{]}\end{tabular}}                             & \multirow{-2}{*}{\begin{tabular}[c]{@{}c@{}}Number \\ of \\ channels\end{tabular}} & \multirow{-2}{*}{\begin{tabular}[c]{@{}c@{}}Channel \\ bandwidth\\ {[}MHz{]}\end{tabular}} & \multirow{-2}{*}{Device types}                                               & \multirow{-2}{*}{\begin{tabular}[c]{@{}c@{}}EIRP limit\\ over channel\\ {[}dBm{]}\end{tabular}} & \multicolumn{1}{c|}{\begin{tabular}[c]{@{}c@{}}Access \\ frequency\end{tabular}} & \multicolumn{1}{c|}{\begin{tabular}[c]{@{}c@{}}Location\\  accuracy\end{tabular}} & \begin{tabular}[c]{@{}c@{}}Time of\\  validity\\ {[}h{]}\end{tabular} \\ \hline

USA                                               & \begin{tabular}[c]{@{}c@{}}Fixed:\\ 54 - 72\\ 76 - 88\\ 174 - 216\\ Fixed and \\ portable:\\ 470 - 698,\\ except \\ 614 - 617\end{tabular} & 50                                                                                                         & 6                                                                                                                  & \begin{tabular}[c]{@{}c@{}}Fixed,\\ Mode 1,\\ Mode 2,\\ Sensing,\\ Mobile,\\ Narrowband\end{tabular} & \begin{tabular}[c]{@{}c@{}}Fixed: \\ up to 40\\ Portable: \\ up to 20\end{tabular}                                      & \multicolumn{1}{c|}{60 min}                                                      & \multicolumn{1}{c|}{50 m}                                                         & 2                                                                     \\ \hline
Canada                                            & \begin{tabular}[c]{@{}c@{}}54 - 72\\ 76 - 88\\ 174 - 216,\\ 470 - 608,\\ 657 - 663\end{tabular}                                                & 49                                                                                                         & 6                                                                                                                  & \begin{tabular}[c]{@{}c@{}}Fixed,\\ Mode 1,\\ Mode 2,\\ Mobile\end{tabular}                          & \begin{tabular}[c]{@{}c@{}}Fixed:\\ 36\\ Portable: \\ up to 20\end{tabular}                                             & \multicolumn{1}{c|}{24 hours}                                                                            & \multicolumn{1}{c|}{\begin{tabular}[c]{@{}c@{}}95\%\\ confidence\end{tabular}}                            & 48                                                                    \\ \hline

Singapore                                         & \begin{tabular}[c]{@{}c@{}}174 - 188\\ 195 - 202\\ 209 - 230\\ 470 - 534\\ 614 - 694\end{tabular}                                             & 13                                                                                                         & 7 or 8                                                                                                             & \begin{tabular}[c]{@{}c@{}}Fixed,\\ Mode 1,\\ Mode 2\end{tabular}                                    & \begin{tabular}[c]{@{}c@{}}Fixed: \\ 36\\ Mode 1,2: \\ 20\end{tabular}                                                  & \multicolumn{1}{c|}{6 hours}                                                     & \multicolumn{1}{c|}{50 m}                                                         & 6                                                                     \\ \hline
South Africa                                      & \begin{tabular}[c]{@{}c@{}}470 - 694\\ except \\ 606 - 614\end{tabular}                                                                       & 27                                                                                                         & 8                                                                                                                  & \begin{tabular}[c]{@{}c@{}}Fixed,\\ Nomadic\end{tabular}                                             & \begin{tabular}[c]{@{}c@{}}Fixed:\\ 41.2 (Rural)\\ 36 (Urban)\\ Nomadic: \\ 20\end{tabular}                             & \multicolumn{1}{c|}{\begin{tabular}[c]{@{}c@{}}Fixed:\\ 24 hours\\ Nomadic:\\ 12 hours\end{tabular}}     & \multicolumn{1}{c|}{\begin{tabular}[c]{@{}c@{}}95\%\\ confidence\end{tabular}}                            & 48                                                                    \\ \hline
\end{tabular}
\end{table*}

The \ac{FCC} regulations for \ac{TVWS} devices are specified in Chapter 1, Subchapter A, Part 15, Subpart H \cite{FCCTVWS}. These rules classify devices into four main categories: fixed, mobile, portable (Mode 1 and Mode 2), and sensing. Each category is subject to specific limits on transmit power, allowable frequencies, and database access requirements. For instance, fixed and Mode 2 devices, when operating in a \ac{BS} role, must connect to the \ac{WSDB}, whereas Mode 1 devices cannot function as a \ac{BS} and do not require \ac{WSDB} access. The sensing category applies to devices that use spectrum sensing to access the spectrum. In recent discussions, the \ac{FCC} has also introduced a new category of narrowband devices \cite{NBWSDFCC}. These operate at a maximum bandwidth of 100 kHz and are limited to a 1 percent duty cycle, equivalent to 36 seconds of transmission per hour. They must also follow a channelization scheme that divides each 5.5 MHz block into 55 narrowband carriers, with 6 MHz spacing between channels. Such devices are primarily intended for \ac{IoT} applications.

The frequency range of 470–614 MHz (channels 14–37) is authorized for both fixed and portable devices. In contrast, only mobile devices and fixed devices communicating with other fixed or mobile devices are permitted to operate within the 54–72 MHz (channels 2–4), 76–88 MHz (channels 5–6), and 174–216 MHz (channels 7–13) bands. The 617–699 MHz range is generally reserved for licensed use, although it may be accessed opportunistically when no licensed devices are present. It is also important to note that narrowband and mobile devices are restricted to operation below 602 MHz.

The Canadian TVWS regulations \cite{CANADATVWS} largely mirror those of the US, with some distinctions, such as different access requirements and the time of validity, as well as the absence of a sensing device category.

In Singapore, the TVWS requirements are outlined in the Telecommunication Standards Advisory Committee Technical Specification \cite{SINGAPORETVWS}. This framework categorizes devices into three types: fixed, mode 2, and mode 1. The first two categories can function as master devices and must have geolocation capabilities, whereas mode 1 devices operate without such capabilities.

South Africa's regulatory framework for TVWS \cite{SAFRICATVWS} features two types of devices: fixed and nomadic. This framework offers a relatively limited number of available channels and has the highest potential \ac{EIRP} among the standards under consideration. Deployment of \ac{TVWS} in South Africa has been notably successful, primarily because the country relies heavily on this technology to support data delivery.

The UK regulator Ofcom issued a document \cite{OFCOMTVWS} describing and discussing TVWS in 2016. However, as of 2024, Ofcom has announced that the device authorization framework is no longer available due to the low number of users \cite{OFCOMSTOPPED, OFCOMSTOPPED2}.

\subsubsection{Database spectrum management}

Private companies operate WSDB under the control of an official regulator. There may be multiple databases within a single country. In this case, information exchange between databases is mandatory. TVWS devices must register and regularly consult with the WSDB, which is aligned with their geolocation, to identify accessible channels and the maximum transmit power. Despite the unlicensed use of spectrum, device registration in the WSDB incurs a fee. The database includes details on TV broadcasting stations, coverage, and operating TVWS devices to assign channels to prevent interference. To calculate potential interference to TV broadcasts from unlicensed users, WSDB utilizes propagation models defined by the regulator. To explore the feasibility of deploying TVWS devices across the United States and gain a deeper understanding of white space availability, visit \cite{USCHECKAVAILABILITY}.

\begin{figure} [h]
    \centering
    \begin{tikzpicture}
        \def\spacing{1.5} 
        \def\stemheight{0.55} 
        \def\armwidth{0.3} 

        \node[draw, fill=gray, minimum size=1cm, anchor=center] (square) at (-5,2) {};
        \coordinate (ant1) at (-5, 2.5+\stemheight);
        \draw[thick] (ant1) -- ++(0, -\stemheight); 
        \draw[thick] (ant1) -- ++(-\armwidth, \armwidth); 
        \draw[thick] (ant1) -- ++(\armwidth, \armwidth);  
        \node at (-5,1.2) {\small Node 1};

        \draw[red, thick] (-4.6,2.7+\stemheight) -- (-3.4,3.2+\stemheight);
        
        \node[draw, fill=red!50, minimum size=1cm, anchor=center] (square) at (-3,2.5) {};
        \coordinate (ant1) at (-3, 3+\stemheight);
        \draw[thick] (ant1) -- ++(0, -\stemheight); 
        \draw[thick] (ant1) -- ++(-\armwidth, \armwidth); 
        \draw[thick] (ant1) -- ++(\armwidth, \armwidth);  
        \node at (-3,1.7) {\small Relay};

        \draw[red, thick] (-2.6,3.2+\stemheight) -- (-0.9, 2.2+\stemheight);

        \node[draw, fill=gray, minimum size=1cm, anchor=center] (square) at (-2.5,0) {};
        \coordinate (ant1) at (-2.5, 0.5+\stemheight);
        \draw[thick] (ant1) -- ++(0, -\stemheight); 
        \draw[thick] (ant1) -- ++(-\armwidth, \armwidth); 
        \draw[thick] (ant1) -- ++(\armwidth, \armwidth);  
        \node at (-2.5,-0.8) {\small Node 2};

        \draw[red, thick] (-2.1, 0.8+\stemheight) -- (-0.9, 2.2+\stemheight);

         \node[draw, fill=blue!50, minimum size=1cm, anchor=center] (square) at (-0.5,1.5) {};
        \coordinate (ant1) at (-0.5, 2+\stemheight);
        \draw[thick] (ant1) -- ++(0, -\stemheight); 
        \draw[thick] (ant1) -- ++(-\armwidth, \armwidth); 
        \draw[thick] (ant1) -- ++(\armwidth, \armwidth);  
        \node at (-0.5,0.7) {\small Main Node};

        \draw[blue, thick] (0.1,1.5) -- (0.66, 1.5);

        \node at (1.5,1.5) {\includegraphics[width=1.27cm]{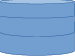}};
        \node at (1.5,0.8) {\small WSDB};

    \end{tikzpicture}
    \caption{Example of a \ac{TVWS} network utilizing a \ac{WSDB}}
    \label{fig:tvws_network}
\end{figure}

Fig.~\ref{fig:tvws_network} illustrates a typical \ac{WSDB} usage scenario in which multiple nodes are connected. Each node queries the \ac{WSDB} through a master node with internet access.

The operation and requirements of the WSDB are critical components in the TVWS regulatory framework. In the United States, these specifications are provided by the FCC in \cite{FCCTVWSDB}. Additionally, the European Electronic Communications Committee has issued guidance for the national implementation of regulatory frameworks related to TVWS databases \cite{ECCREPORTTVWSDB}. The \ac{ETSI} also addresses the topic of TVWS databases in its standard \cite{ETSITVWS}, offering guidelines for their management and operation. The Internet Engineering Task Force developed an open-source protocol for WSDB operation called \ac{PAWS} \cite{PAWSSOURCE}. \ac{PAWS} was successfully used to implement WSDB in several countries, such as India \cite{PAWSINDIA}.

As stated in the WSDB survey paper \cite{DATABASEPAPER16}, database mechanisms face several challenges, primarily related to the unification of access and testing procedures, query languages, and inter-database cooperation both within and between countries.

\subsubsection{Standards}

Two standards are primarily used for TVWS. A comparison is presented in Table \ref{tab:tvws_standard_comparison}:
\begin{itemize}
    \item IEEE 802.22 WRAN: In 2004, the IEEE working group started the development of the first standard tailored for TVWS \cite{IEEE80222FIRST}. It offers a point-to-multipoint architecture between a BS and \ac{UE} and operates over a range of up to 100 km at frequencies from 54 to 862 MHz. The physical layer is optimized for long channel response times and highly frequency-selective channels, supporting signal bandwidths of 6, 7, and 8 MHz. It incorporates \ac{OFDMA}, cognitive radio techniques such as spectrum sensing, and spectrum management \cite{IEEE80222OVERVIEW, IEEE80222STANDARD}.
    \item IEEE 802.11af (Super Wi-Fi): Approved in 2014, this standard allows for both point-to-point and point-to-multipoint networks, using channel reservation and distributed beaconing. Based on the PHY layer of IEEE 802.11ac, it requires downclocking from 40 MHz bandwidth to 6, 7, or 8 MHz, depending on the required channel bandwidth \cite{BOOKIEEE80211af}. It supports up to 256-QAM modulation and initially targets a coverage area of 100 m, extendable to a few kilometers. This standard supports both spectrum sensing and database mechanisms \cite{BELLALTA20161}. 
\end{itemize}

\begin{table}[ht]
\caption{TVWS specific standards}
\label{tab:tvws_standard_comparison}
\centering
\begin{tabular}{|c|c|c|}
\hline
Standard                                                               & IEEE 802.22                                                                                                                                                                        & IEEE 802.11af                                                                                                           \\ \hline
Modulation                                                             & up to 64 QAM                                                                                                                                                                       & up to 256 QAM                                                                                                           \\ \hline
Multiple access                                                        & OFDMA                                                                                                                                                                              & CSMA                                                                                                                    \\ \hline
\begin{tabular}[c]{@{}c@{}}Similar \\ technologies\end{tabular}        & WiMAX or LTE                                                                                                                                                                       & Wi-Fi                                                                                                                   \\ \hline
Typical range                                                          & 5 km                                                                                                                                                                               & 1 km                                                                                                                    \\ \hline
\begin{tabular}[c]{@{}c@{}}Maximum \\ data rate\end{tabular}           & 300 Mbps                                                                                                                                                                           & 26.7 Mbps                                                                                                               \\ \hline
\begin{tabular}[c]{@{}c@{}}Maximum \\ transmit power\end{tabular}      & 27 dBm                                                                                                                                                                             & 20 dBm                                                                                                                  \\ \hline
\begin{tabular}[c]{@{}c@{}}BS and Client \\ equipment\end{tabular}     & \begin{tabular}[c]{@{}c@{}}Hardware for BS \\ and Client is different.\\ Spectrum mask \\ conformance for Client \\ is challenging\end{tabular}                                    & \begin{tabular}[c]{@{}c@{}}Hardware for BS \\ and Client are \\ the same. \\ The difference\\  in software\end{tabular} \\ \hline
\begin{tabular}[c]{@{}c@{}}Upper layers \\ implementation\end{tabular} & \begin{tabular}[c]{@{}c@{}}Existence of security \\ gateway and EPC.\\ It provides \\ authorization and \\ mobility services \\ for the users.\end{tabular} & \begin{tabular}[c]{@{}c@{}}No EPC and \\ security gateway\end{tabular}                                                  \\ \hline
\end{tabular}
\end{table}

The industry predominantly uses the IEEE 802.11af standard for actual device implementation. To the best of our knowledge, Adaptrum Company is the sole developer of the IEEE 802.22 standard modem for their device implementation, whereas others predominantly rely on the IEEE 802.11af standard. For example, modems developed by Mediatek and NICT. Therefore, a valid question concerns the pros and cons of IEEE 802.22 and IEEE 802.11af \cite{TWVSSTANDARDCOMPARISON}. 

\textbf{\underline{In conclusion}}, the appeal of the IEEE 802.11af standard for both customers and device manufacturers primarily lies not in its standard parameters but in its cost-effective and user-friendly solution for expanding wireless network coverage and capabilities.

Instead of utilizing specific standards, some companies have adopted existing technologies, such as LTE, to operate in the TVWS frequency region \cite{WIFROST}. For that, they utilized an LTE modem along with carrier frequency conversion.

\subsubsection{Product overview and applications}

Initially, TVWS technology has garnered interest from various organizations and technology companies, highlighting its potential to enhance connectivity and spectrum utilization. For example, it was promoted by the Dynamic Spectrum Alliance \cite{DYNAMICSPECALLIANCE} - a global organization that advocates for policies, regulations, and practices that promote more efficient spectrum utilization. Microsoft provided several installations to expand internet access in rural villages using TVWS through the Microsoft Airband initiative \cite{MICROSOFTAIRBAND}. In 2013, Google's WSBD implementation was approved for operation in the US. Other applications of TVWS are internet access, connectivity, IoT, sensing and monitoring, and smart cities \cite{COMPREHENSIVEPAPER19}. 

To address the niche in TVWS technology, many vendors have developed specialized devices \cite{FEASIBILITYPAPER19}. A summary of these devices, collected from datasheets, is provided in Table \ref{tab:tvws_product_summary}. The main applications of the considered devices are long-range connectivity, internet backhaul, and IoT.

\begin{table*}[ht]
\caption{TVWS products characteristics summary}
\label{tab:tvws_product_summary}
\centering
\begin{tabular}{|c|c|c|c|c|c|c|}
\hline

\begin{tabular}[c]{@{}c@{}}Distance Covered \\ {[}km{]}\end{tabular} & \begin{tabular}[c]{@{}c@{}}Tx Power \\ {[}dBm{]}\end{tabular} & \begin{tabular}[c]{@{}c@{}}Data Rate \\ (throughput)\\  {[}Mbps{]}\end{tabular}  & MIMO      & \begin{tabular}[c]{@{}c@{}}Channel \\ Bandwidth \\ {[}MHz{]}\end{tabular}         & \begin{tabular}[c]{@{}c@{}}Modulation\\ Coding Rate\end{tabular}  & Topology                                                                     \\ \hline
2 – 40                                                               & 23 – 36                                                       & \begin{tabular}[c]{@{}c@{}}13 – 300\\ (with carrier \\ aggregation)\end{tabular} & 1x1 – 2x2 & \begin{tabular}[c]{@{}c@{}}6 / 7 / 8\\ (with carrier \\ aggregation)\end{tabular} & \begin{tabular}[c]{@{}c@{}}BPSK – 256 QAM\\ 1/2 – 5/6\end{tabular} & \begin{tabular}[c]{@{}c@{}}Point to point\\ Point to multipoint\end{tabular} \\ \hline
\end{tabular}
\end{table*}

Some of these characteristics raise questions about real-world deployment. For example, the authors of \cite{BROADBANDMEASTVWS} conducted a series of experiments to establish a connection between a BS and a client using Adaptrum equipment. The average throughput on a 10 MHz channel at a distance of 6.32 km was 16.27 Mbps. They also reported that the actual downlink throughput was about 70\% of the maximum value specified in the datasheet. Another study reported DL/UL rates of 14/2.2 Mbps at 1 km and 5/1.6 Mbps at 5 km.


\subsubsection{\textbf{\underline{Conclusion}}}
After more than 14 years of development, some conclusions can be drawn about the feasibility of TVWS. The use of this spectrum has not become as widespread as anticipated around 2015. Although a large amount of spectrum is available, the number of active users remains limited in many countries, as illustrated by the United Kingdom. Some exceptions exist, such as South Africa, where TVWS plays a more central role. In the United States, which was among the first to introduce this technology, the number of users also remains relatively small. The main conclusion is that without large-scale deployment, it is difficult to ensure the long-term success and protection of a particular technology.

From a technological perspective, although the carrier frequencies of TVWS offer promising characteristics, current devices cannot simultaneously provide both high throughput and wide coverage. These factors limit the usability of TVWS for broadband connectivity. Furthermore, research and development efforts remain constrained by the relatively small number of users. Nevertheless, TVWS still shows potential for IoT applications, where low data rates are sufficient and reliability over larger areas is important.

Research topics in the field of TVWS include resource allocation, spectrum access, and management \cite{SURVEYTVWS2018, COMPREHENSIVEPAPER19}. One of the challenges occurring when considering communication over TVWS is atmospheric ducting. This phenomenon occurs when atmospheric layers act as a waveguide, allowing waveforms to travel over extended distances \cite{WIVEYATMDUCTING}. This can lead to interference from distant TV broadcast stations that operate hundreds of kilometers away, thereby affecting the quality of the received signal. Investigating strategies to suppress this interference is a practical as well as open research area \cite{ATMDUCTMIDEAST}. {In order to improve efficiency and reliability of WSDB operation, blockchain technology might be employed \cite{blockchain_spectrum_sharing_survey, DSM_blockchain}. This helps to decentralize spectrum sharing mechanisms, making the system more robust, improving reliability, and enabling seamless \ac{ML} integration.}

\subsection{Industrial Scientific Medical bands}

In general, ISM bands span the frequency range from 6.7 MHz to 246 GHz. Initially, this spectrum was designated for non-communication devices to avoid interference with wireless communication systems \cite{ITUREC1056}. The application domains include medicine, science, and industry, encompassing equipment like microwave ovens, medical sterilizers, and magnetic resonance imaging machines. Over time, various wireless communication devices, for example, \ac{SRD} \cite{SRDWORKSHOP} in Europe, began to utilize these frequencies. Nowadays, communication systems are widely available in that part of the spectrum, for example, Wi-Fi technology. There are decisions about incorporating parts of these radio bands into mobile networks, as indicated in the documents such as \cite{LTEUWORKSHOP_CS, NRUMEETING2019, 3GPPREL16SUMMARY, REPORTUNLICENSED}.

The \ac{ITU} Radio Regulations \cite{ITURREG2020} classify ISM band allowances and restrictions across three regions: Region 1 (Europe, Africa, Mongolia, the Middle East, the Persian Gulf, and the Commonwealth of Independent States), Region 2 (the Americas, Greenland, and some Pacific Islands), and Region 3 (Iran, Asian countries, and most of Oceania). ISM regulations distinguish between Type A communication devices, which require approval from the relevant administration to prevent unwanted interference with working equipment within their operating range, and Type B communication devices, which operate without authorization. Regulations are presented in Table \ref{tab:ism_regulations}.

\begin{table*}[ht]
\caption{ISM regulations worldwide}
\label{tab:ism_regulations}
\centering
\begin{tabular}{|c|c|c|c|c|}
\hline

Frequency band {[}MHz{]}                                                             & 433.05 – 434.79                                                                                      & \begin{tabular}[c]{@{}c@{}}863 – 876,\\ 915 – 921\end{tabular}                                      & 902 – 928                                                                                           & 2400 – 2500                                                                                                                    \\ \hline

Primary users                                                                        & \begin{tabular}[c]{@{}c@{}}Radiolocation, \\ Amateur, \\ Earth exploration \\ satellite\end{tabular} & \begin{tabular}[c]{@{}c@{}}Fixed, Mobile, \\ Broadcasting,\\ Amateur, \\ Radiolocation\end{tabular} & \begin{tabular}[c]{@{}c@{}}Fixed, Mobile, \\ Broadcasting,\\ Amateur, \\ Radiolocation\end{tabular} & \begin{tabular}[c]{@{}c@{}}Fixed, Mobile,\\ Mobile-Satellite,\\ Radiolocation, \\ Radiodetermination-\\ Satellite\end{tabular} \\ \hline
\begin{tabular}[c]{@{}c@{}}Aggregated bandwidth\\  {[}MHz{]}\end{tabular}            & 1.74                                                                                                 & 19                                                                                                  & $\leq$26                                                                                            & $\leq$100                                                                                                                      \\ \hline

Number of channels                                                                   & 69                                                                                                   & -                                                                                                   & -                                                                                                   & 14                                                                                                                             \\ \hline
Transmit power {[}dBm{]}                                                             & 10                                                                                                   & 14                                                                                                  & 30                                                                                                  & 36                                                                                                                             \\ \hline

\begin{tabular}[c]{@{}c@{}}Channel occupancy\\ time proportion {[}\%{]}\end{tabular} & 100                                                                                                  & 1                                                                                                   & 2                                                                                                   & 100                                                                                                                            \\ \hline
Device                                                                               & ISM Type A                                                                                           & SRD                                                                                                 & ISM Type B                                                                                          & ISM Type B                                                                                                                     \\ \hline

Availability                                                                         & Region 1                                                                                             & Europe                                                                                              & Region 2                                                                                            & Worldwide                                                                                                                      \\ \hline
Standards                                                                            & \begin{tabular}[c]{@{}c@{}}Custom RF switches\\ and IoT\end{tabular}                                 & \begin{tabular}[c]{@{}c@{}}Zigbee, Sigfox, \\ LoRa\end{tabular}                                     & \begin{tabular}[c]{@{}c@{}}Zigbee, Sigfox, \\ LoRa, \\ IEEE 802.11ah\end{tabular}                   & \begin{tabular}[c]{@{}c@{}}IEEE 802.11b/g/n/ac, \\ Bluetooth, Zigbee\end{tabular}                                              \\ \hline
\end{tabular}
\end{table*}

The 433.05–434.79 MHz ISM band, accessible in Region 1, is notable for its use in embedded applications such as smart homes, radio-controlled switches, toys, and wireless sensors. The specific regulations for this band can be found in \cite{ETSIEMC25to1000}.

In Region 2, frequencies from 902 to 928 MHz are predominantly utilized for \ac{LPWAN}. These frequency allocations are governed by the FCC, as outlined in \cite{FCCISMPOWER}. In Europe, a comparable yet distinct band known as the SRD band ranges from 863 to 876 MHz and from 915 to 921 MHz. ETSI has established regulations for this band \cite{ETSIEMC25to1000}.

The 2.4–2.5 GHz ISM band is notably the most heavily utilized due to the widespread adoption of Wi-Fi, Bluetooth, Zigbee technologies, and cordless phones. This frequency range is divided into 14 channels, each separated by 5 MHz \cite{IEEE80211}. Globally, most countries utilize 13 channels. In the US, channels 12 and 13 are restricted to low-power mode to avoid interference with ground-to-satellite communication systems \cite{GLOBALSTAR}, while Canada permits the use of 12 channels, and Japan uses all 14. 

\textbf{\underline{In conclusion}}, ISM is of particular interest to IoT devices because its limited bandwidth is suitable primarily for low-throughput applications. The transmission power and channel occupation time for such devices are minimal, reducing the need for extensive coordination similar to that of the WSDB. Instead, coexistence issues are primarily addressed at the physical layer of each standard.

\subsection{The 3.55 to 3.7 GHz spectrum: Civil Broadband Radio Services}
CBRS is a 150 MHz unlicensed spectrum band in the US, ranging from 3.55 to 3.7 GHz, and aligns with the LTE and NR n48 band. Discussion to utilize this band began in 2010, and the first baseline standards were released in 2019 \cite{CBRSSURVEY}. This frequency range is governed by FCC rules \cite{FCCCBRSRULES}, which divide the spectrum into a three-tier structure. The top tier is occupied by Federal \ac{IU}, such as satellite services and naval radars, followed by \ac{PAL}, and the lowest tier is occupied by \ac{GAA}. Each channel has a bandwidth of 10 MHz. In the 3550 – 3650 MHz band, each PAL can use a 10 MHz channel with the possibility of channel bonding. However, no more than seven PALs can be assigned simultaneously. The 3650–3700 MHz band is reserved for General GAA but may be extended to 3550–3700 MHz. Coordination between tiers is managed by \ac{SAS}, similar to TVWS databases. SAS registers and authorizes users based on their location and identity, allocates channels for PAL and GAA, and protects incumbent users from interference from lower tiers using information from the database and online sensors. Maximum EIRP limits are set at 23 dBm for UE and at 30 dBm and 47 dBm for lower and higher-power BS, respectively.

As stated in the comprehensive survey \cite{CBRSSURVEY}, resource allocation in such networks is challenging due to several factors, including:
\begin{itemize}
    \item The need for fast and robust detection of IU.
    \item Timing allocation for fair coexistence and optimal spectrum usage.
    \item Effective bandwidth usage, along with avoidance of IU interference.
    \item Achieving high \ac{QoS} in case of sharing the channel between different users.
    \item Preserving the privacy of all-tier users from adversaries.
\end{itemize}

Today, significant technological advances have been made to address these problems. Numerous vendors have developed BS and UE solutions tailored to CBRS \cite{WIACBRS}, which benefit from alignment with cellular network bands. This frequency range supports various technologies, including LTE, 5G-NR, and specialized FWA technologies \cite{TARANAPRODUCT}.

Key applications of CBRS include \cite{WIACBRS}:
\begin{itemize}
    \item Internet Service Providers, who benefit from enabling mobility and providing cheaper last-mile connectivity.
    \item Neutral host networks can achieve extended (and possibly temporary) capacity in stadiums, centers, and hotels.
    \item Industry, which can benefit from deploying private networks in this band or using IoT applications in factories, warehouses or farms.
\end{itemize}

A notable example of the CBRS application is described in the white paper \cite{CBRSHOSPITALITY}, which discusses the process of deploying a private network in the hospitality sector using CBRS bands. The white paper proposes establishing a private LTE network for this scenario. When using PAL bands, benefits include high reliability, high \ac{QoS}, high performance, wide coverage, cost-effectiveness, and mobility support. 

Furthermore, a comparative study in \cite{COMPARECBRSWIFI} analyzes commercial Wi-Fi versus CBRS systems, concluding that LTE/NR technology over CBRS ensures more robust and reliable indoor coverage than Wi-Fi 5, with fewer BS required.

{In order to improve \ac{QoS} for SAS systems, several ideas based on Optimization, \ac{ML} was proposed \cite{CBRS_SAS_optimization, RL_spectrum_sharing}. In addition, such systems support anomaly and malicious attack detection \cite{CBRS_SAS_optimization}. To improve IU detection capabilities, the authors of \cite{IU_detection_AI} proposed a generalized AI-based framework. Articles \cite{ORAN_ML_Radar_detection, ORAN_CBRS_sensing} introduced an Open RAN-based \ac{ML}-assisted framework for fast and robust UI signal detection, for instance, radars.}

The Wireless Innovation Forum \cite{WINNFORUMSTD} plays a key role in supporting the CBRS frequency band. As a standards-developing organization, it not only crafts CBRS standards but also issues informative reports and infographics and oversees device registration. Furthermore, the OnGo Alliance \cite{ONGOALLIANCE}, previously known as the CBRS Alliance, promotes the interests of its members, including implementers and operators, in the use of CBRS band solutions. This alliance consists of manufacturers and end-user companies, promotes CBRS-based technologies, and offers certification.

\begin{table*}[hb]
\caption{Worldwide regulations on 5150 - 5925 MHz spectrum}
\label{tab:unii_1}
\centering
\begin{tabular}{|c|c|ccccclcl|}
\hline

\begin{tabular}[c]{@{}c@{}}Frequency\\  band {[}MHz{]}\end{tabular}        & 5150 – 5250                                                                     & \multicolumn{1}{c|}{5250 – 5350}                                                                     & \multicolumn{1}{c|}{5350 – 5470}                                                             & \multicolumn{1}{c|}{5470 – 5650}       & \multicolumn{1}{c|}{5650 – 5725}      & \multicolumn{2}{c|}{5725 – 5850}                                                             & \multicolumn{2}{c|}{5850 – 5925}                                  \\ \hline
\begin{tabular}[c]{@{}c@{}}Total \\ Bandwidth\\ {[}MHz{]}\end{tabular}     & 100                                                                           & \multicolumn{1}{c|}{100}                                                                                                   & \multicolumn{1}{c|}{120}                                                                                           & \multicolumn{1}{c|}{180}                                     & \multicolumn{1}{c|}{75}                                     & \multicolumn{2}{c|}{125}                                                                                           & \multicolumn{2}{c|}{75}                                                                 \\ \hline

\begin{tabular}[c]{@{}c@{}}EIRP (for US)\\ {[}dBm{]}\end{tabular}          & \begin{tabular}[c]{@{}c@{}}23 /\\ 36\end{tabular}                        & \multicolumn{1}{c|}{23}                                                                         & \multicolumn{1}{c|}{\begin{tabular}[c]{@{}c@{}}Not \\ available\end{tabular}}              & \multicolumn{2}{c|}{30}                                                                         & \multicolumn{2}{c|}{\begin{tabular}[c]{@{}c@{}}33 /\\ 36\end{tabular}}                & \multicolumn{2}{c|}{36}                                      \\ \hline
                                                                           &                                                                               & \multicolumn{8}{c|}{Federal Radiolocation}                                                                                                                                                                                                                                                                                                                                                                                                                                                                                                                                                  \\ \cline{2-10} 
                                                                           & \begin{tabular}[c]{@{}c@{}}Fixed satellite\\  service\end{tabular}            & \multicolumn{3}{c|}{Active spaceborn sensors}                                                                                                                                                                                                                                                                  & \multicolumn{3}{c|}{}                                                                                                                                                            & \multicolumn{2}{c|}{\begin{tabular}[c]{@{}c@{}}Fixed satellite \\ service\end{tabular}} \\ \cline{2-10} 
\multirow{-3}{*}{\begin{tabular}[c]{@{}c@{}}Licensed\\ users\end{tabular}} & \begin{tabular}[c]{@{}c@{}}Amateur Radio \\ Networking \\ system\end{tabular} & \multicolumn{1}{c|}{}                                                                                                      & \multicolumn{2}{c|}{Radionavigation}                                                                                                                                              & \multicolumn{5}{c|}{Amateur radio/satellite}                                                                                                                                                                                                                               \\ \hline

US                                                                         & \begin{tabular}[c]{@{}c@{}}Indoor/Outdoor\\ (U-NII-1)\end{tabular}            & \multicolumn{1}{c|}{\begin{tabular}[c]{@{}c@{}}Indoor/Outdoor\\ DFS/TPC\\ (U-NII-2A)\end{tabular}} & \multicolumn{1}{c|}{\begin{tabular}[c]{@{}c@{}}Not \\ available\\ (U-NII-2B)\end{tabular}} & \multicolumn{2}{c|}{\begin{tabular}[c]{@{}c@{}}Indoor/Outdoor\\ DFS/TPC\\ (U-NII-2C)\end{tabular}} & \multicolumn{2}{c|}{\begin{tabular}[c]{@{}c@{}}Indoor/\\ Outdoor\\ (U-NII-3)\end{tabular}} & \multicolumn{2}{c|}{(U-NII-4)}                                  \\ \hline
Europe                                                                     & Indoor                                                                        & \multicolumn{1}{c|}{\begin{tabular}[c]{@{}c@{}}Indoor/Outdoor\\ DFS/CCA/\\ TPC/CAC\end{tabular}}                           & \multicolumn{1}{c|}{\begin{tabular}[c]{@{}c@{}}Not \\ available\end{tabular}}                                      & \multicolumn{2}{c|}{\begin{tabular}[c]{@{}c@{}}Indoor/Outdoor\\ DFC/CCA/TPC/CAC\end{tabular}}                              & \multicolumn{4}{c|}{Under discussion}                                                                                                                                                                        \\ \hline

Japan                                                                      & Indoor                                                                        & \multicolumn{1}{c|}{\begin{tabular}[c]{@{}c@{}}Indoor\\ DFS/TPC\end{tabular}}                      & \multicolumn{1}{c|}{\begin{tabular}[c]{@{}c@{}}Not \\ available\end{tabular}}              & \multicolumn{2}{c|}{DFS/TPC}                                                                       & \multicolumn{2}{c|}{TPC}                                                                   & \multicolumn{2}{c|}{Not available}                              \\ \hline
\end{tabular}
\end{table*}

\textbf{\underline{In conclusion}}, the future of CBRS is promising. According to the Wireless Innovation Forum infographic \cite{WINNINFOGR}, hundreds of thousands of devices are already being successfully deployed. When comparing this band with TVWS, several differences emerge.
\begin{itemize}
    \item {Technological advantage:} CBRS supports existing LTE and NR bands, allowing easy use of existing equipment. TVWS, however, requires a long and complex product design and development process.
    \item {Market interest:} Strong industry support means more effort is invested in promoting CBRS.
    \item {Bandwidth:} CBRS offers 10 MHz channels, which are better suited for high data rates than smaller TVWS channels.
    \item {PAL bands:} Licensed PAL bands in CBRS ensure \ac{QoS} for critical communication, unlike TVWS.
    \item {Geographic scope:} CBRS was developed particularly for the needs of the US, whereas TVWS has been tested in multiple countries. For example, it is successful in South Africa but not applied in the UK.
\end{itemize}

\subsection{The 5150 to 7125 MHz spectrum: Unlicensed National Information Infrastructure}

This frequency range has recently gained attention for unlicensed use because it offers nearly 2 GHz of spectrum, allowing channel bandwidths of up to 160 MHz. A clear example is the U-NII bands in the US, which cover the frequency range from 5.150 to 7.125 GHz and are divided into eight parts, each with specific rules \cite{FCCUNIIGENERALGUID}. Around the world, rules for these bands vary, but they typically follow the FCC's approach of splitting the spectrum, as discussed later in this section.

Before looking at the regulations, it is important to know some specific rules and mechanisms \cite{DFS_AFC, ETSIBRAN301893} that devices in the U-NII band must follow. These mechanisms protect the main users, such as radar and satellite systems.

\begin{itemize}
    \item \ac{AFC}: A database system that allows devices to operate only on approved frequencies and power levels in certain areas. This protects existing users and is similar to WSDB and SAS.
    \item \ac{CCA}: Devices must first sense the channel before transmitting to avoid using occupied frequencies. This is the carrier-sense part of \ac{CSMA}, for example, used in Wi-Fi. It is mandatory in unlicensed bands in Europe \cite{ETSIBRAN301893} and Japan.
    \item \ac{CAC}: Devices must listen to the channel before sending to check if radar signals are present.
    \item \ac{DFS}: If a radar signal is found, devices must stop using that channel for 30 minutes to avoid interference.
    \item \ac{TPC}: Devices adjust their transmit power to reduce interference, for example, between two nearby Wi-Fi networks. In Europe, devices must be able to reduce power by at least 3~dB.
\end{itemize}

More detailed rules for this part of the spectrum are shown in Tables \ref{tab:unii_1} and \ref{tab:unii_2}.

The device operating modes in this spectrum can be divided into three categories:
\begin{itemize}
    \item Low Power Indoor devices with built-in antennas: For indoor use only. EIRP is limited to 24~dBm for an 80~MHz channel. Example: creating a LAN (e.g., Wi-Fi).
    \item Standard Power devices: For both indoor and outdoor use. They must use geolocation and work with AFC. Example: internet access and general connectivity.
    \item Very Low Power devices: For mobile indoor and outdoor use. EIRP is limited to 14~dBm. Example: devices in vehicles or personal gadgets.
\end{itemize}

The 5G Release~17 specification identifies band n46 (from 5150 to 5925~MHz) for shared spectrum access, treating it as unlicensed \cite{REL17TECHSPEC}. Release~18 adds bands n46, n96, and n102 to the list of unlicensed bands \cite{REL18TECHSPEC}. Detailed global rules for the 5150--5925~MHz band, including spectral masks, power limits, and mobile network restrictions, are given in \cite{3GPP38889}.

In Japan, the 5150--5350~MHz and 5470--5830~MHz ranges are available for unlicensed low-power devices, with power limits depending on the occupied bandwidth \cite{JAPANFREQALLOC, JAPANANNEX}. In Europe, similar bands can be used for local networks \cite{ETSIBRAN301893, EUROPEFREQALLOC}, while Canadian rules are mostly the same as those in the US \cite{CANADAFREQREG}. A key ETSI rule \cite{ETSIBRAN301893} requires the occupied channel bandwidth to be at least 80\% of the nominal channel bandwidth. This is especially relevant for systems with variable bandwidth, such as OFDMA in the uplink.

\begin{table}[hb]
\caption{Worldwide regulations on 5925 – 7125 MHz spectrum}
\label{tab:unii_2}
\centering
\begin{tabular}{|c|cc|}
\hline

Frequency band {[}MHz{]}                                 & \multicolumn{1}{c|}{5925 – 6425}            & 6425 – 7125     \\ \hline
Total Bandwidth {[}MHz{]}                                & \multicolumn{1}{c|}{500}                                            & 700           \\ \hline

                                 & \multicolumn{1}{c|}{Microwave links}        & Mobile        \\ \cline{2-3} 

\multirow{-2}{*}{Licensed users} & \multicolumn{2}{c|}{Fixed - Satellite}                      \\ \hline

US                                                       & \multicolumn{2}{c|}{14 dBm EIRP AFC}                                           \\ \hline

Kingdom Saudi Arabia                                             & \multicolumn{2}{c|}{Indoor/Outdoor AFC}                    \\ \hline
Europe                                                   & \multicolumn{1}{c|}{Indoor/Outdoor}                                 & Not available \\ \hline

Japan                                                    & \multicolumn{1}{c|}{23 dBm EIRP for 20 MHz} & Not available \\ \hline
\end{tabular}
\end{table}

The 5925--7125~MHz band, important for Wi-Fi~6E, has different levels of adoption across countries, with a detailed list available on the official website \cite{COUNTRIESENABLEDWIFI6}. In 5G Release~18, bands n96 (5925--7125~MHz) and n102 (5925--6425~MHz) are also defined for unlicensed access, but their use depends on national rules \cite{REL18TECHSPEC}.  

In the US, the entire band is open for unlicensed low-power use, with a limit of 14~dBm EIRP \cite{USNEW6GHBANDANNOUNC}. A higher power of up to 36~dBm EIRP is allowed when devices connect through AFC \cite{US6GRULES}. Saudi Arabia has also opened the n96 band for unlicensed use and requires AFC to protect primary users such as microwave links \cite{SAUDIAFC6GH}. In Europe, only the 5925--6425~MHz range is approved for unlicensed 5G and local networks \cite{3GPPREL16SUMMARY, EUROPEFREQALLOC}. Japan permits the same range for low-power use, with a limit of about 23~dBm EIRP for a 20~MHz channel \cite{JAPANFREQALLOC}.

{
\textbf{\underline{In conclusion}}, the 6 GHz U-NII band is highly promising, which naturally attracts strong attention from multiple stakeholders. These interests often diverge, as the Wi-Fi community and the cellular industry hold different expectations and priorities for how the band should be used.
}

\subsection{Unified spectrum coordination system} 
\label{sec:unified_spectrum_coordination}
In the previous sections, we introduced three database-based sharing systems: WSDB for TV White Space, SAS for CBRS, and AFC for the 6~GHz U-NII bands. These systems have similar goals: protecting primary users, monitoring the spectrum, registering devices, and assigning resources. Because of this, Ofcom proposed combining them into a single Database Sharing Management System \cite{OFCOMEXTENDEDSHARING}.  

The 2023 report from the Dynamic Spectrum Alliance \cite{DSADSMS} reviews different spectrum-sharing models, including:
\begin{itemize}
    \item Manual Licensed Access: The traditional approach used in most systems.  
    \item Semi-Automated Database-Assisted Coordination: For example, the US approach for 70/80/90~GHz links. First, the database checks if the new user causes interference, and then the FCC manually handles the request.  
    \item Licensed Shared Access: A two-tier authorization model in which the government permits controlled access to part of the spectrum within predefined geographic areas, ensuring no harmful interference to the primary services. Such frameworks are used in several countries \cite{CST2024SpectrumLightLicensing}, most commonly in the 3.7--4.2\,GHz band.

    \item Fully Automated Database Coordination: The WSDB was the first system of this type.  
    \item Dynamic Coordination Databases: Example is the SAS for CBRS, which extends WSDB by adding a three-tier sharing policy.  
\end{itemize}

The report also considers the development of a generalized spectrum coordination system. The main benefits of such a system include simplifying spectrum access and making it more uniform, establishing clear access policies, enhancing spectrum utilization through detailed data, and integrating inputs from various sensors. The report also highlights its potential for unlicensed 5G and 6G use.

Further details on implementing such a system are given in the Communications, Space, and Technology Commission whitepaper \cite{CSTSPECTRUMSHARING}. The authors emphasize the importance of advanced sharing methods for addressing local and time-specific needs, such as event-driven connectivity or seasonal fluctuations in demand. They review current initiatives and present a roadmap for future frameworks, outlining the roles of both local stakeholders (e.g., the private sector and government agencies) and global organizations (e.g., the ITU). A key challenge is building effective cooperation among these stakeholders, since spectrum sharing is not yet a common practice.   

{The authors of \cite{Spectrum_sharing_regulatory} examine licensed access and SAS-based spectrum sharing from a regulatory perspective and propose a dynamic approach that combines the strengths of both regulation and flexibility. This and previous works conclude that future dynamic spectrum sharing systems should dynamically adapt to the market.}

{
To improve the performance and reliability of database-driven spectrum sharing, researchers have proposed several advanced approaches. These include blockchain-based mechanisms for secure and transparent coordination \cite{blockchain_spectrum_sharing_survey}, ML-driven resource management and anomaly detection \cite{CBRS_SAS_optimization, ml_resource_allocation}, and optimization-based resource allocation \cite{CBRS_SAS_optimization}. Spectrum-sharing auctions can also be automated using AI agents, with a corresponding implementation framework presented in \cite{AI_spectrum_auction}. Moreover, with the advent of Large Language Models (LLMs), mobile networks can now adapt to high-level operator commands, as demonstrated in \cite{MX-AI}. LLMs have also shown the ability to analyze network conditions based on operational logs \cite{LLM_analyse_logs}. Together, these capabilities may contribute to more efficient and intelligent spectrum-sharing mechanisms.
}

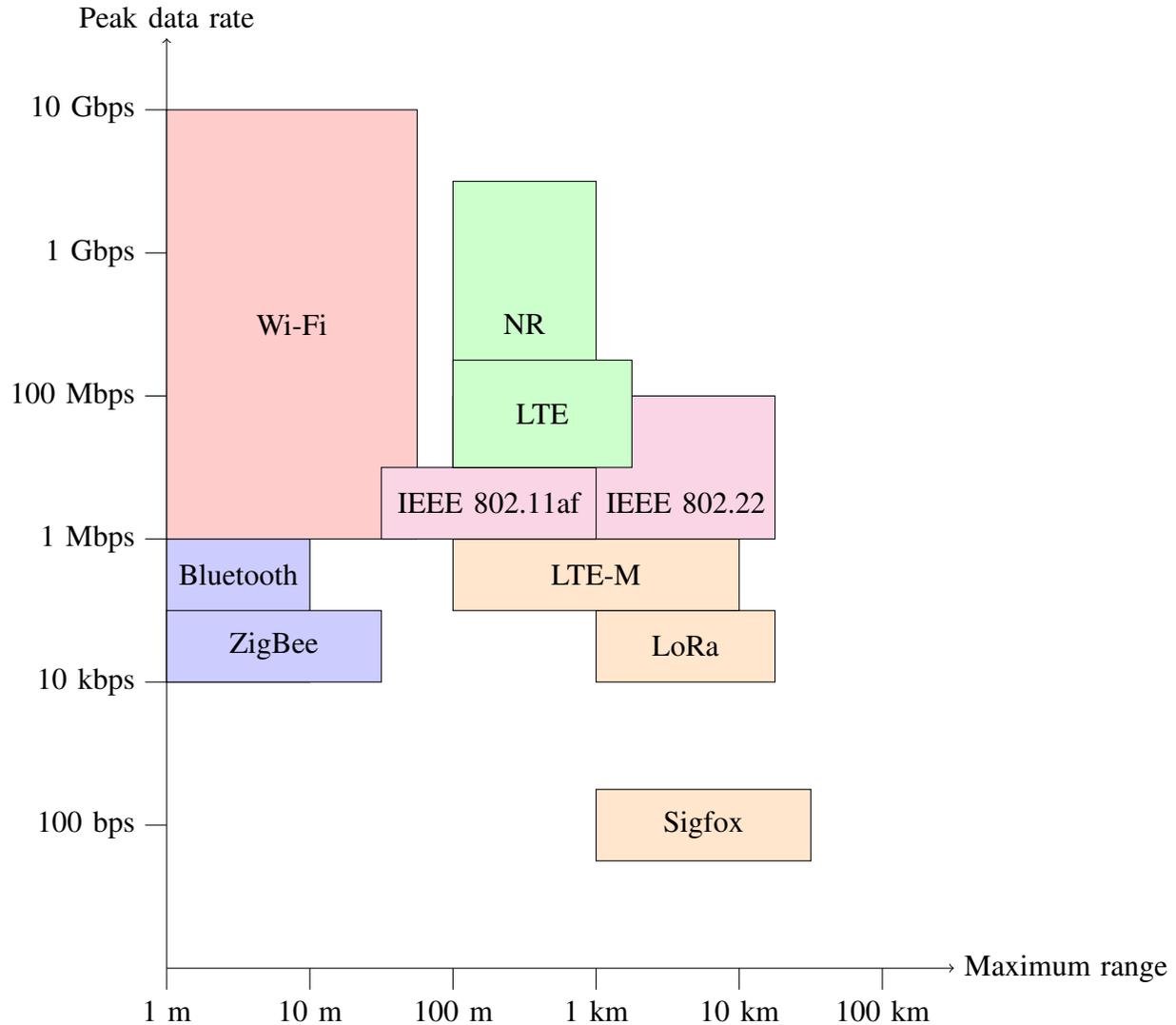
\begin{figure*}[hb]
\centering
\begin{tikzpicture}[scale=0.84]
    \draw[->] (0,0) -- (11,0) node[right] {Maximum range};
    \draw[->] (0,0) -- (0,13) node[above] {Peak data rate};

    \foreach \x/\label in {0/1 m, 2/10 m, 4/100 m, 6/ 1 km, 8/10 km, 10/100 km} {
        \draw[draw=black] (\x, 0) -- (\x, -0.3) node[below] {\label};
    }
    \foreach \y/\label in {2/ 100 bps, 4/ 10 kbps, 6/1 Mbps, 8/100 Mbps, 10/1 Gbps, 12/10 Gbps} {
        \draw[draw=black] (0, \y) -- (-0.3, \y) node[left] {\label};
    }

    \draw[fill=blue!20, draw=black] (0,4) rectangle (2,6);
    \node at (1,5.5) {Bluetooth};
    \draw[fill=blue!20, draw=black] (0,4) rectangle (3,5);
    \node at (1.5,4.5) {ZigBee};

    \draw[fill=red!20, draw=black] (0, 6) rectangle (3.5, 12);
    \node at (1.75, 9) {Wi-Fi};

    \draw[fill=orange!20, draw=black] (6,1.5) rectangle (9,2.5);
    \node at (7.5,2) {Sigfox};
    \draw[fill=orange!20, draw=black] (6,4) rectangle (8.5,5);
    \node at (7.25,4.5) {LoRa};
    \draw[fill=orange!20, draw=black] (4,5) rectangle (8,6);
    \node at (6,5.5) {LTE-M};

    \draw[fill=magenta!20, draw=black] (4, 6) rectangle (8.5, 8);
    \node at (7.25, 6.5) {IEEE 802.22};
    \draw[fill=magenta!20, draw=black] (3, 6) rectangle (6, 7);
    \node at (4.5, 6.5) {IEEE 802.11af};

    \draw[fill=green!20, draw=black] (4, 7) rectangle (6, 11);
    \node at (5, 9) {NR}; 
    \draw[fill=green!20, draw=black] (4, 7) rectangle (6.5, 8.5);
    \node at (5.25, 7.75) {LTE};
\end{tikzpicture} 
\caption{A rough estimate of technology use cases by the range and required throughput. Green rectangles represent cellular technologies, and red and magenta indicate Wi-Fi-based technologies. Blue rectangles correspond to short-range devices, while orange rectangles represent long-range, low-throughput devices.}
\label{fig:techologies}
\end{figure*}

\textbf{\underline{In conclusion}}, automated database coordination systems are well-developed and widely used in practice. However, their ability to improve spectrum efficiency remains limited, as their primary goal is to protect primary users from interference. When primary users are absent, and multiple networks coexist in the same area, additional PHY and MAC-layer coexistence techniques are needed. These techniques will be discussed in Section~\ref{coexistence_subsection}.

\subsection{{Extension for licensed spectrum}}
{
Extending the proposed unlicensed spectrum-sharing framework to licensed environments is conceptually straightforward. Core components such as spectrum sensing, database-assisted operation, and system maintenance remain directly applicable without modification. Moreover, existing SAS deployments already demonstrate the feasibility of managing multi-tier access with strict protection mechanisms, which aligns naturally with licensed spectrum scenarios.

The research community has also shown that short-term and dynamic licensing is achievable within tiered systems, as demonstrated in several works on online spectrum allocation and platform-based or auction-driven sharing models \cite{online_spectrum_sharing, platform_spectrum_sharing, auction_spectrum_sharing}.

Consequently, 6G systems are expected to integrate multiple spectrum-sharing regimes. These mechanisms will need to differentiate QoS guarantees for diverse classes of users, including governmental or primary users, long-term licensed users, short-term licensed users, and unlicensed users. Such developments would indicate a gradual convergence of licensed and unlicensed spectrum regimes within a unified 6G sharing framework.
}

\section{An Overview of Wireless Technologies Employing Unlicensed Spectrum}
\subsection{Introduction}

The frequency spectrum is the foundation of wireless communication, with various technologies serving as the structures built upon it. The market provides a wide variety of these technologies. This section provides an introductory overview, focusing on key aspects of unlicensed spectrum use, including operating frequency, throughput, and range.  
Fig.~\ref{fig:techologies} shows a rough estimate of the two most important practical parameters.

As shown, current technologies already cover most deployment scenarios. The next step is to focus on their coexistence to maximize their full potential. The final subsection will discuss coexistence issues related to interference. Such interference may occur within a single technology in the same network (intra-cell or inter-cell) or when two different standards share the same band, for example, Wi-Fi and NR-U.

\subsection{Low rate technologies}

\begin{table*}[hb]
\centering
\caption{Summary of LPWAN Device Characteristics}
\label{tab:lpwan_summary}
\begin{tabular}{|c|c|c|c|c|c|c|c|}
\hline

                                                                             & Bluetooth                                                          & Zigbee                                                                            & LoRa                                                                                          & Sigfox                                                            & NB-IoT                                                 & LTE-M                                                  & Wi-Fi HaLow                                                             \\ \hline
Supported by                                                                 & \begin{tabular}[c]{@{}c@{}}Special\\ Interest\\ Group\end{tabular} & \begin{tabular}[c]{@{}c@{}}Connectivity \\ Standards \\ Alliance\end{tabular}     & LoRa Alliance                                                                                 & SIGFOX                                                            & 3GPP                                                   & 3GPP                                                   & IEEE                                                                    \\ \hline

\begin{tabular}[c]{@{}c@{}}Operating \\ Frequencies\\ {[}MHz{]}\end{tabular} & \begin{tabular}[c]{@{}c@{}}2400 – \\ 2480\end{tabular}             & \begin{tabular}[c]{@{}c@{}}868 – 868.6  \\ 902 – 928  \\ 2400 – 2483\end{tabular} & \begin{tabular}[c]{@{}c@{}}863 – 870 \\ 902, 915 – 928  \\ 915 – 928 \\ 2400 – 2500\end{tabular} & \begin{tabular}[c]{@{}c@{}}868 – 868.6  \\ 902 – 928\end{tabular} & 450 – 1980                                             & 450 – 1980                                             & \begin{tabular}[c]{@{}c@{}}863 – 876\\ 902 – 928\\ 915 – 921\end{tabular} \\ \hline
\begin{tabular}[c]{@{}c@{}}Licensed or \\ Unlicensed\end{tabular}             & Unlicensed                                                         & Unlicensed                                                                        & Unlicensed                                                                                    & Unlicensed                                                        & Licensed                                               & Licensed                                               & Unlicensed                                                              \\ \hline

\begin{tabular}[c]{@{}c@{}}Range \\ (Rural) {[}m{]}\end{tabular}             & 100                                                                & 20                                                                                & 18k                                                                                           & 50k                                                               & 15k                                                    & 11k                                                    & \textgreater 1k                                                         \\ \hline
\begin{tabular}[c]{@{}c@{}}Range \\ (Urban) {[}m{]}\end{tabular}             & 100                                                                & 20                                                                                & 5k                                                                                            & 10k                                                               & 10k                                                    & \textless 11k                                          & 1k                                                                      \\ \hline

\begin{tabular}[c]{@{}c@{}}Peak data rate \\ (UL) {[}bps{]}\end{tabular}          & 2M                                                                 & 250k                                                                              & 50k                                                                                           & 600                                                               & 79k                                                    & 1M                                                     & 7.8M                                                                  \\ \hline
\begin{tabular}[c]{@{}c@{}}Peak data rate \\ (DL) {[}bps{]}\end{tabular}          & 2M                                                                 & 250k                                                                              & 50k                                                                                           & 600                                                               & 106k                                                   & 1M                                                     & 7.8M                                                                  \\ \hline

\begin{tabular}[c]{@{}c@{}}Encryption \\ standard\end{tabular}               & AES-128                                                            & AES-128                                                                           & AES-128                                                                                       & AES-128                                                           & \begin{tabular}[c]{@{}c@{}}3GPP\\ 128 – 256\end{tabular} & \begin{tabular}[c]{@{}c@{}}3GPP\\ 128 – 256\end{tabular} & WPA3                                                                    \\ \hline
\begin{tabular}[c]{@{}c@{}}Organization\\ or standard\end{tabular}             & IEEE 802.15.1                                                      & IEEE 802.15.4                                                                     & Proprietary                                                                                   & Proprietary                                                       & 3GPP                                                   & 3GPP                                                   & IEEE 802.11ah                                                           \\ \hline
\end{tabular}
\end{table*}

This section reviews LPWAN technologies that enable long-range, low-throughput communication for up to 10 years on a single battery \cite{LRWRTSURVEY}. Most of these technologies operate in ISM bands between 433~MHz and 2.5~GHz. However, as noted in \cite{BOOKLPWANFORIOTANDM2M}, TVWS can also be used for LPWAN, particularly with LoRa and Sigfox. Although Bluetooth and Zigbee support much shorter ranges, they are included here for completeness. 

Today, the leading LPWAN technologies are NB-IoT, LoRa, and Sigfox \cite{STATISTAMARKETSHARE}. Therefore, we will not cover less common options such as Ingenu, Weightless, Telensa, DASH7, NB-Fi, or IQRF in detail. For a broader overview, see \cite{LPWANBROADSURVEY}, which provides comparison tables and application discussions for most LPWAN technologies. The key features of unlicensed use are summarized in Table~\ref{tab:lpwan_summary}.

Bluetooth \cite{BLUETOOTHSTANDARD} is a widely used technology for short-range wireless connections, offering data rates up to 2~Mbps over distances of about 10~meters. It is mainly used in devices such as headphones, wireless mice, and other peripherals. The standard is maintained by the Bluetooth Special Interest Group \cite{BLUETOOTHALLIANCE}.

Zigbee \cite{ZIGBEESTANDARD} is an alternative to Bluetooth, designed for different use cases. It is mainly used to build small radio mesh networks for automation and short-range IoT applications. Thanks to its mesh structure and longer range, Zigbee can reach up to 100~meters, compared to the shorter range of Bluetooth. Zigbee networks can also scale to thousands of devices, while Bluetooth typically supports about twenty \cite{COMPAREBLZIG}. This makes Zigbee more suitable for industrial applications in limited areas. The Connectivity Standards Alliance (formerly the Zigbee Alliance) manages certification and promotes adoption of the standard \cite{ZIGBEEALLIANCE}.

LoRa, defined in the LoRaWAN standard by ITU \cite{LORAPATENTNEW}, enables low-power, long-distance communication, with ranges up to 15~km in rural areas. It provides sufficient throughput for IoT use cases and supports connectivity in rural communities. LoRa offers a balance between Zigbee’s shorter range and Sigfox’s very low data rate. Semtech Corporation is one of the founding members of the LoRa Alliance \cite{LORAALLIANCE}.

Sigfox \cite{SIGFOXPATENT} provides the longest communication range among LPWAN technologies, reaching up to 40~km, but with the lowest data rate of about 600~bps. Unlike mesh-based systems, Sigfox utilizes a cellular-like network structure, where base stations, managed by local providers, serve specific areas. It has extensive global coverage, with more details available on its official website \cite{SIGFOXWEB}.

NB-IoT and LTE-M (or LTE-MTC) \cite{EMTCLTE} use licensed spectrum and are included here for completeness, as they form the basis for enhanced Mobile Broadband (eMTC) in 5G \cite{GSMANB5G}. Developed by 3GPP in LTE Release~13, they enable LPWAN operation over existing cellular networks. While both offer higher throughput than unlicensed LPWAN technologies, they also involve higher deployment costs. NB-IoT provides broader coverage and supports very dense device deployments, while LTE-M is better suited for less dense areas where higher throughput is required \cite{COVCOMPNBIOTLTEM}.

IEEE~802.11ah, also known as Wi-Fi HaLow, was developed for IoT devices and supports low-power connections over distances of about 1~km. The study in \cite{IEEE80211ahHALOPAPER} analyzes the scalability, throughput, latency, and energy efficiency of the system under bidirectional traffic. A broader overview and comparison with related technologies are provided by the Wi-Fi Alliance \cite{WIFIHALOWOVERVIEW}. Overall, Wi-Fi HaLow offers higher data rates and stronger security than most LPWAN technologies.

To estimate LPWAN deployment costs, see the report by the Communication, Space \& Technology Commission of Saudi Arabia \cite{CITCREPORT}. The report reviews LPWAN deployment in the Kingdom and provides detailed cost estimates for different technologies, from network setup to device manufacturing and deployment. It concludes that Sigfox is the cheapest option, followed by LoRa, while LTE-IoT is the most expensive.  

\textbf{\underline{In summary}}, LPWAN technologies serve specific niches. Operating mainly in traditional ISM bands, they are not suited for modern high-throughput applications such as FWA, and their use cases generally do not overlap. As a result, there is little need for a unified spectrum access framework.

\subsection{Wi-Fi}
\subsubsection{Introduction and historical perspective}

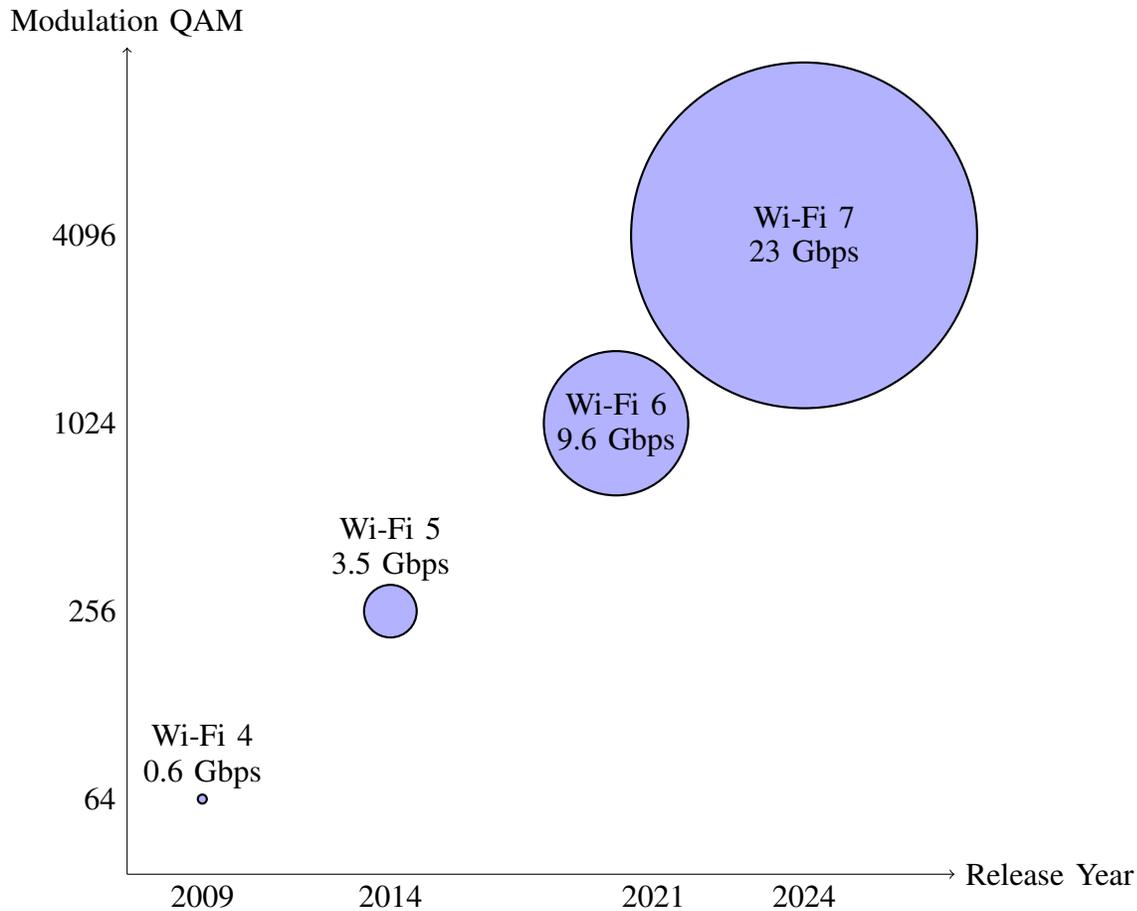
\begin{figure}[ht]
\centering
\begin{tikzpicture}[scale = 0.55]
\draw[->] (0,0) -- (11,0) node[above] {Release Year};
\draw[->] (0,0) -- (0,11) node[above] {Modulation QAM};

\foreach \x/\lbl in {1/2009, 3.5/2014, 7/2021, 9/2024} {
    \node[below] at (\x,0) {\lbl};
}
\foreach \y/\lbl in {1/64,3.5/256,6/1024,8.5/4096} {
    \node[left] at (0,\y) {\lbl};
}

\draw[thick, fill=blue!30] (9,8.5) circle (2.3);
\node at (9,8.75) {Wi-Fi 7};
\node at (9,8.25) {23 Gbps};

\draw[thick, fill=blue!30] (6.5,6) circle (0.96);
\node at (6.5,4.75) {Wi-Fi 6};
\node at (6.5,4.25) {9.6 Gbps};

\draw[thick, fill=blue!30] (3.5,3.5) circle (0.35);
\node at (3.5,4.6) {Wi-Fi 5};
\node at (3.5,4.1) {3.5 Gbps};

\draw[thick, fill=blue!30] (1,1) circle (0.06);
\node at (1,1.85) {Wi-Fi 4};
\node at (1,1.335) {0.6 Gbps};

\end{tikzpicture}
\caption{Evolution of Wi-Fi standards. Circle radius corresponds to the peak throughput.}
\label{fig:wifievolution}
\end{figure}

Wi-Fi, based on the IEEE~802.11 standard, is widely used in setting up \ac{WLAN}. Studying its evolution is crucial to understanding how unlicensed spectrum sharing has developed, how current systems operate, and what the future of license-exempt wireless networking may look like. Fig.~\ref{fig:wifievolution} illustrates the evolution of the Wi-Fi standard.

Wi-Fi standards began with 802.11b, 802.11a, and 802.11g, and later advanced to 802.11n and 802.11ac, also known as Wi-Fi~4 and Wi-Fi~5. The evolution continued with IEEE~802.11ax \cite{WIFI6TUTORIAL}, leading to the current IEEE~802.11be standard \cite{WIFIORG7GEN}. Several specialized variants also exist, including IEEE~802.11af for TVWS, IEEE~802.11ah for LPWAN, IEEE~802.11ay for millimeter-wave (WiGig), and IEEE~802.11bb for light-based (Li-Fi) communication.  

In this section, the focus will be on standards designed for WLAN. For context, we briefly review Wi-Fi~4 and Wi-Fi~5, and then discuss Wi-Fi~6 and Wi-Fi~7 in more detail. The other standards mentioned are either covered in earlier sections or are beyond the scope of this paper. Table~\ref{tab:wifi_comparison} summarizes the key characteristics of Wi-Fi standards.

\begin{table*}[ht]
\caption{Wi-Fi standards comparison}
\label{tab:wifi_comparison}
\centering
\begin{tabular}{|c|c|c|c|c|}
\hline

Name                                                                      & Wi-Fi 4                                                                                                                      & Wi-Fi 5                                                                                                           & Wi-Fi 6                                                                                & Wi-Fi 7                                                                                                                  \\ \hline
Standard                                                                  & IEEE802.11n                                                                                                                  & IEEE 802.11ac                                                                                                     & IEEE 802.11ax                                                                          & IEEE 802.11be                                                                                                            \\ \hline

Release                                                                   & Ocotber, 2009                                                                                                                & January, 2014                                                                                                     & May, 2021                                                                              & January, 2024                                                                                                            \\ \hline
\begin{tabular}[c]{@{}c@{}}Frequency band \\ {[}GHz{]}\end{tabular}       & 2.4, 5                                                                                                                      & 5                                                                                                                 & 2.4, 5, 6                                                                            & 2.4, 5, 6                                                                                                              \\ \hline

\begin{tabular}[c]{@{}c@{}}Bandwidth \\ {[}MHz{]}\end{tabular}            & 20 / 40                                                                                                                      & 20 / 40 / 80 / 160                                                                                                & 20 / 40 / 80 / 160                                                                     & 20 / 40 / 80 / 160 / 320                                                                                                 \\ \hline
\begin{tabular}[c]{@{}c@{}}Maximum \\ data rate\\ {[}Gbps{]}\end{tabular} & 0.6                                                                                                                          & 3.5                                                                                                               & 9.6                                                                                    & \begin{tabular}[c]{@{}c@{}}23\\ (single link over 6 GHz)\\ 36\\ (aggregated links over \\ 2.4, 5 and 6 GHz)\end{tabular} \\ \hline

Modulation                                                                & up to 64 QAM                                                                                                                 & up to 256 QAM                                                                                                     & up to 1024 QAM                                                                         & up to 4096 QAM                                                                                                           \\ \hline
\begin{tabular}[c]{@{}c@{}}Symbol Time \\ {[}us{]}\end{tabular}           & 3.2                                                                                                                          & 3.2                                                                                                               & 12.8                                                                                   & 12.8                                                                                                                     \\ \hline

\begin{tabular}[c]{@{}c@{}}Cyclic Prefix \\ {[}us{]}\end{tabular}         & 0.8                                                                                                                          & 0.8 / 0.4                                                                                                         & 0.8 / 1.6 / 3.2                                                                        & 0.8 / 1.6 / 3.2                                                                                                          \\ \hline
\begin{tabular}[c]{@{}c@{}}Multiple Antenna \\ Multiuser\end{tabular}     & MIMO                                                                                                                         & \begin{tabular}[c]{@{}c@{}}MU MIMO DL\\ (up to 4 users)\end{tabular}                                              & \begin{tabular}[c]{@{}c@{}}MU MIMO \\ DL and UL\\ (up to 8 users)\end{tabular}         & \begin{tabular}[c]{@{}c@{}}MU MIMO DL and UL\\ (up to 8 users)\end{tabular}                                              \\ \hline

OFDMA                                                                     & No                                                                                                                           & No                                                                                                                & Used                                                                                   & Used                                                                                                                     \\ \hline
\begin{tabular}[c]{@{}c@{}}Introduced PHY \\ features\end{tabular}        & \begin{tabular}[c]{@{}c@{}}Short guard\\  interval (400 ns)\end{tabular}                                                     & \begin{tabular}[c]{@{}c@{}}Dynamic channel \\ bandwidth management, \\ unified beamforming \\ method\end{tabular} & Target Wake Time                                                                       & Triggered uplink access                                                                                                  \\ \hline
\begin{tabular}[c]{@{}c@{}}Introduced MAC\\  mechanisms\end{tabular}      & \begin{tabular}[c]{@{}c@{}}Frame Aggregation, \\ Block Acknowledgment, \\ Reverse Direction, \\ Greenfield mode\end{tabular} & \begin{tabular}[c]{@{}c@{}}Enhanced Frame\\  Aggregation, \\ TXOP sharing\end{tabular}                            & \begin{tabular}[c]{@{}c@{}}Enhanced Frame\\  Aggregation, \\ BSS coloring\end{tabular} & \begin{tabular}[c]{@{}c@{}}512 Compressed block \\ Ack, Multiple resource \\ units to single station\end{tabular}        \\ \hline
\end{tabular}
\end{table*}

Wi-Fi~4 introduced several important enhancements to the MAC layer \cite{IMPACTPHYMACTHROUGHPUT, WIFI6TUTORIAL}. Multi-level frame aggregation combined multiple packets into a single frame, reducing header overhead and improving throughput. The reverse direction mechanism improved the use of \ac{TXOP}, enabling bidirectional communication within the same time slot originally designed for unidirectional traffic.  

For compatibility, the frame structure normally begins with a legacy header, even though it is inefficient. With Greenfield mode, however, this requirement is removed when no legacy devices are present, allowing a more optimized frame structure. Finally, \ac{QoS} was significantly improved by introducing differentiated channel access parameters for traffic categories such as background, best effort, video, and voice, ensuring resources are allocated according to priority.

The MAC layer in Wi-Fi~5 \cite{IMPACTPHYMACTHROUGHPUT, WIFI6TUTORIAL} was enhanced alongside new PHY features such as \ac{MU}~MIMO. Frame aggregation was further improved, allowing larger frame sizes and supporting higher throughput.  

Wi-Fi security has also evolved from the early Wired Equivalent Privacy protocol to successive \ac{WPA} versions, with WPA3 introduced in 2018 \cite{WIFIWPA3WSEC}. WPA3 is mandatory for new devices and fixes many vulnerabilities of earlier standards. However, it still faces challenges, including rogue AP and evil twin attacks, ARP spoofing, downgrade attacks, and side-channel exploits \cite{WIFIWPA3SYSTREVIEW}.

The Wi-Fi~6 protocol, as described in \cite{SURVEYWIFI6}, introduces several important features. One is OFDMA, which extends conventional OFDM by dividing groups of subcarriers among different users within the same band. This makes Wi-Fi more similar to LTE and NR, though working in unlicensed bands still creates challenges for channel estimation and scheduling. Another key feature is MU-MIMO, which increases capacity in both uplink and downlink by applying beamforming to serve multiple users at the same time. Wi-Fi~6 also includes \ac{TWT}, which saves device battery and improves spectral efficiency by letting the BS schedule exact wake-up times. Finally, it brings a MAC improvement called basic service set coloring, where each BS assigns a unique number in the PHY header. By checking this number, a device can quickly see if the frame is relevant, lowering decoding overhead and reducing interference by allowing overlapping transmissions from different BS.

Wi-Fi~6 shows clear differences from earlier standards. In \cite{WIFI6FIRSTLOOK}, the authors carried out practical tests to measure throughput, latency, energy efficiency, and security. The results showed that with OFDMA, Wi-Fi~6 can achieve up to 3.2 times higher throughput for both DL and UL in dense user scenarios, and reduce latency by a factor of five. However, this improvement comes with up to six times higher power consumption. The study also found that unencrypted \ac{TWT} can be exposed to different malicious attacks.

Wi-Fi~7 has recently been introduced with several major improvements. At the PHY level, its features are described in \cite{ROHDESHCWARZ}, while the Wi-Fi Alliance gives an overview in \cite{WIFIORG7GEN}. One key feature is deterministic latency, which is important for industrial uses of unlicensed spectrum. Modern Wi-Fi~7 systems can also support high-mobility scenarios, making them closer to cellular technologies than earlier Wi-Fi systems. To further increase performance, Wi-Fi~7 introduces multi-link operation, where a single user can use up to three links across different frequency bands, similar to carrier aggregation. This can raise throughput to about 36~Gbps and also improve transmission reliability \cite{WIFI7MLO}.

{Wi-Fi~8 is expected to further reduce latency, optimize power consumption, increase spectral efficiency, and enhance AP cooperation to reduce interference \cite{wifi8}.}

{textbf{\underline{In conclusion}}, the evolution of Wi-Fi provides a valuable example of how shared-spectrum technologies can mature over time. The most recent releases demonstrate a clear shift toward cellular-like capabilities, including MU-MIMO, scheduled communication, and enhanced mobility support, while still preserving the unique Wi-Fi characteristics.}

\subsection{LTE in unlicensed spectrum}

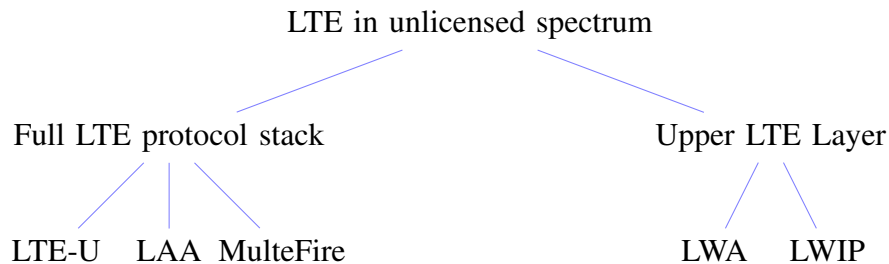
\begin{figure}[h] 
\centering
\begin{tikzpicture}[scale = 0.9]
  \node {LTE in unlicensed spectrum} [sibling distance=40mm, blue!50]
    child {node [black] {Full LTE protocol stack} [sibling distance=15mm]
        child {node [black] {LTE-U}}
        child {node [black] {LAA}}
        child {node [black] {MulteFire}}
    }
    child {node [black] {Upper LTE Layer} [sibling distance=15mm]
      child {node [black] {LWA}}
      child {node [black] {LWIP}}
    };
\end{tikzpicture}
\caption{Types of LTE utilizing unlicensed spectrum}
\label{fig:lte}
\end{figure}

LTE systems, widely deployed globally \cite{BIGCOMM}, support multiple modes of operation in unlicensed spectrum bands. Although Wi-Fi 6 has integrated OFDMA and moved closer to cellular-like performance, important differences remain. Cellular networks primarily function in licensed bands, ensuring exclusive spectrum access and more predictable resource allocation. They also provide wider coverage and stronger security features. In contrast, Wi-Fi networks often struggle to support high mobility and achieve deterministic latency, whereas cellular systems are designed to meet these requirements reliably. The use of LTE in unlicensed spectrum is represented by different technological approaches, as outlined in \cite{WHITEPAPERINTELLTE}. These can be divided into two categories: solutions adopting the complete LTE protocol stack, and those employing only the upper layers. An overview of these variants is shown in Fig.\ref{fig:lte} and Table\ref{tab:lte_variants}.

\begin{table*}[ht]
\caption{LTE Standards Using Unlicensed Spectrum}
\label{tab:lte_variants}
\centering
\begin{tabular}{|c|c|c|c|}
\hline

Name                                                                  & LTE-U                                                                                                                            & LAA                                                                                                            & MulteFire                                                                                                                                                           \\ \hline
\begin{tabular}[c]{@{}c@{}}Introduced in \\ 3GPP release\end{tabular} & 12                                                                                                                               & 13                                                                                                             & Not standardized                                                                                                                                                     \\ \hline

\begin{tabular}[c]{@{}c@{}}Supporting\\ Organization\end{tabular}     & LTE-U Forum                                                                                                                      & 3GPP                                                                                                           & \begin{tabular}[c]{@{}c@{}}Alliance for \\ Private Networks\end{tabular}                                                                                            \\ \hline
Operation Mode                                                        & DL only                                                                                                                          & DL and UL                                                                                                      & DL and UL                                                                                                                                                           \\ \hline

\begin{tabular}[c]{@{}c@{}}Coexistence\\ Mechanisms\end{tabular}      & \begin{tabular}[c]{@{}c@{}}Carrier selection,\\ on-off switching,\\ Carrier Sensing Adaptive \\ Transmission \end{tabular} & \begin{tabular}[c]{@{}c@{}}Carrier selection\\ LBT\end{tabular}                                                & LBT                                                                                                                                                                 \\ \hline
\begin{tabular}[c]{@{}c@{}}Spectrum \\ Utilization\end{tabular}       & Licensed and unlicensed                                                                                                          & Licensed and unlicensed                                                                                        & Unlicensed                                                                                                                                                          \\ \hline

\begin{tabular}[c]{@{}c@{}}Compatibility \\ with 3GPP\end{tabular}    & Yes                                                                                                                              & Yes                                                                                                            & No                                                                                                                                                                  \\ \hline
Purpose                                                               & \begin{tabular}[c]{@{}c@{}}Enhancing the throughput \\ of mobile networks\end{tabular}                                           & \begin{tabular}[c]{@{}c@{}}Enhancing the throughput\\ of mobile networks\end{tabular}                          & \begin{tabular}[c]{@{}c@{}}Establishing\\  standalone network\end{tabular}                                                                                          \\ \hline

Comments                                                              & \begin{tabular}[c]{@{}c@{}}Initially developed \\ technology faced limited \\ global acceptance\end{tabular}                     & \begin{tabular}[c]{@{}c@{}}A globally accepted \\ technology widely \\ prevalent in the 4G market\end{tabular} & \begin{tabular}[c]{@{}c@{}}A fusion of Wi-Fi's simplicity\\  and LTE's performance, \\ served as the foundation \\ for developing \\ solutions in NR-U\end{tabular} \\ \hline
\end{tabular}
\end{table*}

The first category is represented by LTE-U, \ac{LAA}, and MulteFire technologies \cite{EXTLTEINTOUNLICENSED}.

LTE-U and LAA were initially proposed to enhance system capacity by enabling user-plane data transmission over unlicensed spectrum, while maintaining control-plane signaling in licensed bands. The fundamental distinction \cite{LAAVSLTEU} between the two approaches lies in their release versions and regulatory alignment. LTE-U, standardized in Release 12, targeted markets without mandatory \ac{LBT} requirements, such as the United States and China. By contrast, LAA, introduced in Release 13, incorporated LBT functionality to comply with international regulatory frameworks, thereby achieving global applicability. This design decision reflects LAA's primary objective: ensuring fair coexistence with incumbent Wi-Fi systems. To this end, the LTE frame structure was modified \cite{LTELICEXSPECTRUM}, with certain subframes replaced by signals consistent with CSMA-based protocols, including channel sensing and reservation.

Initially, LAA supported only downlink (DL) data transmission in unlicensed bands. With Release 14, uplink (UL) operation was added via scheduled transmissions, in which the eNodeB controls the start of transmission. This enhancement is referred to as extended LAA (eLAA) \cite{3GPPREL1416OVER}. Release 15 further advanced the concept with further enhanced LAA (feLAA), which enables autonomous UL operation in license-exempt spectrum and introduces additional options for starting and ending \ac{PUSCH} transmissions \cite{ETSIFELAA}.

MulteFire \cite{MULTEFIREPAPER} represents another approach, enabling a standalone LTE network entirely within unlicensed spectrum. It achieves at least twice the throughput of Wi-Fi 5 \cite{MULTEFIREQUALPRES}. However, its main limitation is the lack of backward compatibility with existing 3GPP standards.

The second category of LTE use in unlicensed spectrum includes solutions that combine the upper LTE layers with the lower layers of Wi-Fi. Representative technologies in this group are the \ac{LWIP} tunnel and \ac{LWA} \cite{LWALWIP}. \ac{LWIP} performs packet division at the Internet Protocol layer, while \ac{LWA} is more flexible, as it operates at the Packet Data Convergence Protocol layer \cite{PRIVATE5GCONCEPTSARCH}. Compared to LWIP, \ac{LWA} provides higher throughput gains when deployed in existing networks and requires fewer changes to the infrastructure to enable support \cite{LTELWAINTELPAPER}.

{\textbf{\underline{In conclusion}}, the evolution of LTE in unlicensed bands represents an important early step toward cellular technologies operating in license-exempt spectrum. Approaches such as LAA, LTE-U, and MulteFire demonstrated that unlicensed cellular access is feasible and provided practical experience that contributed to the development of 5G NR-U.}

\subsection{5G-NR in unlicensed spectrum}
Building on LTE developments, 5G New Radio (NR) represents the most recent mobile communication standard \cite{SYRVEY5GANDBEYOND}. Release 15 introduced an ultra-lean design that transmits reference and broadcast signals only when needed, in contrast to LTE’s continuous signaling. Another key feature is a flexible frame structure, which allows transmissions to begin at any OFDM symbol rather than only at slot boundaries, thereby improving spectrum efficiency. While originally designed for licensed spectrum, these improvements are equally beneficial in unlicensed deployments.

\begin{figure*}[th]
    \centering
    \includegraphics[width=1\textwidth]{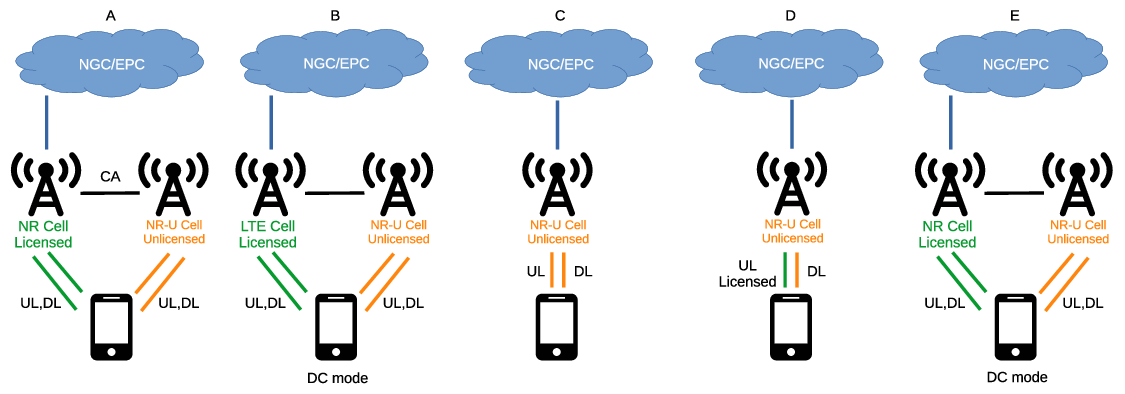}
    \caption{NR-U deployment options}
    \label{fig:nr_deployment}
\end{figure*}

In Release 16, NR-U was introduced as a version of NR specifically adapted for unlicensed spectrum \cite{5GNRREL16OVER}. NR-U integrates LAA features with MulteFire's standalone operation, enabling use across sub-8~GHz and higher frequency bands. Beyond unlicensed operation, it also introduces two methods for mitigating TDD interference: Cross-Link Interference (CLI), addressing short-range issues such as high-power downlink transmissions affecting uplink, and Remote Interference Management (RIM), designed for long-range interference caused by atmospheric ducting \cite{CLIRIM}. Both mechanisms can be applied through beamforming or by adjusting the timing between uplink and downlink transmissions.

Finally, \ac{DSS} supports coexistence between LTE and NR by allowing them to share the same carrier, which further promotes efficient use of shared spectrum resources.

Release 17 added support for \ac{URLLC} in license-exempt spectrum below 7.125~GHz \cite{5GNRURLLC17OVER}. Key enhancements include harmonizing uplink data transfer and feedback procedures between URLLC and NR-U, as well as improving channel access mechanisms to ensure both listen-before-talk (LBT) compliance and low-latency performance \cite{5GNRURLLC151617}. Another milestone in Release 17 is the introduction of non-terrestrial network (NTN) support, which extends NR-U connectivity to satellite and other space-based networks.

Release 18 continues this evolution by targeting sidelink communication — direct device-to-device interaction without base station involvement — in unlicensed spectrum. This extension, designed for scenarios such as vehicle-to-vehicle communication, further broadens the applicability of NR-U \cite{5GNRREL18SIDELINKULIC, 5GNRURLLC18OVER}.

The NR framework defines five deployment scenarios, labeled A through E. These range from standalone unlicensed operation with an NR-U base station (BS) to various configurations that combine NR, NR-U, or LTE BS across licensed and unlicensed spectrum \cite{NRGENERALDESCRIPTION}. To further enhance throughput and reliability, two types of connectivity are supported. The first is \ac{DC}, which enables data exchange between a user equipment (UE) and two BSs, with one designated as the primary. The second is \ac{CA}, which allows data transmission over two carriers, either contiguous or non-contiguous. Unlike DC, carrier aggregation can be applied in all deployment scenarios \cite{NRGENERALDESCRIPTION}. These options are illustrated in Fig.~\ref{fig:nr_deployment}.

\begin{itemize}
    \item Scenario A includes the DC of the user with an NR and an NR-U BS, with CA between them.
    \item Scenario B consists of the DC of the user with LTE BS operating over licensed bands and NR-U over unlicensed bands.
    \item Scenario C is defined by a standalone NR-U operation.
    \item Scenario D features an NR BS operating in the shared spectrum for downlink and exclusively in the licensed spectrum for uplink.
    \item Scenario E involves the DC of the user with NR and NR-U BS, similar to Scenario A. The key difference is that in Scenario E, both BSs are connected to the core network.
\end{itemize}

The NR-U has several challenges \cite{5GNRCHALLENGESEVALUATION}: 

\begin{itemize}
    \item Waveform design must comply with ETSI’s minimum bandwidth rule, which requires the actual signal bandwidth to occupy at least 80\% of the assigned channel \cite{ETSIBRAN301893}. This condition may be challenging to meet in uplink OFDMA scenarios with few active devices, as dedicated carriers may not utilize sufficient bandwidth. While sending dummy data is a simple solution, it is not an energy-efficient approach.
    \item Operating over multiple channels increases the complexity of applying the LBT mechanism and of distinguishing between primary and secondary channels.  
    \item The energy detection threshold must be carefully set to ensure fair coexistence with Wi-Fi (further details are provided in the following section).  
    \item Since NR-U scheduling for downlink and uplink is asynchronous, system timing uncertainties must be taken into account. For example, after data transmission, acknowledgment signals may be delayed if the channel remains occupied.  
    \item The design of \ac{HARQ} is particularly challenging in the unlicensed spectrum due to the probabilistic nature of channel access.  
    \item Initial access procedure parameters, such as the waiting time before a user retries channel access, must be carefully tuned to ensure fair coexistence with Wi-Fi.  
\end{itemize}

{\textbf{\underline{In conclusion}}, NR-U offers a flexible and promising way to operate in unlicensed bands, but its deployment in real products has been delayed. One reason is that mobile operators are cautious about using unlicensed spectrum because sharing it may affect their network performance. Another factor is that NR-U is designed to coexist fairly with Wi-Fi, which means following LBT rules and behaving similar to Wi-Fi devices. This creates additional constraints and reduces the freedom operators usually have in licensed bands \cite{ITU_WRC23_FinalActs_2023}. The situation also differs across regions. For example, in the US the 6\,GHz band is mainly for Wi-Fi, while in Europe the band is split between Wi-Fi and licensed use. These differences, along with concerns from the Wi-Fi community that increased cellular use of unlicensed bands may redirect spectrum opportunities away from Wi-Fi, add to the tension surrounding NR-U and help explain its slower adoption.}

\subsection{Coexistence between technologies in unlicensed bands} \label{coexistence_subsection}
Coexistence between these technologies can be organized into several categories, as shown in Table~\ref{tab:coexistence}, which illustrates how interference challenges can be addressed separately.

\begin{table*}
\caption{Coexistence of Various Technologies Within the Same Spectrum}
\label{tab:coexistence}
\centering
\rotatebox{90}{
\begin{tabular}{|c|c|c|c|c|c|c|c|c|c|c|l|}
\hline
                                                         & \begin{tabular}[c]{@{}c@{}}IEEE \\ 802.11af\end{tabular} & \begin{tabular}[c]{@{}c@{}}IEEE \\ 802.22\end{tabular}                                                    & \begin{tabular}[c]{@{}c@{}}IEEE \\ 802.11ah\end{tabular}                                          & Bluetooth                                                                     & Zigbee                                                                                & LoRa                                                                                  & Sigfox & Wi-Fi                                                                                                 & LTE                                                                                                   & NR                                                                                                    & \multicolumn{1}{c|}{FWA} \\ \hline
\begin{tabular}[c]{@{}c@{}}IEEE \\ 802.11af\end{tabular} &                                                          &                                                                                                           &                                                                                                   &                                                                               &                                                                                       &                                                                                       &        &                                                                                                       &                                                                                                       &                                                                                                       &                          \\ \hline
\begin{tabular}[c]{@{}c@{}}IEEE \\ 802.22\end{tabular}   & \cellcolor[HTML]{9AFF99}470 - 694                        &                                                                                                           &                                                                                                   &                                                                               &                                                                                       &                                                                                       &        &                                                                                                       &                                                                                                       &                                                                                                       &                          \\ \hline
\begin{tabular}[c]{@{}c@{}}IEEE \\ 802.11ah\end{tabular} &                                                          &                                                                                                           &                                                                                                   &                                                                               &                                                                                       &                                                                                       &        &                                                                                                       &                                                                                                       &                                                                                                       &                          \\ \hline
Bluetooth                                                &                                                          & \begin{tabular}[c]{@{}c@{}} \end{tabular} &                                                                                                   &                                                                               &                                                                                       &                                                                                       &        &                                                                                                       &                                                                                                       &                                                                                                       &                          \\ \hline
Zigbee                                                   &                                                          &                                                                                                           & \cellcolor[HTML]{FFCCC9}\begin{tabular}[c]{@{}c@{}}902 – 928\\ 915 – 921\end{tabular}             & \cellcolor[HTML]{FFCE93}\begin{tabular}[c]{@{}c@{}}2400 –\\ 2500\end{tabular} &                                                                                       &                                                                                       &        &                                                                                                       &                                                                                                       &                                                                                                       &                          \\ \hline
LoRa                                                     &                                                          &                                                                                                           & \cellcolor[HTML]{FFCCC9}\begin{tabular}[c]{@{}c@{}}902 – 928\\ 915 – 921\end{tabular}             & \cellcolor[HTML]{FFCE93}\begin{tabular}[c]{@{}c@{}}2400 –\\ 2500\end{tabular} & \cellcolor[HTML]{FFCE93}\begin{tabular}[c]{@{}c@{}}2400 –\\ 2500\end{tabular}         &                                                                                       &        &                                                                                                       &                                                                                                       &                                                                                                       &                          \\ \hline
Sigfox                                                   &                                                          &                                                                                                           & \cellcolor[HTML]{FFCCC9}\begin{tabular}[c]{@{}c@{}}863 – 876\\ 902 – 928\\ 915 – 921\end{tabular} &                                                                               & \cellcolor[HTML]{FFCCC9}\begin{tabular}[c]{@{}c@{}}902 – 928\\ 915 – 921\end{tabular} & \cellcolor[HTML]{FFCCC9}\begin{tabular}[c]{@{}c@{}}902 – 928\\ 915 – 921\end{tabular} &        &                                                                                                       &                                                                                                       &                                                                                                       &                          \\ \hline
Wi-Fi                                                    &                                                          &                                                                                                           &                                                                                                   & \cellcolor[HTML]{CBCEFB}\begin{tabular}[c]{@{}c@{}}2400 –\\ 2500\end{tabular} & \cellcolor[HTML]{CBCEFB}\begin{tabular}[c]{@{}c@{}}2400 –\\ 2500\end{tabular}         & \cellcolor[HTML]{CBCEFB}\begin{tabular}[c]{@{}c@{}}2400 –\\ 2500\end{tabular}         &        &                                                                                                       &                                                                                                       &                                                                                                       &                          \\ \hline
LTE                                                      & \cellcolor[HTML]{9AFF99}470 – 694                        & \cellcolor[HTML]{9AFF99}470 – 694                                                                         &                                                                                                   &                                                                               &                                                                                       &                                                                                       &        & \cellcolor[HTML]{9AFF99}\begin{tabular}[c]{@{}c@{}}470 – 694\\ 3550 – 3700\\ 5150 – 7125\end{tabular} &                                                                                                       &                                                                                                       &                          \\ \hline
NR                                                       & \cellcolor[HTML]{9AFF99}470 – 694                        & \cellcolor[HTML]{9AFF99}470 – 694                                                                         &                                                                                                   &                                                                               &                                                                                       &                                                                                       &        & \cellcolor[HTML]{9AFF99}\begin{tabular}[c]{@{}c@{}}470 – 694\\ 3550 – 3700\\ 5150 – 7125\end{tabular} & \cellcolor[HTML]{9AFF99}\begin{tabular}[c]{@{}c@{}}470 – 694\\ 3550 – 3700\\ 5150 – 7125\end{tabular} &                                                                                                       &                          \\ \hline
FWA                                                      & \cellcolor[HTML]{9AFF99}470 – 694                        & \cellcolor[HTML]{9AFF99}470 – 694                                                                         &                                                                                                   &                                                                               &                                                                                       &                                                                                       &        & \cellcolor[HTML]{9AFF99}\begin{tabular}[c]{@{}c@{}}470 – 694\\ 3550 – 3700\\ 5150 – 7125\end{tabular} & \cellcolor[HTML]{9AFF99}\begin{tabular}[c]{@{}c@{}}470 – 694\\ 3550 – 3700\\ 5150 – 7125\end{tabular} & \cellcolor[HTML]{9AFF99}\begin{tabular}[c]{@{}c@{}}470 – 694\\ 3550 – 3700\\ 5150 – 7125\end{tabular} &                          \\ \hline
\end{tabular}
}
\end{table*}

\subsubsection{Coexistence between short-range devices and LoRa}  

Yellow table entries indicate potential interference between short-range devices and long-range LoRa systems operating at 2.4~GHz. The need to address this issue is debatable, as it becomes significant only when such devices are in close physical proximity; signal attenuation at greater distances generally prevents interference. A straightforward mitigation strategy is to operate LoRa in lower-frequency bands, thereby avoiding overlap with short-range devices.  

The study in \cite{INTERFLORABLUETOOTH} examines the coexistence of Bluetooth and LoRa within the 2.4~GHz band. Based on experimental measurements, the authors suggest that adjusting transmission parameters can improve coexistence between technologies.  

\subsubsection{Coexistence between short-range devices and Wi-Fi}  

The blue table entries indicate interference between Wi-Fi networks and short-range devices, which represents a challenge due to their co-location and the demand for high throughput. Prior studies, such as \cite{WIFIONZIGBEE}, \cite{WIFIBLUETOOTH}, and \cite{WIFILORA}, investigate different aspects of this issue: the effect of Wi-Fi on Zigbee, mutual interference between Wi-Fi and Bluetooth, and coexistence between Wi-Fi and LoRa, respectively. Proposed solutions include adjusting transmission parameters (e.g., code rate or transmission power) or modifying device placement to reduce interference and improve coexistence.  

\subsubsection{Coexistence of LPWAN technologies}  

Table entries highlighted in orange represent interference between LPWAN technologies. Ensuring coexistence within this group is important, as these devices are specifically designed for long-range, low-power communication scenarios. A central question is whether explicit coexistence mechanisms are necessary. Given the typically low duty cycles of LPWAN devices, one might assume that collisions are unlikely. However, in the case of massive deployments, collisions should be expected.  

To quantify this, the authors of \cite{LPWANCOLLISION} analyzed the time–frequency occupation of LoRa, Sigfox, and IQRF devices sharing the same channel. Assuming each device transmits 140 messages per day, they calculated the probability of collisions involving two or more devices. Their results indicate that with approximately 320 devices, the probability of collision is 0.5, and with around 1300 devices, it rises above 0.95. Nevertheless, the study does not examine the impact of these collisions on system performance.  

To enable efficient coexistence, LPWAN devices rely primarily on two modulation techniques: \ac{UNB} and \ac{SS}. The study in \cite{LPWANUNBaSS} analyzes these approaches and shows how each modulation type influences downlink and uplink performance, with the uplink being more strongly affected. The results indicate that UNB networks may provide up to five times higher capacity compared to SS. However, this advantage is difficult to exploit in practice due to the high implementation costs. In contrast, SS offers greater throughput and imposes a substantially lower power spectral density requirement on the transmitter. The authors conclude that fair coexistence between the two modulation techniques is feasible when the number of active devices remains limited.  

A related study in \cite{CHALUNLLPWANAREN} reached similar conclusions by evaluating network performance as a function of the number of devices and the corresponding throughput across a given coverage area. Their findings suggest that LPWAN networks are suitable primarily for low-rate, sparsely distributed sensing applications.  

To address coexistence challenges in LPWAN, the authors of \cite{LPWANCOEXISTENCESURVEY} propose modifications spanning multiple protocol layers. At the intra-technology level, they recommend improvements such as adaptive data rate MAC scheduling with variable slot sizes, along with corresponding PHY-layer enhancements like fast and low-power synchronization. They also argue that the CSMA-CA protocol, while bandwidth-efficient for single-device transmissions, does not fully satisfy LPWAN requirements due to the large number of devices transmitting small data packets. As an alternative, they propose \ac{TDMA}-based protocols or joint time–frequency scheduling.  

For inter-technology coexistence, the authors emphasize the importance of reliable detection techniques, potentially leveraging additional sources of information to identify other technologies. They further suggest mechanisms such as more robust coding schemes, limiting packet sizes, and implementing fine-grained TDMA approaches. However, they note that such mechanisms are not yet widely supported in existing LPWAN devices.  

\textbf{\underline{In conclusion}}, LPWAN technologies encounter distinct coexistence challenges. Their use cases typically involve dense device deployments with wide-area coverage and low data rates. However, the literature indicates that current LPWAN systems can only accommodate a limited number of devices within the same area. Furthermore, the diversity of standards and the absence of effective intra-technology coordination mechanisms make coexistence particularly difficult. As a result, coexistence strategies for LPWAN often differ from, or even contradict, those applied in high-throughput systems—for example, relying on TDMA-based MAC protocols instead of CSMA/CA, or incorporating database-driven regulation. Importantly, LPWAN technologies do not generally overlap with high-throughput systems, as they occupy lower frequency bands. This separation allows their coexistence issues to be addressed independently from those of high-throughput cases.

\subsubsection{Coexistence between high throughput technologies}

The green table entries correspond to interference between devices designed for high-throughput applications, such as Wi-Fi and unlicensed variants of LTE and NR-U. Since high throughput inherently requires maximizing spectrum utilization, collisions are inevitable. One possible mitigation approach is the use of database-driven coordination systems in certain frequency bands. While this aspect has already been discussed in Section~\ref{sec:unified_spectrum_coordination}, the present section focuses on PHY/MAC-level coexistence mechanisms.

The first method of spectrum sharing relies on detecting and utilizing unoccupied frequency channels, a process known as \ac{DCS}. This approach is employed in unlicensed LTE, NR-U, and similarly in Wi-Fi. However, when multiple users must operate on the same channel, additional techniques are required.  

The most widely adopted method to enable coexistence in shared spectrum is channel contention, implemented through mechanisms such as LBT in LTE and NR-U, and CSMA/CA in Wi-Fi. Other techniques, involving adjustments in time, frequency, or power, also exist but are less prominent and are typically specific to cellular-based systems \cite{COEXISTENCECELLULARWIFI}.  

For inter-cell coexistence, early Wi-Fi standards adopted the \ac{DCF}, which integrates CSMA/CA with an exponential backoff algorithm. In this scheme, the CSMA process begins by sensing the channel to determine whether it is occupied. The sensing is performed over a period known as the \ac{DIFS} and relies on two methods: \ac{CS}, which detects and decodes Wi-Fi preambles, and \ac{ED}, which identifies the presence of undecodable signals. Each method applies a distinct detection threshold, with CS typically operating at a sensitivity level about 20~dB lower than that of ED.

If the channel is free, devices perform a backoff, i.e., counting down for a period known as the contention window. The contention window size is randomly sampled from a defined range of minimum and maximum sizes. If the spectrum remains free, transmission begins. If the spectrum is busy at any moment, devices pause their countdown until it is free again. 
If the counter reaches zero and the channel is free, the \ac{RTS} \ac{CTS} mechanism might be employed to avoid collisions. The transmitting device sends an RTS frame to the destination. Upon receiving this frame, the destination responds with a CTS frame, notifying all other devices that the channel is now occupied for a specific duration. Following this, data transmission occurs during the TXOP, and the sender expects to receive an acknowledgment frame. In case of a failed transmission, the contention window size doubles.

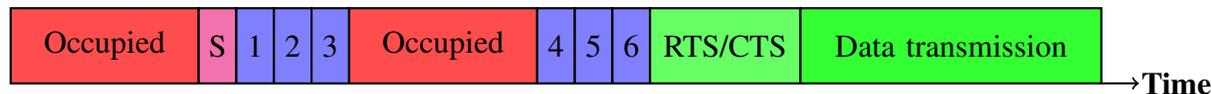
\begin{figure*}[hb]
\centering

\begin{tikzpicture}

\draw[thick,->] (0,0) -- (15,0);
\node at (15.5, 0) {\textbf{Time}};

\fill[red!40] (0,0) rectangle (2.5,1);
\draw[thick] (0,0) rectangle (2.5,1); 
\node at (1.25, 0.5) {Occupied};

\fill[magenta!40] (2.5,0) rectangle (3,1);
\draw[thick] (2.5,0) rectangle (3,1); 
\node at (2.75, 0.5) {S};

\fill[blue!20] (3,0) rectangle (3.5,1);
\draw[thick] (3,0) rectangle (3.5,1); 
\node at (3.25, 0.5) {1};

\fill[blue!20] (3.5,0) rectangle (4,1);
\draw[thick] (3.5,0) rectangle (4,1); 
\node at (3.75, 0.5) {2};

\fill[blue!20] (4,0) rectangle (4.5,1);
\draw[thick] (4,0) rectangle (4.5,1); 
\node at (4.25, 0.5) {3};

\fill[red!40] (4.5,0) rectangle (7,1);
\draw[thick] (4.5,0) rectangle (7,1); 
\node at (5.75, 0.5) {Occupied};

\fill[blue!20] (7,0) rectangle (7.5,1);
\draw[thick] (7,0) rectangle (7.5,1); 
\node at (7.25, 0.5) {4};

\fill[blue!20] (7.5,0) rectangle (8,1);
\draw[thick] (7.5,0) rectangle (8,1); 
\node at (7.75, 0.5) {5};

\fill[blue!20] (8,0) rectangle (8.5,1);
\draw[thick] (8,0) rectangle (8.5,1); 
\node at (8.25, 0.5) {6};

\fill[green!25] (8.5,0) rectangle (10.5,1);
\draw[thick] (8.5,0) rectangle (10.5,1); 
\node at (9.5, 0.5) {RTS/CTS};

\fill[green!30] (10.5,0) rectangle (14.5,1);
\draw[thick] (10.5,0) rectangle (14.5,1); 
\node at (12.5, 0.5) {Data transmission};

\end{tikzpicture}  
\caption{CSMA/CA protocol example}
\label{fig:csmaca}
\end{figure*}

When the counter reaches zero, and the channel remains available, the \ac{RTS}/\ac{CTS} mechanism can be employed to further reduce the risk of collisions. In this procedure, the transmitting device first sends an RTS frame to the intended receiver. Upon reception, the receiver replies with a CTS frame, informing all neighboring devices that the channel will be reserved for a specified duration. Data transmission then occurs within the allocated TXOP, after which the sender expects an acknowledgment frame. If the transmission fails, the contention window size is doubled. An illustration of this procedure is shown in Fig.~\ref{fig:csmaca}. 

In this example, the channel was initially occupied but became free during the sensing interval, indicated by the magenta rectangle. Following this, a random backoff was applied, with six periods selected. During the countdown, the channel became occupied again, causing the countdown to pause and resume only once the channel was free. When the counter reached zero, the RTS/CTS procedure was triggered, after which data transmission began.  

Wi-Fi 4 introduced a newer version of the DCF, called the Hybrid Coordination Function, which utilizes Enhanced Distributed Channel Access to prioritize packets by adjusting access parameters, such as minimum and maximum contention window sizes and transmit times. On top of that, DIFS were replaced by Arbitration Infter-Frame Spacing with variable size. This enhancement improves \ac{QoS} for various data categories, including background, best effort, video, and voice \cite{IMPACTPHYMACTHROUGHPUT}.

In LTE and NR-U, devices equipped with LBT capability perform spectrum sensing for a minimum of 16~µs and determine channel occupancy based on an ED threshold, which is technology-specific and may be subject to regulatory requirements. This procedure is commonly referred to as clear channel assessment (CCA). Once free spectrum is detected, devices execute a random backoff process before initiating transmission, similar to Wi-Fi. In LTE-LAA, when a device is ready to transmit, a channel reservation signal is employed to maintain access until the beginning of the next LTE slot. At this point, data transmission occurs \cite{LTELICEXSPECTRUM}.  

In the process of defining LBT for LTE-LAA \cite{CATLBT, 5GNRCHALLENGESEVALUATION}, four categories (CAT) of LBT were introduced:  
\begin{itemize}
    \item \textbf{CAT1-LBT:} Immediate channel access without performing LBT.  
    \item \textbf{CAT2-LBT:} LBT with carrier sensing but without random backoff.  
    \item \textbf{CAT3-LBT:} LBT with carrier sensing and random backoff, using a fixed contention window size.  
    \item \textbf{CAT4-LBT:} LBT with carrier sensing and exponential random backoff, where the contention window size is variable, similar to the CSMA/CA procedure.  
\end{itemize}

LTE-LAA employs CAT4-LBT, whereas NR-U supports CAT1, CAT2, and CAT4 \cite{NRLBT, 5GNRCHALLENGESEVALUATION}. Different categories can be applied to different types of transmissions within a channel occupancy time (COT). For example, CAT1 is typically used for back-to-back transmissions over short intervals, while frame-based LBT in CAT2 ensures reliable delivery of critical control information, such as discovery frames. CAT4, on the other hand, is employed by both the BS and the UE to initiate data transmissions \cite{OFNOWHITESHEET}. 

We now turn to specific examples of coexistence. It is important to note that in the following discussion, we use the term *fair*; however, there is no universally accepted definition of fairness in the context of coexistence among different technologies \cite{COEXISTENCECELLULARWIFI}. {Below, we summarize several widely used definitions of fairness:
\begin{itemize}

    \item \textbf{Jain's fairness index}~\cite{jain_fairness_definition} is a commonly used measure defined as
    \begin{equation}
        J(x_1, x_2, \ldots, x_n)
        = \frac{\left( \sum_{i=1}^{n} x_i \right)^2}
               {n \sum_{i=1}^{n} x_i^2}
        = \frac{\bar{x}^{\,2}}{\overline{x^2}},
        \label{eq:jain_index}
    \end{equation}
    where $n$ is the number of users and $x_i$ denotes the performance metric
    (typically throughput) of the $i^{\text{th}}$ user. When applied to throughput, it is often referred to as the throughput fairness index. The value
    ranges from $\frac{1}{n}$ (worst case) to $1$ (best case), and reaches its
    maximum when all users receive identical allocations. However, the index
    implicitly assumes that all users require similar throughput levels.
    In coexistence scenarios where networks have heterogeneous throughput demands,
    equalizing throughputs is not meaningful, and the index may therefore
    misleadingly penalize allocations that are optimal for each network.

    \item \textbf{Airtime fairness} is another metric derived from the same
    principle. Instead of throughput, $x_i$ represents each user's share of
    occupied airtime. Airtime fairness addresses the limitation of throughput-based
    fairness for heterogeneous-rate networks and can be combined with throughput
    fairness to strike a balance between channel utilization and overall
    throughput~\cite{oxford_fairness}.

    \item \textbf{3GPP fairness definition} was originally introduced for
    LTE–LAA~\cite{fairness_laa} and later adopted for NR–U~\cite{fairness_nr}.
    This definition ensures that a coexisting system (e.g., a cellular network)
    does not degrade a Wi-Fi network more than an additional Wi-Fi network
    would. In our view, this remains one of the most suitable fairness criteria
    for coexistence among high-throughput technologies.
\end{itemize}
}

{
A variety of approaches have been proposed to analyze, optimize, and enable coexistence in unlicensed bands. The comprehensive survey on high-throughput coexistence \cite{coexistence_6G_survey} provides an in-depth overview of AI-driven techniques. Here, we briefly highlight several widely used methods in the literature. Many works rely on optimization-based formulations \cite{3gpp_fairness_optimization, theorem_fairness_optimization}, game-theoretic tools \cite{GAMETHEORY}, multiarmed bandit models \cite{MULTIARMED}, or Markov-chain-based analyses \cite{WIFIMARKOV, oxford_fairness, markov_coexistence}. Machine-learning approaches \cite{OPTIMIZINGCOEXIST_DLNFR} and reinforcement-learning techniques \cite{RLWIFILTE, NRUDRL} have also been extensively explored.
}

The coexistence of LTE-LAA and Wi-Fi is generally expected to be fair, based on the design principles underlying LAA. Empirical studies support this assumption, showing that LTE-LAA often coexists more effectively with Wi-Fi than Wi-Fi networks do with each other \cite{COEXISTENCELTEWIFI}. Nevertheless, several challenges remain, most notably the disparity in ED thresholds between Wi-Fi and cellular systems. For instance, Wi-Fi employs an ED threshold of -62~dBm, whereas LTE-LAA typically uses a threshold of -72~dBm. This asymmetry has significant implications for the perceived fairness of coexistence \cite{ASSYMENERGYTHRESH}. {On the other hand, LTE-LAA offers higher spectral efficiency, more advanced link adaptation, and better performance stability compared to Wi-Fi 5 \cite{ltelaa_wifi_coexistence_survey}, which may compensate for the asymmetry introduced by different threshold settings}.

The coexistence of LTE-U and Wi-Fi presents unique challenges, as LTE-U does not employ a CSMA-like mechanism but instead relies on the \ac{CSAT} protocol. CSAT regulates spectrum access by alternating between “on” and “off” transmission periods, which are adjusted according to measured Wi-Fi utilization \cite{ADVANCEMENTSCOEXISTENSELTE}. However, this mechanism can complicate the management of acknowledgment signals and may limit Wi-Fi’s ability to access the channel \cite{LTEUWIFICOEX}. Contrary to the common assumption that LTE-U coexists less favorably with Wi-Fi than LAA, experimental results reported in \cite{LAAVSLTEU} indicate that this is not always the case. 

The study in \cite{5GNRCHALLENGESEVALUATION} examines coexistence challenges between NR-U Release~16 and Wi-Fi~5 (IEEE~802.11ac), including simulation-based throughput comparisons for both indoor and outdoor scenarios in the 5~GHz and 6~GHz bands. Overall, NR-U demonstrates higher spectrum utilization in high-throughput scenarios, resulting in higher throughput and lower latency. However, in outdoor high-throughput scenarios, Wi-Fi typically performs better in terms of throughput and latency, primarily due to differences in ED thresholds.   

Further research presented in \cite{ADJCHANNELCOEXISTENCE} focuses on NR and Wi-Fi~6 coexistence in the 6~GHz band. The findings indicate that, because of different detection thresholds, the probability of successful transmission for NR-U is consistently 20–30\% lower than for Wi-Fi~6. In addition, the use of MU-OFDMA for Wi-Fi uplink transmission is shown to improve coexistence between NR-U and Wi-Fi.  

{The authors of \cite{no_reservations_coexistence} proposed eliminating the reservation signal in NR-U to improve overall resource utilization when coexisting with Wi-Fi. Through analytical modeling and simulation, they showed that airtime fairness can be maintained by appropriately configuring the contention window, without relying on reservation signals. Furthermore, they demonstrated that machine-learning-based prediction of optimal contention window settings enables NR-U to coexist efficiently with Wi-Fi, thereby improving spectrum utilization for both technologies.}

Additional research challenges for unlicensed spectrum sharing are outlined in \cite{NEXTGEN6GHOPPCHALL}. These include the optimization of coexistence both within a single technology and across different unlicensed technologies. For instance, one challenge is ensuring fairness in coexistence between Wi-Fi~7 and legacy devices, particularly with advanced features such as MU-MIMO and OFDMA. Another challenge arises in optimizing coexistence between different unlicensed systems, such as NR-U and Wi-Fi, where the absence of a global clock complicates the implementation of a unified frame-based coexistence mechanism.  

{With the advent of open-source full-stack RAN projects such as \ac{OAI} \cite{OAI_core_paper}, it has become feasible to evaluate coexistence on real hardware. Early studies have already demonstrated the implementation of quasi-LBT techniques for NR \cite{OAI_coexistence_peixoto}, and another work investigates coexistence between NR and LBT-based protocol for inter-vehicle communications \cite{OAI_coexistence_itsnr}. Under limited-throughput conditions, the researchers showed that the two networks can cooperate.

The complementary initiative known as Open RAN introduces a novel open-source implementation of the RAN stack that promotes interoperability between different network elements. Open RAN also enables the development of near-real-time and real-time applications for network parameter configuration, supporting dynamic network reconfiguration. In the context of coexistence, these applications can perform spectrum sensing \cite{ORAN_CBRS_sensing, ORAN_ML_Radar_detection, ORAN_sensing} as well as RL-based spectrum sharing \cite{ORAN_RL_spectrum_sharing}.
}

One approach to coexistence extends beyond conventional spectrum management techniques, which are based on time and frequency-domain adjustments. It employs Space–Time Adaptive Processing (STAP) to form beam nulls in the direction of interference. This method can be applied to the entire signal or selectively to individual sub-bands \cite{TARANAPATENT1}. The adoption of MU-MIMO in Wi-Fi and the long-standing use of MIMO in cellular systems illustrate the role of beamforming in current high-throughput technologies. Nevertheless, further research is needed to address spatial spectrum sensing challenges in the context of coexistence.  

The underlying concept is to move away from uniform sensing across all directions and instead perform sensing specifically toward the intended user, an approach commonly referred to as beamform-assisted sensing. This idea originates in cognitive radio research and has been explored in UAV communications \cite{UAVSSS} and secure 5G systems \cite{5GSSS}.

{
In addition, we believe it is necessary to investigate spectrum-sharing possibilities between terrestrial systems and high-Doppler communications, such as satellite links. Existing coexistence approaches primarily rely on sensing \cite{spectrum_sharing_ntn_survey}. However, we argue that radar techniques could be adapted to distinguish signals based on their Doppler characteristics \cite{radar_doppler_isac}, thereby enabling coexistence. Moreover, the Doppler-shift distribution, derived in \cite{doppler_stochastic_geometry}, can be used to further differentiate between space and terrestrial links.
}

\begin{table*}[hb]
\caption{Applications Utilizing Unlicensed Spectrum}
\label{tab:applications}
\centering
\begin{tabular}{|c|c|c|c|}
\hline

Name                                                                        & \begin{tabular}[c]{@{}c@{}}High-throughput \\ applications\end{tabular}                                                                             & \begin{tabular}[c]{@{}c@{}}Privacy and latency\\ demanding applications\end{tabular} & \begin{tabular}[c]{@{}c@{}}Massive connectivity \\ applications\end{tabular} \\ \hline
Use cases                                                                   & \begin{tabular}[c]{@{}c@{}}Increasing throughput of \\ licensed cellular systems\\ FWA \\ Establishing local networks\end{tabular} & Private networks                                                                     & IoT                                                                          \\ \hline

\begin{tabular}[c]{@{}c@{}}Most suitable \\ frequency spectrum\end{tabular} & TVWS, CBRS, UNII                                                                                                                                    & CBRS, UNII                                                                           & TVWS, ISM                                                                    \\ \hline
\begin{tabular}[c]{@{}c@{}}Most suitable \\ technologies\end{tabular}       & Wi-Fi, LTE, NR                                                                                                                                      & LTE, NR                                                                              & LPWAN                                                                        \\ \hline
\end{tabular}
\end{table*}

{\textbf{\underline{In conclusion}}, coexistence among high-throughput technologies differs from the other classes considered in this survey. Advanced PHY and MAC mechanisms, largely based on LBT variants, are generally effective, yet several open challenges remain in achieving truly fair coexistence. Emerging research directions such as beamforming-based sensing and ML-driven methods appear promising. In particular, real-world hardware often exhibits deterministic transmission patterns, and ML techniques are capable of learning these dependencies to predict future "white gaps" in network activity.}
\section{Applications in unlicensed spectrum}

\subsection{Introduction}  

This section examines real-world applications of technologies operating in the unlicensed spectrum. In line with the ITU-defined usage scenarios for 5G \cite{ITUGOALSFOR5G}, these applications are categorized into three groups: \ac{eMBB}, \ac{URLLC}, and \ac{mMTC}, each reflecting distinct requirements and use cases. A summary of this categorization is presented in Table~\ref{tab:applications}.

\subsection{High-throughput applications}

This category includes two main directions: (i) increasing the throughput of licensed systems by offloading data to the unlicensed spectrum, and (ii) fixed wireless access (FWA).  

The first direction focuses on reducing congestion in licensed bands by exploiting unlicensed spectrum resources. This motivation was one of the key drivers behind the development of LTE-U and LAA, and by extension, all mobile networks operating in unlicensed spectrum. According to a report by the Global Mobile Suppliers Association, 4G and 5G operators continue to show sustained interest in unlicensed technologies \cite{GSALTE5GUNLICENSED}. The report further notes that more than half of LAA-capable devices are smartphones, underscoring the significance of unlicensed spectrum for broadband mobile connectivity.  

FWA refers to the use of wireless technologies to deliver internet connectivity or backhaul in areas where traditional wired solutions are not economically feasible. It is most often considered a last-mile access solution. According to ITU data, a significant digital divide exists, with internet usage reaching only about 46\% in rural areas compared to 82\% in urban regions \cite{ITUCONNECTIVITY}. Conventional options, such as fiber or copper cabling, remain prohibitively expensive, particularly in rural and some urban areas. As a result, reliable wireless access offers a cost-effective alternative. The growing interest of both large and small technology providers in 5G-based FWA further underscores its potential, making it a dynamic and competitive market segment \cite{GSA5GFWAOPPORTUNITY}.  

The unique challenges and operational requirements of FWA systems demand customized technological solutions. These include addressing the absence of mobility, enabling long-distance communication over several kilometers, supporting high-volume data transmission with low latency, and mitigating interference from other users and technologies.

{\textbf{\underline{In conclusion}}, high-throughput applications are promising use case for unlicensed spectrum. Insights from FWA deployments indicate that realizing their full potential may depend on the development of specialized hardware tailored to these requirements.}

\subsection{Privacy and latency-demanding applications}  

Representative examples of this category include private networks, which are localized wireless systems dedicated to specific use cases. These networks offer several advantages, such as full control over network resources, predictable \ac{QoS}, low latency for mission-critical operations, guaranteed coverage, and enhanced security \cite{GSMAPRIVATENET}. Their adoption spans a wide range of sectors, including factories, warehouses, ports, hospitals, mines, airports, and university campuses. Numerous real-world implementations of private networks highlight their growing relevance in addressing the demands of privacy- and latency-sensitive applications \cite{PRIVATE5GCONCEPTSARCH}.  

Private networks can operate independently of a \ac{PLMN}, enabling fully standalone deployments that do not require access to a public mobile network. In such cases, the base station databases, along with the control and user planes, are managed entirely within the private domain. Alternative configurations may integrate certain control functions with the PLMN \cite{GSMAPRIVATENET}.  

Numerous vendors currently offer both LTE and 5G-based private network solutions and have already seen widespread adoption. In addition, private networks may also rely on Wi-Fi technology, which is typically chosen for scenarios where requirements for connection reliability and security are less stringent \cite{ATNTSTRATEGICPRIVATE}.

{\textbf{\underline{In conclusion}}, private networks represent a highly promising and practical solution for many enterprises. Vendors and technologies, including Wi-Fi and cellular systems, compete in this space, and the choice between them depends on multiple factors such as cost, coverage requirements, latency constraints, and security needs.}

\subsection{Massive connectivity applications}

This category is primarily addressed by IoT, an emerging technology designed to support a wide range of future applications in domains such as industry, smart cities, agriculture, healthcare, energy, waste management, and transportation \cite{IOTOPPORTISSUECHALL, IOTRECENTSURVEY}.  

Unlicensed spectrum represents a cost-effective option for IoT deployments, enabling both industry and communities to reduce expenses and expand connectivity. However, specific use cases with strict reliability requirements may require the use of licensed spectrum \cite{URLLCEMBBIOTSURVEY}, mainly due to the risk of random access collisions in unlicensed bands. Despite these challenges, the global demand for IoT devices continues to grow rapidly \cite{IOTSTATS}. For many large-scale sensing applications, unlicensed spectrum remains a practical and widely accepted solution \cite{SPECTSHFORIOTSURVEY}.  

Several whitepapers on LoRa deployment are presented in \cite{SEMTECHLORAWP, LORAALLWP}, covering real-world applications such as sensor networks for water, gas, and parking occupancy monitoring. A notable use case of unlicensed spectrum technologies in IoT is agriculture \cite{LORAAGRICULTURE}, where typical scenarios include the transmission of sensor data related to temperature, sunlight intensity, soil moisture, and water levels. LoRa has also been applied to enable remote control of irrigation systems.  

A practical example is described in \cite{LORAROMA}, which reports on the deployment of an IoT network in southwestern Romania. The system covers approximately 25,000 hectares of land, with altitudes ranging from 35 to 151~meters. A total of 25 stations were installed to collect data on eleven parameters. Beyond data transmission, LoRa-based systems can also provide geolocation capabilities. As noted in \cite{LORAGEOLOCATION}, the technology achieves accuracies of up to 20~meters using time-of-arrival methods, and up to 1~kilometer when relying on received signal strength indicators.  

Another notable use of unlicensed spectrum is wildlife monitoring based on the Sigfox network. To study the ecology and behavior of animals, the authors of \cite{WILDMONITOR} deployed a system that collected telemetry data from sensors attached to wildlife, including GPS, accelerometer, thermometer, and barometer measurements. The network achieved a maximum line-of-sight communication range of 280~km, with an average transmission success rate of approximately 68\% for flying species and 54\% for terrestrial species. The location accuracy, based on received signal strength indicators, was around 12~km.  

Sigfox has also been proposed for other applications, such as providing connectivity for fire and smoke sensors, which could be deployed in residential environments to enhance safety \cite{SGFOXHOMELIFESTYLE}.  

Despite the success of LPWAN technologies, their ability to support massive connectivity remains limited, as discussed in the coexistence section. Consequently, 5G emerges as the only viable option for achieving massive and reliable connectivity. The study in \cite{URLLCEMBBIOTSURVEY} addresses the key aspects of reliable communications for IoT, focusing on URLLC and eMBB in NR, and outlines the corresponding requirements for various applications. 

For instance, latency demands may range from below 1~ms to 100~ms, while reliability, defined as the error probability \cite{RELIABILITYDEF}, can vary between $10^{-9}$ and $10^{-2}$. Throughput requirements may reach up to 300~Mbps, and network capacity may span from 10~$\frac{Gbps}{km^2}$ to 1000~$\frac{Gbps}{km^2}$.  

However, 5G still faces several challenges in fully supporting massive IoT. These include limitations in data rate, latency, power consumption, and the inability to support both URLLC and eMBB traffic simultaneously. Additional obstacles involve insufficient connection capacity and the relatively small physical coverage of 5G cells. Reflecting these issues, the authors of \cite{ENABLMASSIVEIOT6G} questioned 5G’s capability to meet the demands of massive IoT deployments and instead outlined requirements for 6G to enable such functionality.  

At the same time, emerging technologies such as tinyML \cite{TINYMLPAPER} and the associated shift toward \ac{AIoT} \cite{AIOTPAPER} may change this outlook. By enabling edge-based inference, AIoT reduces network traffic by transmitting inference results rather than raw data, thereby increasing the number of devices that can be effectively supported. This raises the possibility that 5G, or even existing LPWAN technologies such as LoRa and Sigfox, could prove sufficient for massive AIoT scenarios. However, this hypothesis requires further research and validation.  

{\textbf{\underline{In conclusion}}, massive-connectivity applications represent the best use case for IoT LPWAN technologies. Existing deployments demonstrate the practicality of this approach, although performance in heavily congested and densely deployed scenarios still requires careful attention and remains insufficiently mature.}

\section{Conclusion}

This paper examines the multifaceted aspects of unlicensed sub-8~GHz spectrum technologies, with a particular emphasis on spectrum regulations, technological developments, coexistence management strategies, and application domains. We provided an overview of global unlicensed spectrum regulations, covering ISM bands, TVWS, CBRS, and U-NII bands. While the regulatory landscape for sub-5~GHz bands varies across regions, there is a clear trend toward harmonization, particularly in the 6~GHz band. In the lower frequency ranges, restrictions are primarily related to transmit power, whereas spectrum access mechanisms become increasingly important at higher frequencies. Approaches to address these issues, including database-based coexistence schemes, were briefly discussed.  

Our technology review spanned a broad range of unlicensed solutions, from low-rate systems designed for IoT, through TVWS-based technologies for diverse use cases, to high-throughput solutions operating in the 5~GHz and 6~GHz bands. A central theme has been the challenge of coexistence among these technologies, which demands tailored strategies for different categories. For example, LPWAN technologies in IoT scenarios require lightweight coexistence mechanisms, whereas high-throughput systems such as LTE, NR, and Wi-Fi depend on more robust approaches. Although existing mechanisms rely primarily on spectrum access protocols, future research should advance towards more sophisticated interference mitigation techniques, including beamforming-assisted sensing.  

Finally, we highlighted the broad application potential of unlicensed spectrum technologies, with growing importance in areas such as private networks, fixed wireless access, licensed–unlicensed spectrum aggregation, and IoT. These applications highlight the crucial role of unlicensed spectrum in meeting diverse connectivity demands and motivate ongoing research into coexistence and optimization strategies.  

\bibliographystyle{ieeetr}
\bibliography{References}

@book{FR1REF, title="38.101-1 NR; User Equipment (UE) radio transmission and reception; Part 1: Range 1 Standalone 2018", url={https://portal.3gpp.org/desktopmodules/Specifications/SpecificationDetails.aspx?specificationId=3283}, institution={3GPP}, year={2018}, month=jan }

@book{ITUREC1056, title="SM.1056 : Limitation of radiation from industrial, scientific and medical (ISM) equipment", url={https://www.itu.int/rec/R-REC-SM.1056-1-200704-I/en}, institution={ITU}, year={2007}, month=apr }

@misc{SRDWORKSHOP,
  title        = {{ITU} Workshop on Short Range Devices and Ultra Wide Band},
  year         = {2014},
  url          = {https://www.itu.int/en/ITU-R/study-groups/workshops/RWP1B-SRD-UWB-14/Pages/Program.aspx},
}

@misc{MX-AI,
      title={MX-AI: Agentic Observability and Control Platform for Open and AI-RAN}, 
      author={Ilias Chatzistefanidis and Andrea Leone and Ali Yaghoubian and Mikel Irazabal and Sehad Nassim and Lina Bariah and Merouane Debbah and Navid Nikaein},
      year={2025},
      eprint={2508.09197},
      archivePrefix={arXiv},
      primaryClass={cs.NI},
      url={https://arxiv.org/abs/2508.09197}, 
}

@misc{AI_spectrum_auction,
      title={Agoran: An Agentic Open Marketplace for 6G RAN Automation}, 
      author={Ilias Chatzistefanidis and Navid Nikaein and Andrea Leone and Ali Maatouk and Leandros Tassiulas and Roberto Morabito and Ioannis Pitsiorlas and Marios Kountouris},
      year={2025},
      eprint={2508.09159},
      archivePrefix={arXiv},
      primaryClass={cs.NI},
      url={https://arxiv.org/abs/2508.09159}, 
}

@INPROCEEDINGS{ml_resource_allocation,
  author={Tikar, Smita and Patil, Rajendrakumar},
  booktitle={2024 IEEE 9th International Conference for Convergence in Technology (I2CT)}, 
  title={Machine Learning Methods for Allocating Resources in Wireless Communication}, 
  year={2024},
  volume={},
  number={},
  pages={1-6},
  doi={10.1109/I2CT61223.2024.10544278}}

@ARTICLE{blockchain_spectrum_sharing_survey,
  author={Perera, Lavan and Ranaweera, Pasika and Kusaladharma, Sachitha and Wang, Shen and Liyanage, Madhusanka},
  journal={IEEE Open Journal of the Communications Society}, 
  title={A Survey on Blockchain for Dynamic Spectrum Sharing}, 
  year={2024},
  volume={5},
  number={},
  pages={1753-1802},
  doi={10.1109/OJCOMS.2024.3376233}}

@INPROCEEDINGS{Spectrum_sharing_regulatory,
  author={Dludla, Gcina and Mekuria, Fisseha},
  booktitle={2021 IST-Africa Conference (IST-Africa)}, 
  title={Dynamic Spectrum Sharing for Future Wireless Networks: Regulators Perspective}, 
  year={2021},
  volume={},
  number={},
  pages={1-7},
  doi={}}

@ARTICLE{ltelaa_wifi_coexistence_survey,
  author={Chen, Bolin and Chen, Jiming and Gao, Yuan and Zhang, Jie},
  journal={IEEE Communications Surveys and Tutorials}, 
  title={Coexistence of LTE-LAA and Wi-Fi on 5 GHz With Corresponding Deployment Scenarios: A Survey}, 
  year={2017},
  volume={19},
  number={1},
  pages={7-32},
  doi={10.1109/COMST.2016.2593666}}

@article{wifi8,
author = {Cox, Landon and Tang, Fengxiao and Zhao, Ming and Kato, Nei},
title = {Wi-Fi 8: The Next Generation of Wireless Connectivity},
year = {2025},
issue_date = {September 2025},
publisher = {Association for Computing Machinery},
address = {New York, NY, USA},
volume = {29},
number = {3},
issn = {2375-0529},
url = {https://doi.org/10.1145/3774505.3774513},
doi = {10.1145/3774505.3774513},
journal = {GetMobile: Mobile Comp. and Comm.},
month = oct,
pages = {18–23},
numpages = {6}
}

@inproceedings{LLM_analyse_logs,
author = {Kamatani, Osamu and Saruwatari, Shunsuke and Seshasayee, Sushila and Mamaghani, Ali and Bharadia, Dinesh},
title = {LLM-5GMAC: Performance Optimization in O-RAN Split 7.2 Using LLM-Based MAC-Layer Log Analysis},
year = {2025},
isbn = {9798400719776},
publisher = {Association for Computing Machinery},
address = {New York, NY, USA},
url = {https://doi.org/10.1145/3737900.3770164},
doi = {10.1145/3737900.3770164},
booktitle = {Proceedings of the 2nd ACM Workshop on Open and AI RAN},
pages = {1–7},
numpages = {7},
location = {
},
series = {OpenRan '25}
}

@ARTICLE{coexistence_6G_survey,
  author={Cao, Xuelin and Yang, Bo and Wang, Kaining and Li, Xinghua and Yu, Zhiwen and Yuen, Chau and Zhang, Yan and Han, Zhu},
  journal={Proceedings of the IEEE}, 
  title={AI-Empowered Multiple Access for 6G: A Survey of Spectrum Sensing, Protocol Designs, and Optimizations}, 
  year={2024},
  volume={112},
  number={9},
  pages={1264-1302},
  doi={10.1109/JPROC.2024.3417332}}

@INPROCEEDINGS{auction_spectrum_sharing,
  author={Benedetto, Francesco and Mastroeni, Loretta and Quaresima, Greta},
  booktitle={2021 44th International Conference on Telecommunications and Signal Processing (TSP)}, 
  title={Auction-based Theory for Dynamic Spectrum Access: a Review}, 
  year={2021},
  volume={},
  number={},
  pages={146-151},
  doi={10.1109/TSP52935.2021.9522600}}

@inproceedings{no_reservations_coexistence,
author = {Tinnirello, Ilenia and Lo Valvo, Alice and Szott, Szymon and Kosek-Szott, Katarzyna},
title = {No Reservations Required: Achieving Fairness between Wi-Fi and NR-U with Self-Deferral Only},
year = {2021},
isbn = {9781450390774},
publisher = {Association for Computing Machinery},
address = {New York, NY, USA},
url = {https://doi.org/10.1145/3479239.3485680},
doi = {10.1145/3479239.3485680},
booktitle = {Proceedings of the 24th International ACM Conference on Modeling, Analysis and Simulation of Wireless and Mobile Systems},
pages = {115–124},
numpages = {10},
location = {Alicante, Spain},
series = {MSWiM '21}
}

@misc{jain_fairness_definition,
      title={A Quantitative Measure Of Fairness And Discrimination For Resource Allocation In Shared Computer Systems}, 
      author={R. Jain and D. Chiu and W. Hawe},
      year={1998},
      eprint={cs/9809099},
      archivePrefix={arXiv},
      primaryClass={cs.NI},
      url={https://arxiv.org/abs/cs/9809099}, 
}

@ARTICLE{markov_coexistence,
  author={Cao, Xuelin and Song, Zuxun and Yang, Bo and Qian, Lijun and Han, Zhu},
  journal={IEEE Transactions on Wireless Communications}, 
  title={Full-Duplex MAC in LAA/ Wi-Fi Coexistence Networks: Design, Modeling, and Analysis}, 
  year={2020},
  volume={19},
  number={8},
  pages={5531-5546},
  doi={10.1109/TWC.2020.2994278}}

@INPROCEEDINGS{oxford_fairness,
  author={Tuladhar, Sudat and Cao, Lei and Viswanathan, Ramanarayanan},
  booktitle={2021 IEEE International Symposium on Dynamic Spectrum Access Networks (DySPAN)}, 
  title={Fair Coexistence of LAA and WiFi in Multi-Carrier LBT based on Joint Throughput and Airtime Fairness}, 
  year={2021},
  volume={},
  number={},
  pages={147-152},
  doi={10.1109/DySPAN53946.2021.9677088}}

@INPROCEEDINGS{theorem_fairness_optimization,
  author={Li, Yuwei and Gao, Yayu},
  booktitle={2024 5th Information Communication Technologies Conference (ICTC)}, 
  title={Fairness-Constrained Throughput Optimization for Coexistence of WiFi and Duty-Cycle 5G NR in the Unlicensed Spectrum}, 
  year={2024},
  volume={},
  number={},
  pages={106-111},
  doi={10.1109/ICTC61510.2024.10601795}}

@ARTICLE{3gpp_fairness_optimization,
  author={Kakkad, Yashraj and Patel, Dhaval K and Kavaiya, Sagar and Sun, Sumei and López-Benítez, Miguel},
  journal={IEEE Transactions on Vehicular Technology}, 
  title={Optimal 3GPP Fairness Parameters in 5G NR Unlicensed (NR-U) and WiFi Coexistence}, 
  year={2023},
  volume={72},
  number={4},
  pages={5373-5377},
  doi={10.1109/TVT.2022.3222964}}

@techreport{fairness_laa,
  author       = {{3GPP}},
  title        = {{Technical Specification Group Radio Access Network; Study on Licensed-Assisted Access to Unlicensed Spectrum (Release 13)}},
  institution  = {3rd Generation Partnership Project (3GPP)},
  type         = {Technical Report},
  number       = {36.889},
  version      = {V13.0.0},
  year         = {2015},
}

@techreport{fairness_nr,
  author       = {{3GPP}},
  title        = {{Study on NR-based Access to Unlicensed Spectrum (Release 16)}},
  institution  = {3rd Generation Partnership Project (3GPP)},
  type         = {Technical Report},
  number       = {38.889},
  version      = {V16.0.0},
  year         = {2018},
}

@ARTICLE{spectrum_sharing_ntn_survey,
  author={Heydarishahreza, Navid and Han, Tao and Ansari, Nirwan},
  journal={IEEE Communications Surveys Tutorials}, 
  title={Spectrum Sharing and Interference Management for 6G LEO Satellite-Terrestrial Network Integration}, 
  year={2025},
  volume={27},
  number={5},
  pages={2794-2825},
  doi={10.1109/COMST.2024.3507019}}

@INPROCEEDINGS{doppler_stochastic_geometry,
  author={Khan, Talha Ahmed and Afshang, Mehtnaz},
  booktitle={ICC 2020 - 2020 IEEE International Conference on Communications (ICC)}, 
  title={A Stochastic Geometry Approach to Doppler Characterization in a LEO Satellite Network}, 
  year={2020},
  volume={},
  number={},
  pages={1-6},
  doi={10.1109/ICC40277.2020.9148880}}

@ARTICLE{radar_doppler_isac,
  author={Sarac, Ugur B. and Guvensen, Gokhan M.},
  journal={IEEE Transactions on Communications}, 
  title={Knowledge-Aided and Adaptive Beam-Squint Aware MIMO-OFDM Radar Detectors for ISAC}, 
  year={2025},
  volume={73},
  number={11},
  pages={10246-10261},
  doi={10.1109/TCOMM.2025.3568219}}

@INPROCEEDINGS{ORAN_CBRS_sensing,
  author={Gangula, Rajeev and Lacava, Andrea and Polese, Michele and D’Oro, Salvatore and Bonati, Leonardo and Kaltenberger, Florian and Johari, Pedram and Melodia, Tommaso},
  booktitle={MILCOM 2024 - 2024 IEEE Military Communications Conference (MILCOM)}, 
  title={Listen-While-Talking: Toward dApp-based Real-Time Spectrum Sharing in O-RAN}, 
  year={2024},
  volume={},
  number={},
  pages={651-652},
  doi={10.1109/MILCOM61039.2024.10773685}}

@INPROCEEDINGS{ORAN_RL_spectrum_sharing,
  author={Gopal, Sneihil and Griffith, David and Rouil, Richard A. and Liu, Chunmei},
  booktitle={2025 IEEE 22nd Consumer Communications Networking Conference (CCNC)}, 
  title={AdapShare: An RL-Based Dynamic Spectrum Sharing Solution for O-RAN}, 
  year={2025},
  volume={},
  number={},
  pages={1-7},
  doi={10.1109/CCNC54725.2025.10976195}}

@INPROCEEDINGS{ORAN_sensing,
  author={Rego, Iago and Medeiros, Lucas and Alves, Pedro and Goldbarg, Mateus and Lopes, Vitor and Flor, Daniel and Barros, Wysterlanya and Filho, Vinícius and Sousa, Vicente and Aranha, Eduardo and Martins, Allan and Fernandes, Marcelo and Fontes, Ramon and Neto, Augusto},
  booktitle={2022 IEEE Conference on Network Function Virtualization and Software Defined Networks (NFV-SDN)}, 
  title={Prototyping near-real time RIC O-RAN xApps for Flexible ML-based Spectrum Sensing}, 
  year={2022},
  volume={},
  number={},
  pages={137-142},
  doi={10.1109/NFV-SDN56302.2022.9974940}}

@INPROCEEDINGS{OAI_coexistence_itsnr,
  author={Peixoto, Filipe and Figueiredo, Andreia and Rito, Pedro and Luís, Miguel and Aguiar, Ana},
  booktitle={2025 IEEE Vehicular Networking Conference (VNC)}, 
  title={Coexistence of ITS-G5 and 5G NR-U in the 5.9 GHz Band}, 
  year={2025},
  volume={},
  number={},
  pages={1-8},
  doi={10.1109/VNC64509.2025.11054093}}

@inproceedings{OAI_coexistence_peixoto,
author = {Peixoto, Filipe Lemos and Aguiar, Ana and Goes, Adriano},
title = {Non-public 5G-NR Unlicensed: Design, Implementation and Evaluation},
year = {2024},
isbn = {9798400704895},
publisher = {Association for Computing Machinery},
address = {New York, NY, USA},
url = {https://doi.org/10.1145/3636534.3697315},
doi = {10.1145/3636534.3697315},
booktitle = {Proceedings of the 30th Annual International Conference on Mobile Computing and Networking},
pages = {1874–1881},
numpages = {8},
location = {Washington D.C., DC, USA},
series = {ACM MobiCom '24}
}

@article{OAI_core_paper,
title = {Driving innovation in 6G wireless technologies: The OpenAirInterface approach},
journal = {Computer Networks},
volume = {269},
pages = {111410},
year = {2025},
issn = {1389-1286},
doi = {https://doi.org/10.1016/j.comnet.2025.111410},
url = {https://www.sciencedirect.com/science/article/pii/S1389128625003779},
author = {Florian Kaltenberger and Tommaso Melodia and Irfan Ghauri and Michele Polese and Raymond Knopp and Tien Thinh Nguyen and Sakthivel Velumani and Davide Villa and Leonardo Bonati and Robert Schmidt and Sagar Arora and Mikel Irazabal and Navid Nikaein}
}

@INPROCEEDINGS{platform_spectrum_sharing,
  author={Zhan, Shun-Cheng and Chang, Shi-Chung and Chou, Chun-Ting and Tsai, Zsehong},
  booktitle={2017 Wireless Days}, 
  title={Spectrum Sharing Auction platform for Short-term licensed shared access}, 
  year={2017},
  volume={},
  number={},
  pages={184-187},
  doi={10.1109/WD.2017.7918140}}

@ARTICLE{online_spectrum_sharing,
  author={Saha, Gourav and Abouzeid, Alhussein A. and Matinmikko-Blue, Marja},
  journal={IEEE/ACM Transactions on Networking}, 
  title={Online Algorithm for Leasing Wireless Channels in a Three-Tier Spectrum Sharing Framework}, 
  year={2018},
  volume={26},
  number={6},
  pages={2623-2636},
  doi={10.1109/TNET.2018.2877184}}

@misc{CST2024SpectrumLightLicensing,
  author       = "{Communications, Space \& Technology Commission (CST)}",
  title        = "{Spectrum Light Licensing Regulations}",
  howpublished = "\url{https://www.cst.gov.sa/en/regulations-and-licenses/regulations/Document-1603}",
  year         = {2024},
  note         = "Decision No. 557/1446, issued 15 August 2024"
}

@ARTICLE{DSM_blockchain,
  author={Cuellar, David and Sallal, Muntadher and Williams, Christopher},
  journal={IEEE Access}, 
  title={BSM-6G: Blockchain-Based Dynamic Spectrum Management for 6G Networks: Addressing Interoperability and Scalability}, 
  year={2024},
  volume={12},
  number={},
  pages={59643-59664},
  doi={10.1109/ACCESS.2024.3393288}}

@techreport{ITU_WRC23_FinalActs_2023,
  title        = {Final Acts WRC-23},
  institution  = {{International Telecommunication Union}},
  year         = {2023},
  type         = {Report},
  url          = {http://handle.itu.int/11.1002/pub/8225d4fb-en},
  note         = {Full document (PDF) available from ITU},
}

@ARTICLE{IU_detection_AI,
  author={Camelo, Miguel and Mennes, Ruben and Shahid, Adnan and Struye, Jakob and Donato, Carlos and Jabandzic, Irfan and Giannoulis, Spilios and Mahfoudhi, Farouk and Maddala, Prasanthi and Seskar, Ivan and Moerman, Ingrid and Latre, Steven},
  journal={IEEE Wireless Communications}, 
  title={An AI-Based Incumbent Protection System for Collaborative Intelligent Radio Networks}, 
  year={2020},
  volume={27},
  number={5},
  pages={16-23},
  doi={10.1109/MWC.001.2000032}}

@ARTICLE{ORAN_ML_Radar_detection,
  author={Reus-Muns, Guillem and Upadhyaya, Pratheek S. and Demir, Utku and Stephenson, Nathan and Soltani, Nasim and Shah, Vijay K. and Chowdhury, Kaushik R.},
  journal={IEEE Journal on Selected Areas in Communications}, 
  title={SenseORAN: O-RAN-Based Radar Detection in the CBRS Band}, 
  year={2024},
  volume={42},
  number={2},
  pages={326-338},
  doi={10.1109/JSAC.2023.3336152}}

@INPROCEEDINGS{RL_spectrum_sharing,
  author={Akbari, Elham and Zeng, Kai},
  booktitle={2025 IEEE International Symposium on Dynamic Spectrum Access Networks (DySPAN)}, 
  title={Federated Deep Reinforcement Learning for Privacy-Preserving Spectrum Sharing among MNOs}, 
  year={2025},
  volume={},
  number={},
  pages={1-8},
  doi={10.1109/DySPAN64764.2025.11115932}}

@ARTICLE{CBRS_SAS_optimization,
  author={Abbass, Waseem and Ahmad Khan, Muzammil and Hussain Farooqi, Ashfaq and Nawaz, Waqas and Abbas, Nasim and Ali, Zulfiqar},
  journal={IEEE Access}, 
  title={Optimizing Spectrum Utilization and Security in SAS-Enabled CBRS Systems for Enhanced 5G Performance}, 
  year={2024},
  volume={12},
  number={},
  pages={165992-166010},
  doi={10.1109/ACCESS.2024.3495972}}

@misc{LTEUWORKSHOP_CS,
  title        = {{3GPP} Workshop LTE in unlicensed spectrum},
  year         = {2014},
  url          = {https://www.3gpp.org/news-events/3gpp-news/lte-in-unlicensed},
}

@misc{NRUMEETING2019,
  title        = {{3GPP} {TSG} {RAN} Meeting 86. {RP}-201834 {S}itges},
  year         = {2019},
  url          = {https://www.3gpp.org/dynareport?code=TDocExMtg--RP-86--19645.htm},
}

@misc{3GPPREL16SUMMARY,
  title        = {3rd Generation Partnership Project; Technical Specification Group Services and System Aspects; Release 16 Description; Summary of Rel-16 Work Items (Release 16)},
  year         = {2020},
  url          = {https://www.3gpp.org/specifications-technologies/releases/release-16},
}

@misc{ITURREG2020,
  author       = {},
  title        = {{ITU} Radio Regulations, edition of 2020: Volume 1: Articles},
  year         = {2020},
  url          = {https://search.itu.int/history/HistoryDigitalCollectionDocLibrary/1.44.48.en.101.pdf},
}

@misc{ETSIEMC25to1000,
  author       = {},
  title        = {{ETSI} {EN} 300 220-1 V2.4.1 Electromagnetic compatibility and Radio spectrum Matters (ERM); Short Range Devices (SRD); Radio equipment to be used in the 25 MHz to 1 000 MHz frequency range with power levels ranging up to 500 mW; Part 1: Technical characteristics and test methods},
  year         = {2020},
  url          = {https://search.itu.int/history/HistoryDigitalCollectionDocLibrary/1.44.48.en.101.pdf},
}

@patent{SIGFOXPATENT,
  author      = {Jorge Perdomo},
  title       = {System and method for digital communication between computing devices},
  year        = {2015},
  number      = {WO2015139026A2},
  url         = {https://patents.google.com/patent/WO2015139026A2/en},
  note        = {Accessed: 2024-09-23}
}

@misc{FCCISMPOWER,
  title        = {Title 47: Telecommunication, Part 15: Radio Frequency Devices, Section 15.247: Operation within the bands 902-928 MHz, 2400-2483.5 MHz, and 5725-5850 MHz},
  year         = {2024},
  url          = {https://www.ecfr.gov/current/title-47/chapter-I/subchapter-A/part-15/subpart-C/subject-group-ECFR2f2e5828339709e/section-15.247},
  note         = {Accessed: 2024-09-23}
}

@misc{WIFIHALOWOVERVIEW,
  title       = {{Wi-Fi CERTIFIED HaLow™ Technology Overview}},
  author      = {{Wi-Fi Alliance}},
  year        = {2021},
  url         = {https://www.wi-fi.org/system/files/Wi-Fi_CERTIFIED_HaLow_Technology_Overview_20211102.pdf},
  note        = {Accessed: 2024-09-23}
}

@article{IEEE80211ahHALOPAPER,
	journal = {Sensors},
	doi = {10.3390/s18020325},
	issn = {1424-8220},
	number = {2},
	language = {en},
	publisher = {MDPI AG},
	title = {{Performance Evaluation of IEEE 802.11ah Networks With High-Throughput Bidirectional Traffic}},
	url = {http://dx.doi.org/10.3390/s18020325},
	volume = {18},
	author = {Šljivo, Amina and Kerkhove, Dwight and Tian, Le and Famaey, Jeroen and Munteanu, Adrian and Moerman, Ingrid and Hoebeke, Jeroen and De Poorter, Eli},
	pages = {325},
	date = {2018-01-23},
	year = {2018},
	month = {1},
	day = {23},
}

@standard{IEEE80211,
  title       = {IEEE Std 802.11-2020: Standard for Information Technology—Telecommunications and Information Exchange between Systems—Local and Metropolitan Area Networks—Specific Requirements—Part 11},
  author      = {{IEEE}},
  year        = {2020},
  url         = {https://standards.ieee.org/ieee/802.11/7028/},
  note        = {Accessed: 2024-09-23}
}

@misc{FCCUNIIGENERALGUID,
  title       = {Part 15 Subpart E U-NII 6 GHz: General Guidance Bands 5, 6, 7, 8},
  author      = {{Federal Communications Commission}},
  year        = {2024},
  institution = {Office of Engineering and Technology, Laboratory Division},
  url         = {https://apps.fcc.gov/kdb/GetAttachment.html?id=Kz5fs0wPvr321yO4Yausgw%3D%3D&desc=987594%20D01%20U-NII%206GHz%20General%20Requirements%20v02r02&tracking_number=277034},
  note        = {Accessed: 2024-09-23}
}

@standard{ETSIBRAN301893,
  title       = {{ETSI EN 301 893 V2.1.1: 5 GHz RLAN; Harmonised Standard Covering the Essential Requirements of Article 3.2 of Directive 2014/53/EU}},
  author      = {{European Telecommunications Standards Institute}},
  year        = {2017},
  month       = {May},
  url         = {https://www.etsi.org/deliver/etsi_en/301800_301899/301893/02.01.01_60/en_301893v020101p.pdf},
  note        = {Accessed: 2024-09-23}
}

@misc{DFS_AFC,
  title       = {FCC 15.407: General Technical Requirements},
  author      = {{Federal Communications Commission}},
  url         = {https://www.ecfr.gov/current/title-47/chapter-I/subchapter-A/part-15/subpart-E/section-15.407},
  note        = {Accessed: 2024-09-23}
}

@misc{JAPANFREQALLOC,
  title       = {Japan Frequency Allocation Table},
  author      = {{Ministry of Internal Affairs and Communications}},
  url         = {https://www.tele.soumu.go.jp/resource/e/search/share/pdf/t2.pdf},
  note        = {Accessed: 2024-09-23}
}

@misc{JAPANANNEX,
  title       = {Japan Annex 8.5},
  author      = {{Ministry of Internal Affairs and Communications}},
  url         = {https://www.tele.soumu.go.jp/resource/e/search/share/pdf/a8.pdf},
  note        = {Accessed: 2024-09-23}
}

@misc{EUROPEFREQALLOC,
  title       = {European Frequency Allocation Table},
  author      = {{European Conference of Postal and Telecommunications Administrations (CEPT)}},
  year        = {2023},
  url         = {https://docdb.cept.org/download/4316},
  note        = {Accessed: 2024-09-23}
}

@misc{CANADAFREQREG,
  title       = {RSS-247: Digital Transmission Systems (DTSs), Frequency Hopping Systems (FHSS) and Licence-Exempt Local Area Network (LE-LAN) Devices},
  author      = {{Innovation, Science and Economic Development Canada}},
  year        = {2023},
  month       = {August},
  issue       = {3},
  url         = {https://ised-isde.canada.ca/site/spectrum-management-telecommunications/en/devices-and-equipment/radio-equipment-standards/radio-standards-specifications-rss/rss-247-digital-transmission-systems-dtss-frequency-hopping-systems-fhss-and-licence-exempt-local},
  note        = {Accessed: 2024-09-23}
}

@misc{COUNTRIESENABLEDWIFI6,
  title       = {Countries Enabling Wi-Fi in 6 GHz: Wi-Fi 6E},
  author      = {{Wi-Fi Alliance}},
  year        = {2024},
  url         = {https://www.wi-fi.org/countries-enabling-wi-fi-in-6-ghz-wi-fi-6e},
  note        = {Accessed: 2024-09-23}
}

@standard{REL17TECHSPEC,
  title       = {ETSI TS 138 104 V17.5.0: 5G; NR; Base Station (BS) Radio Transmission and Reception (3GPP TS 38.104 Version 17.5.0 Release 17)},
  author      = {{European Telecommunications Standards Institute}},
  year        = {2022},
  month       = {April},
  url         = {https://www.etsi.org/deliver/etsi_ts/138100_138199/138104/17.05.00_60/ts_138104v170500p.pdf},
  note        = {Accessed: 2024-09-23}
}

@techreport{REL18TECHSPEC,
  title       = {TR 21.918 V1.1.0: 3rd Generation Partnership Project; Technical Specification Group Services and System Aspects; Release 18 Description; Summary of Rel-18 Work Items},
  author      = {{3rd Generation Partnership Project}},
  year        = {2024},
  month       = {August},
  url         = {https://www.3gpp.org/ftp/Specs/archive/21_series/21.918/},
  note        = {Accessed: 2024-09-23}
}

@report{REPORTUNLICENSED,
  title       = {{LTE in Unlicensed and Shared Spectrum: Trials, Deployments and Devices}},
  author      = {{GSA}},
  year        = {2019},
  month       = {January},
  url         = {https://gsacom.com/paper/lte-unlicensed-shared-spectrum-2/},
  note        = {Accessed: 2024-09-23}
}

@misc{USNEW6GHBANDANNOUNC,
  title       = {FCC Adopts New Rules for the 6 GHz Band, Unleashing 1,200 Megahertz of Spectrum for Unlicensed Use},
  author      = {{Federal Communications Commission}},
  year        = {2020},
  month       = {April},
  url         = {https://docs.fcc.gov/public/attachments/DOC-363945A1.pdf},
  note        = {Accessed: 2024-09-23}
}

@misc{SAUDIAFC6GH,
  title       = {CITC Performed the Globally First Live Demo of the AFC System to Enable Wi-Fi 6E Technology},
  author      = {{Communications and Information Technology Commission (CITC)}},
  year        = {2022},
  month       = {August},
  day         = {7},
  url         = {https://www.cst.gov.sa/en/mediacenter/pressreleases/Pages/2022080701.aspx},
  note        = {Accessed: 2024-09-23}
}

@report{US6GRULES,
  title       = {Unlicensed Use of the 6 GHz Band: Expanding Flexible Use in Mid-Band Spectrum Between 3.7 and 24 GHz},
  author      = {{Federal Communications Commission}},
  year        = {2020},
  month       = {April},
  day         = {23},
  docket      = {ET Docket No. 18-295, GN Docket No. 17-183},
  url         = {https://docs.fcc.gov/public/attachments/FCC-20-51A1.pdf},
  note        = {Accessed: 2024-09-23}
}

@techreport{GLOBALSTAR,
  title       = {Globalstar Licensed 2.4 GHz Technical Review},
  author      = {{Globalstar}},
  year        = {2017},
  month       = {January},
  url         = {https://www.globalstar.com/getmedia/5443c399-447c-439f-8535-529e93569970/globalstar2-4_ghz.pdf},
  note        = {Accessed: 2024-09-23}
}

@misc{OFCOMEXTENDEDSHARING,
  title       = {A Framework for Spectrum Sharing},
  author      = {{Ofcom}},
  year        = {2015},
  month       = {July},
  day         = {31},
  url         = {https://www.ofcom.org.uk/spectrum/innovative-use-of-spectrum/spectrum-sharing-framework/},
  note        = {Accessed: 2024-09-23}
}

@misc{DSADSMS,
  title       = {{Dynamic Spectrum Alliance} Solving the Spectrum Crunch: Dynamic Spectrum Management System},
  author      = {Michael Calabrese },
  year        = {2023},
  url         = {https://www.dynamicspectrumalliance.org/solving-the-spectrum-crunch.pdf},
  note        = {Accessed: 2024-09-23}
}

@report{FCCSECONDMEMOR2008,
  title       = {{FCC} Second Report and Order and Memorandum Opinion and Order},
  author      = {{Federal Communications Commission}},
  year        = {2008},
  month       = {November},
  day         = {4},
  release     = {November 14, 2008},
  url         = {https://docs.fcc.gov/public/attachments/FCC-08-260A1.pdf},
  note        = {Accessed: 2024-09-23}
}

@report{FCCSECONDMEMOR2010,
  title       = {{FCC} Second Memorandum Opinion and Order},
  author      = {{Federal Communications Commission}},
  year        = {2010},
  month       = {September},
  day         = {23},
  url         = {https://docs.fcc.gov/public/attachments/FCC-10-174A1.pdf},
  note        = {Accessed: 2024-09-23}
}

@techreport{TESTSTVWS1,
  title       = {Initial Evaluation of the Performance of Prototype {TV}-Band White Space Devices},
  author      = {{Federal Communications Commission}},
  year        = {2007},
  month       = {July},
  day         = {31},
  institution = {Technical Research Branch, Laboratory Division, Office of Engineering and Technology},
  url         = {https://docs.fcc.gov/public/attachments/DOC-275666A1.pdf},
  note        = {Accessed: 2024-09-23}
}

@techreport{TESTSTVWS2,
  title       = {{Evaluation of the Performance of Prototype TV-Band White Space Devices: Phase II}},
  author      = {{Federal Communications Commission}},
  year        = {2008},
  month       = {October},
  day         = {15},
  institution = {Technical Research Branch, Laboratory Division, Office of Engineering and Technology},
  url         = {https://docs.fcc.gov/public/attachments/DA-08-2243A3.pdf},
  note        = {Accessed: 2024-09-23}
}

@misc{WIFROST,
  title       = {{WiFROST} {LT100E} {D}atasheet},
  author      = {{WiFROST}},
  year        = {2023},
  url         = {https://www.wifrost.com/product},
  note        = {Accessed: 2024-09-23}
}

@INPROCEEDINGS{TVWSOVERVIEW2014,
  author={Oh, Ser Wah and Ma, Yugang and Tao, Ming-Hung and Peh, Edward Chu Yeow},
  booktitle={International Conference on Frontiers of Communications, Networks and Applications (ICFCNA 2014 - Malaysia)}, 
  title={An overview and comparison of TV White Space regulations worldwide}, 
  year={2014},
  volume={},
  number={},
  pages={1-6},
  doi={10.1049/cp.2014.1398}}

@misc{FCCTVWS,
  author = {{Federal Communications Commission}},
  title       = {{FCC} {C}hapter 1, {S}ubchapter {A}, {P}art 15, {S}ubpart {H} — {W}hite {S}pace {D}evices},
  author      = {},
  url         = {https://www.ecfr.gov/current/title-47/chapter-I/subchapter-A/part-15#subpart-H},
  note        = {Accessed: 2024-09-23}
}

@misc{INNONET,
  title        = {Disaster Safety Services},
  year         = {2023},
  howpublished = {Innonet},
  url          = {https://innonet.net/en/p0303.php},
  note         = {Accessed: 2024-10-16}
}

@misc{BOOKLPWANFORIOTANDM2M,
	doi = {10.1016/c2018-0-04787-8},
	isbn = {9780128188804},
	language = {en},
	publisher = {Elsevier},
	title = {LPWAN Technologies for IoT and M2M Applications},
	url = {http://dx.doi.org/10.1016/C2018-0-04787-8},
	date = {2020},
	year = {2020},
}

@INPROCEEDINGS{LWALWIP,
  author={Määttanen, Helka-Liina and Masini, Gino and Bergström, Mattias and Ratilainen, Antti and Dudda, Torsten},
  booktitle={2017 IEEE Conference on Standards for Communications and Networking (CSCN)}, 
  title={LTE-WLAN aggregation (LWA) in 3GPP Release 13 and Release 14}, 
  year={2017},
  volume={},
  number={},
  pages={220-226},
  doi={10.1109/CSCN.2017.8088625}}

@ARTICLE{LTEUWIFICOEX,
  author={Sathya, Vanlin and Mehrnoush, Morteza and Ghosh, Monisha and Roy, Sumit},
  journal={IEEE Access}, 
  title={Wi-Fi/LTE-U Coexistence: Real-Time Issues and Solutions}, 
  year={2020},
  volume={8},
  number={},
  pages={9221-9234},
  doi={10.1109/ACCESS.2020.2964210}}

@article{IOTOPPORTISSUECHALL,
	journal = {International Journal of Sensors, Wireless Communications and Control},
	doi = {10.2174/2210327913666221216160435},
	issn = {2210-3279},
	number = {9},
	language = {en},
	publisher = {Bentham Science Publishers Ltd.},
	title = {A Concise Review on Internet of Things: Architecture, Enabling Technologies,
Challenges, and Applications},
	url = {http://dx.doi.org/10.2174/2210327913666221216160435},
	volume = {12},
	author = {Saqib, Manasha and Moon, Ayaz Hassan},
	pages = {629--650},
	date = {2022-11},
	year = {2022},
	month = {11},
}

@misc{RELIABILITYDEF,
  author       = {Eiman Mohyeldin},
  title        = {Requirements for IMT-2020},
  year         = {2020},
  howpublished = {ITU-R Workshop on IMT-2020 Terrestrial Radio Interfaces},
  url          = {https://www.itu.int/en/ITU-R/study-groups/rsg5/rwp5d/imt-2020/Documents/S01-1_Requirements%20for%20IMT-2020_Rev.pdf},
  note         = {Accessed: [Insert date of access here]}
}

@article{TINYMLPAPER,
	journal = {Multimedia Tools and Applications},
	doi = {10.1007/s11042-023-16740-9},
	issn = {1573-7721},
	number = {10},
	language = {en},
	publisher = {Springer Science and Business Media LLC},
	title = {TinyML: Tools, applications, challenges, and future research directions},
	url = {http://dx.doi.org/10.1007/s11042-023-16740-9},
	volume = {83},
	author = {Kallimani, Rakhee and Pai, Krishna and Raghuwanshi, Prasoon and Iyer, Sridhar and López, Onel L. A.},
	pages = {29015--29045},
	date = {2023-09-09},
	year = {2023},
	month = {9},
	day = {9},
}

@ARTICLE{AIOTPAPER,
  author={Zhang, Jing and Tao, Dacheng},
  journal={IEEE Internet of Things Journal}, 
  title={Empowering Things With Intelligence: A Survey of the Progress, Challenges, and Opportunities in Artificial Intelligence of Things}, 
  year={2021},
  volume={8},
  number={10},
  pages={7789-7817},
  doi={10.1109/JIOT.2020.3039359}}

@techreport{SINGAPORETVWS,
  title       = {{Television White Space Devices: IMDA TS WSD}},
  author      = {{Info-communications Media Development Authority}},
  year        = {2016},
  month       = {October},
  day         = {1},
  issue       = {1},
  address     = {10 Pasir Panjang Road, \#10-01 Mapletree Business City, Singapore 117438},
  url         = {https://www.imda.gov.sg/~/media/imda/files/regulation%20licensing%20and%20consultations/ict%20standards/telecommunication%20standards/radio-comms/imda%20ts%20wsd.pdf},
  note        = {Accessed: 2024-09-23}
}

@article{OFCOMSTOPPED2,
  title        = {Ofcom removes TV white space device authorisation framework},
  author       = {{Telecompaper}},
  year         = {2024},
  journal      = {Telecompaper},
  url          = {https://www.telecompaper.com/news/ofcom-removes-tv-white-space-device-authorisation-framework--1490921},
  note         = {Accessed: 2024-09-26}
}

@misc{OFCOMSTOPPED,
  author       = {Pierre-Jean Muller},
  title        = {Ofcom {UK} has today given notice that the TV white space device (WSD) authorisation framework is no longer available},
  year         = {2024},
  url = {\url{https://www.linkedin.com/posts/pierre-jean-muller-0396521_ofcom-uk-has-today-given-notice-that-the-activity-7161693887076511744-QHzr/}},
  note         = {Accessed: 2024-09-26}
}

@misc{CANADATVWS,
  title       = {White Space Policies: Technical and Operational Requirements},
  author      = {{Innovation, Science and Economic Development Canada}},
  year        = {},
  url         = {https://ised-isde.canada.ca/site/spectrum-management-telecommunications/en/devices-and-equipment/white-space},
  note        = {Accessed: 2024-09-23}
}

@misc{SAFRICATVWS,
  title       = {Regulations on the Use of Television White Spaces},
  author      = {{Independent Communications Authority of South Africa (ICASA)}},
  year        = {2018},
  url         = {https://www.icasa.org.za/uploads/files/Regulations-on-the-use-of-Television-White-Spaces-2018.pdf},
  note        = {Accessed: 2024-09-23}
}

@standard{ETSITVWS,
  title       = {{ETSI EN 301 598 V1.1.1: White Space Devices; Wireless Access Systems Operating in the 470 MHz to 790 MHz TV Broadcast Band; Harmonized EN Covering the Essential Requirements of Article 3.2 of the Directive}},
  author      = {{European Telecommunications Standards Institute}},
  year        = {2014},
  month       = {April},
  url         = {https://www.etsi.org/deliver/etsi_en/301500_301599/301598/01.01.01_60/en_301598v010101p.pdf},
  note        = {Accessed: 2024-09-23}
}

@misc{USCHECKAVAILABILITY,
  title       = {Channel Search},
  author      = {{Red Technologies}},
  year        = {2024}, 
  url         = {https://usa.wavedb.com/channelsearch/tvws},
  note        = {Accessed: 2024-09-23}
}

@INPROCEEDINGS{FEASIBILITYPAPER19,
  author={Alejandrino, Jonnel D. and Concepcion, Ronnie S. and Laugico, Sandy C. and Trinidad, Emmanuel T. and Dadios, Elmer P.},
  booktitle={2019 IEEE 11th International Conference on Humanoid, Nanotechnology, Information Technology, Communication and Control, Environment, and Management ( HNICEM )}, 
  title={Feasibility of Television White Space Spectrum Technologies for Wide Range Wireless Sensor Network: A survey}, 
  year={2019},
  volume={},
  number={},
  pages={1-6},
  doi={10.1109/HNICEM48295.2019.9072794}}

@ARTICLE{DATABASEPAPER16,
  author={Anabi, Kelechi Hilary and Nordin, Rosdiadee and Abdullah, Nor Fadzilah},
  journal={IEEE Access}, 
  title={Database-Assisted Television White Space Technology: Challenges, Trends and Future Research Directions}, 
  year={2016},
  volume={4},
  number={},
  pages={8162-8183},
  doi={10.1109/ACCESS.2016.2621178}}

@article{COMPREHENSIVEPAPER19,
title = {A comprehensive survey on networking over TV white spaces},
journal = {Pervasive and Mobile Computing},
volume = {59},
pages = {101072},
year = {2019},
issn = {1574-1192},
doi = {https://doi.org/10.1016/j.pmcj.2019.101072},
url = {https://www.sciencedirect.com/science/article/pii/S1574119219300306},
author = {Mahbubur Rahman and Abusayeed Saifullah}
}

@article{SURVEYTVWS2018,
author = {Zhang, Wenjie and Yang, Jingmin and Zhang, Guanglin and Yang, Liwei and Kiat Yeo, Chai},
title = {TV white space and its applications in future wireless networks and communications: a survey},
journal = {IET Communications},
volume = {12},
number = {20},
pages = {2521-2532},
doi = {https://doi.org/10.1049/iet-com.2018.5009},
url = {https://ietresearch.onlinelibrary.wiley.com/doi/abs/10.1049/iet-com.2018.5009},
eprint = {https://ietresearch.onlinelibrary.wiley.com/doi/pdf/10.1049/iet-com.2018.5009},
year = {2018}
}

@Article{ATMDUCTMIDEAST,
AUTHOR = {Constantinides, Antonios and Najat, Saam and Haralambous, Haris},
TITLE = {Atmospheric Ducting Interference on DAB, DAB+ Radio in Eastern Mediterranean},
JOURNAL = {Electronics},
VOLUME = {11},
YEAR = {2022},
NUMBER = {24},
ARTICLE-NUMBER = {4183},
URL = {https://www.mdpi.com/2079-9292/11/24/4183},
ISSN = {2079-9292},
DOI = {10.3390/electronics11244183}
}

@misc{ROHDESHCWARZ,
  author       = {},
  title        = {Rohde \& Schwarz: Wi-Fi 7 Technology Introduction},
  howpublished = {White Paper},
  year         = {2023},
  url          = {https://www.rohde-schwarz.com/us/solutions/wireless-communications-testing/wireless-standards/wlan-wi-fi/white-paper-ieee-802-11be-technology-introduction-registration_255637.html},
  note         = {Accessed: 2024-10-02}
}

@misc{CSTSPECTRUMSHARING,
  title        = {Future Paradigms for Spectrum Sharing: Towards A Common Framework For Advanced Spectrum Management},
  author       = {{CST}},
  year         = {2024},
  url          = {https://www.cst.gov.sa/en/researchs-studies/research-innovation/Documents/Future_paradigm_for_spectrum_sharing.pdf},
  note         = {Accessed: 2024-09-29}
}

@ARTICLE{LORAAGRICULTURE,
  author={Aldhaheri, Lameya and Alshehhi, Noor and Manzil, Irfana Ilyas Jameela and Khalil, Ruhul Amin and Javaid, Shumaila and Saeed, Nasir and Alouini, Mohamed-Slim},
  journal={IEEE Internet of Things Journal}, 
  title={{LoRa Communication for Agriculture 4.0: Opportunities, Challenges, and Future Directions}}, 
  year={2024},
  volume={},
  number={},
  pages={1-1},
  doi={10.1109/JIOT.2024.3486369}}

@misc{SEMTECHLORAWP,
  author       = {},
  title        = {LoRa Technology White Papers},
  year         = {2023},
  howpublished = {\url{https://www.semtech.com/lora/resources/lora-white-papers}},
  note         = {Accessed: 2024-10-06}
}

@misc{LORAGEOLOCATION,
  author       = {},
  title        = {LoRa Alliance Geolocation White Paper},
  year         = {2023},
  howpublished = {\url{https://resources.lora-alliance.org/whitepapers/lora-alliance-geolocation-whitepaper}},
  note         = {Accessed: 2024-10-06}
}

@misc{LORAALLWP,
  author       = {},
  title        = {LoRa Alliance White Papers},
  year         = {2023},
  howpublished = {\url{https://resources.lora-alliance.org/whitepapers}},
  note         = {Accessed: 2024-10-06}
}

@INPROCEEDINGS{LORAROMA,
  author={Zarnescu, Adrian and Ungurelu, Razvan and Secere, Mihai and Varzaru, Gaudentiu and Mihailescu, Bogdan},
  booktitle={2020 7th International Conference on Energy Efficiency and Agricultural Engineering (EE and AE)}, 
  title={Implementing a large LoRa network for an agricultural application}, 
  year={2020},
  volume={},
  number={},
  pages={1-5},
  doi={10.1109/EEAE49144.2020.9279019}}

@misc{GSMANB5G,
  author       = {GSMA},
  title        = {{Mobile IoT in the 5G Future - NB-IoT and LTE-M in the Context of 5G}},
  year         = {2018},
  url          = {https://www.ericsson.com/4ac64d/assets/local/reports-papers/5g/doc/gsma-5g-mobile-iot.pdf},
  note         = {Accessed: 2024-10-06}
}

@misc{SGFOXHOMELIFESTYLE,
  author       = {Sigfox},
  title        = {Home Lifestyle Use Cases},
  year         = {2023},
  howpublished = {\url{https://www.sigfox.com/use-cases/home-lifestyle/}},
  note         = {Accessed: 2024-10-06}
}

@article{WILDMONITOR,
	journal = {Animal Biotelemetry},
	doi = {10.1186/s40317-023-00326-1},
	issn = {2050-3385},
	number = {1},
	language = {en},
	publisher = {Springer Science and Business Media LLC},
	title = {A multi-species evaluation of digital wildlife monitoring using the Sigfox IoT network},
	url = {http://dx.doi.org/10.1186/s40317-023-00326-1},
	volume = {11},
	author = {Wild, Timm A. and van Schalkwyk, Louis and Viljoen, Pauli and Heine, Georg and Richter, Nina and Vorneweg, Bernd and Koblitz, Jens C. and Dechmann, Dina K. N. and Rogers, Will and Partecke, Jesko and Linek, Nils and Volkmer, Tamara and Gregersen, Troels and Havmøller, Rasmus W. and Morelle, Kevin and Daim, Andreas and Wiesner, Miriam and Wolter, Kerri and Fiedler, Wolfgang and Kays, Roland and Ezenwa, Vanessa O. and Meboldt, Mirko and Wikelski, Martin},
	date = {2023-03-25},
	year = {2023},
	month = {3},
	day = {25},
}

@ARTICLE{WIVEYATMDUCTING,
  author={Liu, Fangfang and Pan, Jiaxi and Zhou, Xiangwei and Li, Geoffrey Ye},
  journal={Journal of Communications and Information Networks}, 
  title={Atmospheric Ducting Effect in Wireless Communications: Challenges and Opportunities}, 
  year={2021},
  volume={6},
  number={2},
  pages={101-109},
  doi={10.23919/JCIN.2021.9475120}}

@INPROCEEDINGS{IEEE80222FIRST,
  author={Cordeiro, C. and Challapali, K. and Birru, D. and Sai Shankar},
  booktitle={First IEEE International Symposium on New Frontiers in Dynamic Spectrum Access Networks, 2005. DySPAN 2005.}, 
  title={IEEE 802.22: the first worldwide wireless standard based on cognitive radios}, 
  year={2005},
  volume={},
  number={},
  pages={328-337},
  doi={10.1109/DYSPAN.2005.1542649}}

@techreport{IEEE80222OVERVIEW,
  title        = {IEEE P802.22 Wireless RANs},
  author       = {Apurva N. Mody and Gerald Chouinard},
  institution  = {BAE Systems and Communications Research Center, Canada},
  date         = {2010-06-15},
  address      = {P. O. Box 868, MER 15-2350, Nashua, NH 03061},
  phone        = {1-404-819-0314, 1-603-885-2621, 1-613-998-2500},
  email        = {apurva.mody@baesystems.com, apurva_mody@yahoo.com, gerald.chouinard@crc.ca},
  url          = {https://www.ieee802.org/22/Technology/22-10-0073-03-0000-802-22-overview-and-core-technologies.pdf}
}

@standard{IEEE80222STANDARD,
  title       = {ISO/IEC/IEEE 8802-22-2021: Telecommunications and Information Exchange Between Systems—Wireless Regional Area Networks (WRAN)—Specific Requirements—Part 22: Cognitive Wireless RAN Medium Access Control (MAC) and Physical Layer (PHY) Specifications: Policies and Procedures for Operation in the Bands that Allow Spectrum Sharing},
  author      = {{IEEE}},
  year        = {2021},
  month       = {June},
  day         = {13},
  url         = {https://standards.ieee.org/ieee/8802-22/10947/},
  note        = {Accessed: 2024-09-23}
}

@techreport{SIGNALATTENUATION,
  title     = {{Recommendation ITU-R P.676-13: Attenuation by Atmospheric Gases and Related Effects}},
  author    = {{International Telecommunication Union}},
  year      = {2022},
  month     = {August},
  series    = {P Series},
  note      = {Radiowave propagation},
  url       = {https://www.itu.int/dms_pubrec/itu-r/rec/p/R-REC-P.676-13-202208-I!!PDF-E.pdf}
}

@techreport{SIGNALATTENUATION2,
  author      = {W. C. Stone},
  title       = {NIST Construction Automation Program Report No. 3: Electromagnetic Signal Attenuation in Construction Materials},
  institution = {National Institute of Standards and Technology (NIST)},
  year        = {1997},
  month       = {October},
  note        = {Tech. rep.},
    url = {https://nvlpubs.nist.gov/nistpubs/Legacy/IR/nistir6055.pdf}
}

@incollection{BOOKIEEE80211af,
	booktitle = {Handbook of Cognitive Radio},
	doi = {10.1007/978-981-10-1394-2_53},
	isbn = {9789811013935},
	language = {en},
	publisher = {Springer Singapore},
	title = {IEEE 802.11af Wi-Fi in TV White Space},
	url = {http://dx.doi.org/10.1007/978-981-10-1394-2_53},
	author = {Ishizu, Kentaro and Mizutani, Keiichi and Matsumura, Takeshi and Lan, Zhou and Harada, Hiroshi},
	pages = {1509--1535},
	date = {2019},
	year = {2019},
}

@article{BELLALTA20161,
title = {Next generation IEEE 802.11 Wireless Local Area Networks: Current status, future directions and open challenges},
journal = {Computer Communications},
volume = {75},
pages = {1-25},
year = {2016},
issn = {0140-3664},
doi = {https://doi.org/10.1016/j.comcom.2015.10.007},
url = {https://www.sciencedirect.com/science/article/pii/S0140366415003874},
author = {Boris Bellalta and Luciano Bononi and Raffaele Bruno and Andreas Kassler}
}

@INPROCEEDINGS{TWVSSTANDARDCOMPARISON,
  author={George, Divya Sara and Rao, Sethuraman N},
  booktitle={2017 IEEE International Conference on Computational Intelligence and Computing Research (ICCIC)}, 
  title={Enabling Rural Connectivity:Long Range Wi-Fi Versus Super Wi-Fi}, 
  year={2017},
  volume={},
  number={},
  pages={1-4},
  doi={10.1109/ICCIC.2017.8524142}}

@ARTICLE{PROFSLIMAPER,
  author={Matracia, Maurilio and Saeed, Nasir and Kishk, Mustafa A. and Alouini, Mohamed-Slim},
  journal={IEEE Open Journal of the Communications Society}, 
  title={Post-Disaster Communications: Enabling Technologies, Architectures, and Open Challenges}, 
  year={2022},
  volume={3},
  number={},
  pages={1177-1205},
  doi={10.1109/OJCOMS.2022.3192040}}

@misc{MICROSOFTAIRBAND,
  title       = {Microsoft Airband Initiative},
  author      = {{Microsoft}},
  url         = {https://www.microsoft.com/en-us/corporate-responsibility/airband-initiative#coreui-areaheading-rqsc7tl},
  note        = {Accessed: 2024-09-23}
}

@misc{DYNAMICSPECALLIANCE,
  title       = {Dynamic Spectrum Alliance},
  author      = {{Dynamic Spectrum Alliance}},
  url         = {https://www.dynamicspectrumalliance.org/},
  note        = {Accessed: 2024-09-23}
}

@inproceedings{RURALAFRICA,
address = {Calgary},
author = {Miquel Oliver and Sudip Majumder},
copyright = {http://www.econstor.eu/dspace/Nutzungsbedingungen},
language = {eng},
publisher = {International Telecommunications Society (ITS)},
series = {2nd Europe - Middle East - North African Regional Conference of the International Telecommunications Society (ITS): "Leveraging Technologies For Growth", Aswan, Egypt, 18th-21st February, 2019},
title = {{Motivation for TV white space: An explorative study on Africa for achieving the rural broadband gap}},
url = {https://hdl.handle.net/10419/201733},
year = {2019}
}

@misc{FCCTVWSDB,
  title       = {FCC White Space Database Administration},
  author      = {{Federal Communications Commission}},
  url         = {https://www.fcc.gov/general/white-space-database-administration},
  note        = {Accessed: 2024-09-23}
}

@techreport{ECCREPORTTVWSDB,
  title       = {ECC Report 236: Guidance for National Implementation of a Regulatory Framework for TV WSD Using Geo-Location Databases},
  author      = {{European Conference of Postal and Telecommunications Administrations}},
  year        = {2015},
  month       = {May},
  url         = {https://docdb.cept.org/download/1213},
  note        = {Approved: May 2015. Accessed: 2024-09-23}
}

@ARTICLE{BROADBANDMEASTVWS,
  author={Islam, Md Zobaer and O’Hara, John F. and Shadoan, Dylan and Ibrahim, Mostafa and Ekin, Sabit},
  journal={IEEE Open Journal of the Communications Society}, 
  title={TV White Space Based Wireless Broadband Internet Connectivity: A Case Study With Implementation Details and Performance Analysis}, 
  year={2021},
  volume={2},
  number={},
  pages={2449-2462},
  doi={10.1109/OJCOMS.2021.3123939}}

@INPROCEEDINGS{PAWSINDIA,
  author={Ghosh, Soumik and Naik, Gaurang and Kumar, Animesh and Karandikar, Abhay},
  booktitle={2015 Twenty First National Conference on Communications (NCC)}, 
  title={{OpenPAWS: An open source PAWS and UHF TV White Space database implementation for India}}, 
  year={2015},
  volume={},
  number={},
  pages={1-6},
  doi={10.1109/NCC.2015.7084890}}

@misc{OFCOMTVWS,
  title       = {Implementing TV White Spaces: Statement},
  author      = {{Ofcom}},
  year        = {2015},
  month       = {February},
  day         = {12},
  url         = {https://www.ofcom.org.uk/siteassets/resources/documents/consultations/uncategorised/8017-white-space-coexistence/statement/tvws-statement.pdf?v=334070},
  note        = {Accessed: 2024-09-23}
}

@misc{PAWSSOURCE,
  title       = {{Protocol to Access White-Space (PAWS) Databases}},
  author      = {Vincent Chen and Subir Das and Lei Zhu and John Malyar and Pete McCann},
  year        = {2018},
  month       = {December},
  day         = {20},
  note        = {RFC 7545, Internet Engineering Task Force },
  url         = {https://datatracker.ietf.org/doc/rfc7545/},
  accessdate  = {2024-09-23}
}

@misc{NBWSDFCC,
  title       = {{Unlicensed White Space Device Operations in the Television Bands}},
  author      = {{Federal Communications Commission}},
  year        = {2023},
  month       = {May},
  day         = {22},
  note        = {Federal Register, 47 CFR Parts 15, [ET Docket Nos. 20-36 and 14-165; FCC 23-24; FRS 139311]},
  url         = {https://www.federalregister.gov/documents/2023/05/22/2023-10166/unlicensed-white-space-device-operations-in-the-television-bands},
  accessdate  = {2024-09-23}
}

@misc{FCCCBRSRULES,
  title       = {{Part 96—Citizens Broadband Radio Service}},
  author      = {{Federal Communications Commission}},
  url         = {https://www.ecfr.gov/current/title-47/chapter-I/subchapter-D/part-96},
  note        = {Authority: 47 U.S.C. 154(i), 303, and 307. Source: 80 FR 36222, June 23, 2015, unless otherwise noted. Accessed: 2024-09-23}
}

@techreport{WIACBRS,
  title       = {The CBRS Opportunity: New Spectrum, Stakeholders, Networks and Devices},
  author      = {{WIA Innovation \& Technology Council}},
  year        = {2020},
  month       = {March},
  day         = {23},
  url         = {https://wia.org/wp-content/uploads/CBRSOpportunity_paper-web-NEW.pdf},
  note        = {Accessed: 2024-09-23}
}

@misc{WINNINFOGR,
  title       = {{CBRS Ecosystem Infographic}},
  author      = {{Wireless Innovation Forum}},
  year        = {2024},
  url         = {https://cbrs.wirelessinnovation.org/assets/images/CBRS%20Ecosystem%20aug%202024.png},
  note        = {Accessed: 2024-09-23}
}

@INPROCEEDINGS{COMPARECBRSWIFI,
  author={Sathya, Vanlin and Zhang, Lyutianyang and Goyal, Mohit and Yavuz, Mehmet},
  booktitle={2023 International Conference on Computing, Networking and Communications (ICNC)}, 
  title={{Warehouse Deployment: A Comparative Measurement Study of Commercial Wi-Fi and CBRS Systems}}, 
  year={2023},
  volume={},
  number={},
  pages={242-248},
  doi={10.1109/ICNC57223.2023.10074584}}

@article{CBRSSURVEY,
	journal = {Electronics},
	doi = {10.3390/electronics11233985},
	issn = {2079-9292},
	number = {23},
	language = {en},
	publisher = {MDPI AG},
	title = {{A Survey on Citizens Broadband Radio Service (CBRS)}},
	url = {http://dx.doi.org/10.3390/electronics11233985},
	volume = {11},
	author = {Agarwal, Pranay and Manekiya, Mohammedhusen and Ahmad, Tahir and Yadav, Ashish and Kumar, Abhinav and Donelli, Massimo and Mishra, Saurabh Tarun},
	pages = {3985},
	date = {2022-12-01},
	year = {2022},
	month = {12},
	day = {1},
}

@misc{WINNFORUMSTD,
  title       = {Wireless Innovation Forum},
  author      = {{Wireless Innovation Forum}},
  url         = {https://cbrs.wirelessinnovation.org/},
  note        = {Accessed: 2024-09-23}
}

@techreport{CBRSHOSPITALITY,
  title       = {{Citizen’s Broadband Radio Service (CBRS) White Paper}},
  author      = {{Hospitality Technology Next Generation}},
  year        = {2018},
  month       = {September},
  url         = {https://www.ahla.com/sites/default/files/CBRS%20for%20Hospitality%20White%20Paper.pdf},
  note        = {Accessed: 2024-09-23}
}

@misc{ONGOALLIANCE,
  title       = {ON.GO Alliance},
  author      = {{ON.GO Alliance}},
  url         = {https://ongoalliance.org/},
  note        = {Accessed: 2024-09-23}
}

@article{LRWRTSURVEY,
	journal = {Future Internet},
	doi = {10.3390/fi12010013},
	issn = {1999-5903},
	number = {1},
	language = {en},
	publisher = {MDPI AG},
	title = {Long-Range Wireless Radio Technologies: A Survey},
	url = {http://dx.doi.org/10.3390/fi12010013},
	volume = {12},
	author = {Foubert, Brandon and Mitton, Nathalie},
	pages = {13},
	date = {2020-01-14},
	year = {2020},
	month = {1},
	day = {14},
}

@misc{BLUETOOTHSTANDARD,
  title       = {Core Specification},
  author      = {{Bluetooth Special Interest Group}},
  url         = {https://www.bluetooth.com/specifications/specs/},
  note        = {Accessed: 2024-09-23}
}

@misc{BLUETOOTHALLIANCE,
  title       = {Bluetooth Special Interest Group},
  author      = {{Bluetooth Special Interest Group}},
  url         = {https://www.bluetooth.com/},
  note        = {Accessed: 2024-09-23}
}

@misc{ZIGBEESTANDARD,
  title       = {Zigbee Standard Specifications Download Request},
  author      = {{Connectivity Standards Alliance}},
  url         = {https://csa-iot.org/developer-resource/specifications-download-request/},
  note        = {Accessed: 2024-09-23}
}

@misc{ZIGBEEALLIANCE,
  title       = {Connectivity Standards Alliance},
  author      = {{Connectivity Standards Alliance}},
  url         = {https://csa-iot.org/},
  note        = {Accessed: 2024-09-23}
}

@techreport{LORAPATENTNEW,
  title       = {Y.4480: Low Power Protocol for Wide Area Wireless Networks},
  author      = {{International Telecommunication Union}},
  year        = {2021},
  month       = {November},
  day         = {29},
  url         = {https://www.itu.int/rec/T-REC-Y.4480-202111-I/en},
  note        = {Approved: 2021-11-29. Accessed: 2024-09-23}
}

@misc{LORAALLIANCE,
  title       = {{LoRa} {A}lliance},
  author      = {{LoRa Alliance}},
  url         = {https://lora-alliance.org/},
  note        = {Accessed: 2024-09-23}
}

@misc{SIGFOXWEB,
  title       = {Sigfox},
  author      = {{Sigfox}},
  url         = {https://www.sigfox.com/},
  note        = {Accessed: 2024-09-23}
}

@misc{EMTCLTE,
  title       = {{LTE-M (Cat M1)}},
  author      = {{Halberdbastion}},
  url         = {https://halberdbastion.com/technology/iot/iot-protocols/emtc-lte-cat-m1},
  note        = {Accessed: 2024-09-23}
}

@INPROCEEDINGS{COVCOMPNBIOTLTEM,
  author={Lauridsen, Mads and Kovacs, Istvan Z. and Mogensen, Preben and Sorensen, Mads and Holst, Steffen},
  booktitle={2016 IEEE 84th Vehicular Technology Conference (VTC-Fall)}, 
  title={{Coverage and Capacity Analysis of LTE-M and NB-IoT in a Rural Area}}, 
  year={2016},
  volume={},
  number={},
  pages={1-5},
  doi={10.1109/VTCFall.2016.7880946}}

@techreport{CITCREPORT,
  title       = {{Internet of Things: LPWAN Overview and Strategic Perspective}},
  year        = {2021},
  month       = {February},
  url         = {https://www.cst.gov.sa/en/researchs-studies/research-innovation/Documents/CITC-IoT_LPWAN.pdf},
  note        = {Accessed: 2024-09-23}
}

@ARTICLE{LPWANBROADSURVEY,
  author={Chilamkurthy, Naga Srinivasarao and Pandey, Om Jee and Ghosh, Anirban and Cenkeramaddi, Linga Reddy and Dai, Hong-Ning},
  journal={IEEE Access}, 
  title={Low-Power Wide-Area Networks: A Broad Overview of Its Different Aspects}, 
  year={2022},
  volume={10},
  number={},
  pages={81926-81959},
  doi={10.1109/ACCESS.2022.3196182}}

@misc{STATISTAMARKETSHARE,
  title       = {LPWA Market Share by Technology},
  author      = {{Statista}},
  url         = {https://www.statista.com/statistics/1244778/lpwa-market-share-by-technology/},
  note        = {Accessed: 2024-09-23}
}

@inproceedings{CHALUNLLPWANAREN,
	journal = {The 25th Annual International Conference on Mobile Computing and Networking},
	organization = {MobiCom '19: The 25th Annual International Conference on Mobile Computing and Networking},
	doi = {10.1145/3300061.3345444},
	publisher = {ACM},
	title = {Challenge: Unlicensed LPWANs Are Not Yet the Path to Ubiquitous Connectivity},
	url = {http://dx.doi.org/10.1145/3300061.3345444},
	author = {Ghena, Branden and Adkins, Joshua and Shangguan, Longfei and Jamieson, Kyle and Levis, Philip and Dutta, Prabal},
	date = {2019-10-11},
	year = {2019},
	month = {10},
	day = {11},
}

@INPROCEEDINGS{LPWANUNBaSS,
  author={Naik, Nitin},
  booktitle={2018 IEEE International Systems Engineering Symposium (ISSE)}, 
  title={LPWAN Technologies for IoT Systems: Choice Between Ultra Narrow Band and Spread Spectrum}, 
  year={2018},
  volume={},
  number={},
  pages={1-8},
  doi={10.1109/SysEng.2018.8544414}}

@INPROCEEDINGS{INTERFLORABLUETOOTH,
  author={Polak, Ladislav and Paul, Filip and Simka, Marek and Zedka, Radim and Kufa, Jan and Sotner, Roman},
  booktitle={2022 32nd International Conference Radioelektronika (RADIOELEKTRONIKA)}, 
  title={On the Interference between LoRa and Bluetooth in the 2.4 GHz Unlicensed Band}, 
  year={2022},
  volume={},
  number={},
  pages={1-4},
  doi={10.1109/RADIOELEKTRONIKA54537.2022.9764912}}

@INPROCEEDINGS{COMPAREBLZIG,
  author={Danbatta, Salim Jibrin and Varol, Asaf},
  booktitle={2019 7th International Symposium on Digital Forensics and Security (ISDFS)}, 
  title={{Comparison of Zigbee, Z-Wave, Wi-Fi, and Bluetooth Wireless Technologies Used in Home Automation}}, 
  year={2019},
  volume={},
  number={},
  pages={1-5},
  doi={10.1109/ISDFS.2019.8757472}}

@INPROCEEDINGS{LPWANCOLLISION,
  author={Krupka, Lukas and Vojtech, Lukas and Neruda, Marek},
  booktitle={2016 17th International Conference on Mechatronics - Mechatronika (ME)}, 
  title={The issue of LPWAN technology coexistence in IoT environment}, 
  year={2016},
  volume={},
  number={},
  pages={1-8},
  doi={}}

@article{LPWANCOEXISTENCESURVEY,
	journal = {Wireless Personal Communications},
	doi = {10.1007/s11277-017-4419-5},
	issn = {0929-6212},
	number = {1},
	language = {en},
	publisher = {Springer Science and Business Media LLC},
	title = {Sub-GHz LPWAN Network Coexistence, Management and Virtualization: An Overview and Open Research Challenges},
	url = {http://dx.doi.org/10.1007/s11277-017-4419-5},
	volume = {95},
	author = {De Poorter, Eli and Hoebeke, Jeroen and Strobbe, Matthias and Moerman, Ingrid and Latré, Steven and Weyn, Maarten and Lannoo, Bart and Famaey, Jeroen},
	pages = {187--213},
	date = {2017-06-01},
	year = {2017},
	month = {6},
	day = {1},
}

@ARTICLE{UAVSSS,
  author={Shang, Bodong and Marojevic, Vuk and Yi, Yang and Abdalla, Aly Sabri and Liu, Lingjia},
  journal={IEEE Vehicular Technology Magazine}, 
  title={Spectrum Sharing for UAV Communications: Spatial Spectrum Sensing and Open Issues}, 
  year={2020},
  volume={15},
  number={2},
  pages={104-112},
  doi={10.1109/MVT.2020.2980020}}

@article{5GSSS,
	journal = {IET Cyber-Physical Systems: Theory and Applications},
	doi = {10.1049/iet-cps.2017.0010},
	issn = {2398-3396},
	number = {1},
	language = {en},
	publisher = {Institution of Engineering and Technology (IET)},
	title = {Enabling cyber‐physical communication in 5G cellular networks: challenges, spatial spectrum sensing, and cyber‐security},
	url = {http://dx.doi.org/10.1049/iet-cps.2017.0010},
	volume = {2},
	author = {Atat, Rachad and Liu, Lingjia and Chen, Hao and Wu, Jinsong and Li, Hongxiang and Yi, Yang},
	pages = {49--54},
	date = {2017-04},
	year = {2017},
	month = {4},
}

@ARTICLE{IMPACTPHYMACTHROUGHPUT,
  author={Karmakar, Raja and Chattopadhyay, Samiran and Chakraborty, Sandip},
  journal={IEEE Communications Surveys \& Tutorials}, 
  title={Impact of IEEE 802.11n/ac PHY/MAC High Throughput Enhancements on Transport and Application Protocols—A Survey}, 
  year={2017},
  volume={19},
  number={4},
  pages={2050-2091},
  doi={10.1109/COMST.2017.2745052}}

@article{SURVEYWIFI6,
	journal = {Future Internet},
	doi = {10.3390/fi14100293},
	issn = {1999-5903},
	number = {10},
	language = {en},
	publisher = {MDPI AG},
	title = {A Survey of Wi-Fi 6: Technologies, Advances, and Challenges},
	url = {http://dx.doi.org/10.3390/fi14100293},
	volume = {14},
	author = {Mozaffariahrar, Erfan and Theoleyre, Fabrice and Menth, Michael},
	pages = {293},
	date = {2022-10-14},
	year = {2022},
	month = {10},
	day = {14},
}

@INPROCEEDINGS{WIFIONZIGBEE,
  author={Jiang, Haoran and Liu, Bin and Chen, Chang Wen},
  booktitle={2017 IEEE International Conference on Communications (ICC)}, 
  title={Performance analysis for ZigBee under WiFi interference in smart home}, 
  year={2017},
  volume={},
  number={},
  pages={1-6},
  doi={10.1109/ICC.2017.7997161}}

@INPROCEEDINGS{WIFIBLUETOOTH,
  author={He, Shaowei and Zhang, Boyu and Luo, Jing and Qu, Meijun},
  booktitle={2021 International Conference on Microwave and Millimeter Wave Technology (ICMMT)}, 
  title={Research on coexistence and mutual interference of Wi-Fi and Bluetooth}, 
  year={2021},
  volume={},
  number={},
  pages={1-3},
  doi={10.1109/ICMMT52847.2021.9617988}}

@article{WIFILORA,
	journal = {Telecommunication Systems},
	doi = {10.1007/s11235-020-00658-w},
	issn = {1018-4864},
	number = {3},
	language = {en},
	publisher = {Springer Science and Business Media LLC},
	title = {Performance analysis of LoRa in the 2.4 GHz ISM band: coexistence issues with Wi-Fi},
	url = {http://dx.doi.org/10.1007/s11235-020-00658-w},
	volume = {74},
	author = {Polak, Ladislav and Milos, Jiri},
	pages = {299--309},
	date = {2020-03-09},
	year = {2020},
	month = {3},
	day = {9},
}

@ARTICLE{WIFI6TUTORIAL,
  author={Khorov, Evgeny and Kiryanov, Anton and Lyakhov, Andrey and Bianchi, Giuseppe},
  journal={IEEE Communications Surveys \& Tutorials}, 
  title={A Tutorial on IEEE 802.11ax High Efficiency WLANs}, 
  year={2019},
  volume={21},
  number={1},
  pages={197-216},
  doi={10.1109/COMST.2018.2871099}}

@article{WIFI6FIRSTLOOK,
	journal = {Proceedings of the ACM on Measurement and Analysis of Computing Systems},
	doi = {10.1145/3579451},
	issn = {2476-1249},
	number = {1},
	language = {en},
	publisher = {Association for Computing Machinery (ACM)},
	title = {A First Look at Wi-Fi 6 in Action: Throughput, Latency, Energy Efficiency, and Security},
	url = {http://dx.doi.org/10.1145/3579451},
	volume = {7},
	author = {Liu, Ruofeng and Choi, Nakjung},
	pages = {1--25},
	date = {2023-02-27},
	year = {2023},
	month = {2},
	day = {27},
}

@misc{WIFIORG7GEN,
  title       = {Wi-Fi CERTIFIED 7™ Technology Overview},
  author      = {{Wi-Fi Alliance}},
  year        = {2024},
  url         = {https://www.wi-fi.org/discover-wi-fi/papers},
  note        = {Accessed: 2024-09-23}
}

@article{WIFIWPA3WSEC,
	journal = {Electronics},
	doi = {10.3390/electronics7110284},
	issn = {2079-9292},
	number = {11},
	language = {en},
	publisher = {MDPI AG},
	title = {A Comprehensive Attack Flow Model and Security Analysis for Wi-Fi and WPA3},
	url = {http://dx.doi.org/10.3390/electronics7110284},
	volume = {7},
	author = {Kohlios, Christopher P. and Hayajneh, Thaier},
	pages = {284},
	date = {2018-10-30},
	year = {2018},
	month = {10},
	day = {30},
}

@article{WIFI7MLO,
	journal = {IEEE/ACM Transactions on Networking},
	doi = {10.1109/tnet.2023.3283154},
	issn = {1063-6692},
	number = {1},
	publisher = {Institute of Electrical and Electronics Engineers (IEEE)},
	title = {Wi-Fi Multi-Link Operation: An Experimental Study of Latency and Throughput},
	url = {https://dl.acm.org/doi/pdf/10.1109/TNET.2023.3283154},
	volume = {32},
	author = {Carrascosa-Zamacois, Marc and Geraci, Giovanni and Knightly, Edward and Bellalta, Boris},
	pages = {308--322},
	date = {2024-02},
	year = {2024},
	month = {2},
}

@article{WIFIWPA3SYSTREVIEW,
	journal = {IEEE Access},
	doi = {10.1109/access.2023.3322931},
	issn = {2169-3536},
	publisher = {Institute of Electrical and Electronics Engineers (IEEE)},
	title = {Wireless Security Protocols WPA3: A Systematic Literature Review},
	url = {https://ieeexplore.ieee.org/document/10274082},
	volume = {11},
	author = {Halbouni, Asmaa and Ong, Lee-Yeng and Leow, Meng-Chew},
	pages = {112438--112450},
	date = {2023},
	year = {2023},
}

@misc{ITUCONNECTIVITY,
  title       = {Time Series of ICT Data},
  author      = {{International Telecommunication Union}},
  url         = {https://www.itu.int/en/ITU-D/Statistics/Pages/stat/default.aspx},
  note        = {Accessed: 2024-09-23}
}

@patent{TARANAPATENT1,
  title       = {Non-Line of Sight Wireless Communication System and Method},
  author      = {Dale Branlund},
  assignee     = {Tarana Wireless Inc.},
  number      = {US9456354B2},
  year        = {2016},
  url         = {https://patents.google.com/patent/US9456354},
  note        = {Accessed: 2024-09-23}
}

@misc{TARANAPRODUCT,
  title       = {Tarana Wireless Solutions},
  author      = {{Tarana Wireless}},
  url         = {https://www.taranawireless.com/solution/},
  note        = {Accessed: 2024-09-23}
}

@techreport{GSA5GFWAOPPORTUNITY,
  title       = {The 5G FWA Opportunity: Disrupting the Broadband Market},
  author      = {{GSMA Intelligence}},
  year        = {2021},
  url         = {https://data.gsmaintelligence.com/api-web/v2/research-file-download?id=66289674&file=141021-5G-FWA-Opportunity.pdf},
  note        = {Accessed: 2024-09-23}
}

@article{BIGCOMM,
	journal = {Frontiers in Communications and Networks},
	doi = {10.3389/frcmn.2022.785933},
	issn = {2673-530X},
	publisher = {Frontiers Media SA},
	title = {Big Communications: Connect the Unconnected},
	url = {http://dx.doi.org/10.3389/frcmn.2022.785933},
	volume = {3},
	author = {Zhang, Chuanting and Dang, Shuping and Alouini, Mohamed-Slim and Shihada, Basem},
	date = {2022-02-28},
	year = {2022},
	month = {2},
	day = {28},
}

@ARTICLE{EXTLTEINTOUNLICENSED,
  author={Labib, Mina and Marojevic, Vuk and Reed, Jeffrey H. and Zaghloul, Amir I.},
  journal={IEEE Communications Standards Magazine}, 
  title={Extending LTE into the Unlicensed Spectrum: Technical Analysis of the Proposed Variants}, 
  year={2017},
  volume={1},
  number={4},
  pages={31-39},
  doi={10.1109/MCOMSTD.2017.1700040}}

@ARTICLE{PRIVATE5GCONCEPTSARCH,
  author={Wen, Miaowen and Li, Qiang and Kim, Kyeong Jin and López-Pérez, David and Dobre, Octavia A. and Poor, H. Vincent and Popovski, Petar and Tsiftsis, Theodoros A.},
  journal={IEEE Journal of Selected Topics in Signal Processing}, 
  title={Private 5G Networks: Concepts, Architectures, and Research Landscape}, 
  year={2022},
  volume={16},
  number={1},
  pages={7-25},
  doi={10.1109/JSTSP.2021.3137669}}

@ARTICLE{LTELICEXSPECTRUM,
  author={Zhang, Jingjing and Wang, Mao and Hua, Min and Xia, Tingting and Yang, Wenjie and You, Xiaohu},
  journal={IEEE Communications Surveys \& Tutorials}, 
  title={LTE on License-Exempt Spectrum}, 
  year={2018},
  volume={20},
  number={1},
  pages={647-673},
  doi={10.1109/COMST.2017.2771485}}

@ARTICLE{COEXISTENCECELLULARWIFI,
  author={Saha, Rony Kumer},
  journal={IEEE Open Journal of the Communications Society}, 
  title={Coexistence of Cellular and IEEE 802.11 Technologies in Unlicensed Spectrum Bands -A Survey}, 
  year={2021},
  volume={2},
  number={},
  pages={1996-2028},
  doi={10.1109/OJCOMS.2021.3106502}}

@techreport{WHITEPAPERINTELLTE,
  title       = {Unlicensed LTE: Enhancing Mobile Connectivity},
  author      = {{Intel}},
  url         = {https://www.intel.com/content/dam/www/public/us/en/documents/white-papers/unlicensed-lte-paper.pdf},
  note        = {Accessed: 2024-09-23}
}

@ARTICLE{ADVANCEMENTSCOEXISTENSELTE,
  author={Huang, Yan and Chen, Yongce and Hou, Y. Thomas and Lou, Wenjing and Reed, Jeffrey H.},
  journal={IEEE Network}, 
  title={Recent Advances of LTE/WiFi Coexistence in Unlicensed Spectrum}, 
  year={2018},
  volume={32},
  number={2},
  pages={107-113},
  doi={10.1109/MNET.2017.1700124}}

@techreport{LTELWAINTELPAPER,
  title       = {LTE-WLAN Aggregation (LWA): Benefits and Deployment Considerations},
  author      = {{Intel}},
  url         = {https://www.intel.com/content/dam/www/public/us/en/documents/white-papers/lte-wlan-aggregation-deployment-paper.pdf},
  note        = {Accessed: 2024-09-23}
}

@techreport{MULTEFIREQUALPRES,
  title       = {MulteFire Technology Progress and Benefits, and How It Enables a New Breed of Neutral Hosts},
  author      = {{Qualcomm Technologies, Inc.}},
  year        = {2016},
  month       = {May},
  url         = {https://www.qualcomm.com/content/dam/qcomm-martech/dm-assets/documents/mf_webinar_3.1.pdf},
  note        = {Accessed: 2024-09-23}
}

@ARTICLE{LAAVSLTEU,
  author={Bojović, Biljana and Giupponi, Lorenza and Ali, Zoraze and Miozzo, Marco},
  journal={IEEE Access}, 
  title={Evaluating Unlicensed LTE Technologies: LAA vs LTE-U}, 
  year={2019},
  volume={7},
  number={},
  pages={89714-89751},
  doi={10.1109/ACCESS.2019.2926197}}

@techreport{ETSIFELAA,
  title       = {Digital Cellular Telecommunications System (Phase 2+) (GSM); Universal Mobile Telecommunications System (UMTS); LTE; 5G; Release Description; Release 15},
  author      = {{ETSI}},
  number      = {ETSI TR 121 915 V15.0.0},
  year        = {2018},
  url         = {https://www.etsi.org/deliver/etsi_tr/121900_121999/121915/15.00.00_60/tr_121915v150000p.pdf},
  note        = {Accessed: 2024-09-23}
}

@techreport{3GPPREL1416OVER,
  title       = {Wireless Technology Evolution: Transition from 4G to 5G 3GPP Releases 14 to 16},
  author      = {{5G Americas}},
  year        = {2018},
  month       = {October},
  url         = {https://www.5gamericas.org/wp-content/uploads/2019/07/3GPP_Rel_14-16_10.22-final_for_upload.pdf},
  note        = {Accessed: 2024-09-23}
}

@ARTICLE{MULTEFIREPAPER,
  author={Rosa, Claudio and Kuusela, Markku and Frederiksen, Frank and Pedersen, Klaus I.},
  journal={IEEE Communications Magazine}, 
  title={Standalone LTE in Unlicensed Spectrum: Radio Challenges, Solutions, and Performance of MulteFire}, 
  year={2018},
  volume={56},
  number={10},
  pages={170-177},
  doi={10.1109/MCOM.2018.1701029}}

@ARTICLE{COEXISTENCELTEWIFI,
  author={Wszołek, Jacek and Ludyga, Szymon and Anzel, Wojciech and Szott, Szymon},
  journal={IEEE Communications Magazine}, 
  title={Revisiting LTE LAA: Channel Access, QoS, and Coexistence with WiFi}, 
  year={2021},
  volume={59},
  number={2},
  pages={91-97},
  doi={10.1109/MCOM.001.2000595}}

@techreport{3GPP38889,
  title       = {Study on NR-based Access to Unlicensed Spectrum},
  author      = {{3GPP}},
  number      = {38.889},
  status      = {Under change control},
  type        = {Technical Report},
  planned_release = {Release 15},
  url         = {https://portal.3gpp.org/desktopmodules/Specifications/SpecificationDetails.aspx?specificationId=3235},
  note        = {Accessed: 2024-09-23}
}

@techreport{NRGENERALDESCRIPTION,
  title       = {5G; NR; NR and NG-RAN Overall Description; Stage-2},
  author      = {{ETSI}},
  number      = {ETSI TS 138 300 V16.2.0},
  year        = {2023},
  url         = {https://www.etsi.org/deliver/etsi_ts/138300_138399/138300/16.02.00_60/ts_138300v160200p.pdf},
  note        = {Accessed: 2024-09-23}
}

@ARTICLE{5GNRURLLC18OVER,
  author={Lin, Xingqin},
  journal={IEEE Communications Standards Magazine}, 
  title={An Overview of 5G Advanced Evolution in 3GPP Release 18}, 
  year={2022},
  volume={6},
  number={3},
  pages={77-83},
  doi={10.1109/MCOMSTD.0001.2200001}}

@ARTICLE{5GNRCHALLENGESEVALUATION,
  author={Hirzallah, Mohammed and Krunz, Marwan and Kecicioglu, Balkan and Hamzeh, Belal},
  journal={IEEE Transactions on Cognitive Communications and Networking}, 
  title={5G New Radio Unlicensed: Challenges and Evaluation}, 
  year={2021},
  volume={7},
  number={3},
  pages={689-701},
  doi={10.1109/TCCN.2020.3041851}}

@ARTICLE{NEXTGEN6GHOPPCHALL,
  author={Naik, Gaurang and Park, Jung-Min and Ashdown, Jonathan and Lehr, William},
  journal={IEEE Access}, 
  title={{Next Generation Wi-Fi and 5G NR-U in the 6 GHz Bands: Opportunities and Challenges}}, 
  year={2020},
  volume={8},
  number={},
  pages={153027-153056},
  doi={10.1109/ACCESS.2020.3016036}}

@INPROCEEDINGS{ADJCHANNELCOEXISTENCE,
  author={Naik, Gaurang and Park, Jung-Min Jerry},
  booktitle={IEEE INFOCOM 2021 - IEEE Conference on Computer Communications}, 
  title={Coexistence of Wi-Fi 6E and 5G NR-U: Can We Do Better in the 6 GHz Bands?}, 
  year={2021},
  volume={},
  number={},
  pages={1-10},
  doi={10.1109/INFOCOM42981.2021.9488780}}

@ARTICLE{SYRVEY5GANDBEYOND,
  author={Dogra, Anutusha and Jha, Rakesh Kumar and Jain, Shubha},
  journal={IEEE Access}, 
  title={A Survey on Beyond 5G Network With the Advent of 6G: Architecture and Emerging Technologies}, 
  year={2021},
  volume={9},
  number={},
  pages={67512-67547},
  doi={10.1109/ACCESS.2020.3031234}}

@article{ASSYMENERGYTHRESH,
author = {Mehrnoush, Morteza and Sathya, Vanlin and Roy, Sumit and Ghosh, Monisha},
title = {Analytical Modeling of Wi-Fi and LTE-LAA Coexistence: Throughput and Impact of Energy Detection Threshold},
year = {2018},
issue_date = {August 2018},
publisher = {IEEE Press},
volume = {26},
number = {4},
issn = {1063-6692},
url = {https://doi.org/10.1109/TNET.2018.2856901},
doi = {10.1109/TNET.2018.2856901},
journal = {IEEE/ACM Trans. Netw.},
month = aug,
pages = {1990–2003},
numpages = {14}
}

@article{OPTIMIZINGCOEXIST_DLNFR,
  title={Optimizing Unlicensed Coexistence Network Performance Through Data Learning},
  author={Srikant Manas Kala and Vanlin Sathya and Kunal Dahiya and Teruo Higashino and Hirozumi Yamaguchi},
  journal={ArXiv},
  year={2021},
  volume={abs/2111.07583},
  url={https://api.semanticscholar.org/CorpusID:244117334}
}

@ARTICLE{5GNRREL16OVER,
  author={Parkvall, Stefan and Blankenship, Yufei and Blasco, Ricardo and Dahlman, Erik and Fodor, Gabor and Grant, Stephen and Stare, Erik and Stattin, Magnus},
  journal={IEEE Communications Standards Magazine}, 
  title={5G NR Release 16: Start of the 5G Evolution}, 
  year={2020},
  volume={4},
  number={4},
  pages={56-63},
  doi={10.1109/MCOMSTD.011.1900018}}

@ARTICLE{5GNRURLLC17OVER,
  author={Pocovi, Guillermo and Abreu, Renato and Andres, Pilar and Deghel, Matha and Hugl, Klaus and Jacobsen, Thomas and Jayasinghe, Keeth and Kela, Petteri and Korhonen, Juha and Kuo, Ping-Heng and Kuru, Lauri and Li, Zexian and Lunttila, Timo and Peralta-Calvo, Elena and Rosa, Claudio and Tao, Tao},
  journal={IEEE Communications Standards Magazine}, 
  title={Further Enhanced Urllc And Industrial IoT Support With Release-17 5g New Radio}, 
  year={2023},
  volume={7},
  number={4},
  pages={12-19},
  doi={10.1109/MCOMSTD.0002.2200004}}

@ARTICLE{5GNRURLLC151617,
  author={Le, Trung-Kien and Salim, Umer and Kaltenberger, Florian},
  journal={IEEE Access}, 
  title={An Overview of Physical Layer Design for Ultra-Reliable Low-Latency Communications in 3GPP Releases 15, 16, and 17}, 
  year={2021},
  volume={9},
  number={},
  pages={433-444},
  doi={10.1109/ACCESS.2020.3046773}}

@misc{5GNRREL18SIDELINKULIC,
  title       = {Views on SL Positioning over Unlicensed Spectrum in R18},
  author      = {{3GPP}},
  year        = {2022},
  url         = {https://www.3gpp.org/dynareport?code=TDocExMtg--RP-97-e--60332.htm},
  note        = {Accessed: 2024-09-23}
}

@techreport{CLIRIM,
  title       = {Cross Link Interference (CLI) Handling and Remote Interference Management (RIM) for NR},
  author      = {{3GPP}},
  number      = {38.828},
  status      = {Under change control},
  type        = {Technical Report},
  planned_release = {Release 16},
  url         = {https://portal.3gpp.org/desktopmodules/Specifications/SpecificationDetails.aspx?specificationId=3605},
  note        = {Accessed: 2024-09-23}
}

@techreport{NRLBT,
  title       = {3GPP TS 37.213: Physical Layer Procedures for Shared Spectrum Channel Access (Release 16)},
  author      = {{3GPP}},
  year        = {2020},
  month       = {July},
  url         = {https://www.etsi.org/deliver/etsi_ts/137200_137299/137213/16.02.00_60/ts_137213v160200p.pdf},
  note        = {Accessed: 2024-09-23}
}

@report{CATLBT,
  title       = {Status Report to TSG: Study on Licensed-Assisted Access to Unlicensed Spectrum},
  author      = {{3GPP}},
  number      = {RP-150271},
  month       = {March},
  year        = {2015},
  conference   = {3GPP RAN #67},
  note        = {Accessed: 2024-09-23}
}

@misc{OFNOWHITESHEET,
  title       = {The Unlicensed Journey},
  author      = {Nazanin Rastegardoost},
  publisher   = {OFINNO},
  url         = {https://ofinno.com/wp-content/uploads/2020/11/OFNO_White_Sheet_112320.pdf},
  note        = {Accessed: 2024-09-23}
}

@ARTICLE{WIFIMARKOV,
  author={Rastegardoost, Nazanin and Jabbari, Bijan},
  journal={IEEE Communications Letters}, 
  title={Statistical Characterization of WiFi White Space}, 
  year={2017},
  volume={21},
  number={12},
  pages={2674-2677},
  doi={10.1109/LCOMM.2017.2744642}}

@misc{GSALTE5GUNLICENSED,
  title       = {Unlicensed Spectrum: August 2020 – LTE \& 5G Executive Summary},
  author      = {{GSA}},
  year        = {2020},
  url         = {https://gsacom.com/paper/unlicensed-spectrum-august-2020-lte-5g-executive-summary/},
  note        = {Accessed: 2024-09-23}
}

@ARTICLE{ENABLMASSIVEIOT6G,
  author={Guo, Fengxian and Yu, F. Richard and Zhang, Heli and Li, Xi and Ji, Hong and Leung, Victor C. M.},
  journal={IEEE Internet of Things Journal}, 
  title={Enabling Massive IoT Toward 6G: A Comprehensive Survey}, 
  year={2021},
  volume={8},
  number={15},
  pages={11891-11915},
  doi={10.1109/JIOT.2021.3063686}}

@ARTICLE{URLLCEMBBIOTSURVEY,
  author={Khan, Benish Sharfeen and Jangsher, Sobia and Ahmed, Ashfaq and Al-Dweik, Arafat},
  journal={IEEE Open Journal of the Communications Society}, 
  title={URLLC and eMBB in 5G Industrial IoT: A Survey}, 
  year={2022},
  volume={3},
  number={},
  pages={1134-1163},
  doi={10.1109/OJCOMS.2022.3189013}}

@article{SPECTSHFORIOTSURVEY,
	journal = {Computer Networks},
	doi = {10.1016/j.comnet.2018.12.008},
	issn = {1389-1286},
	language = {en},
	publisher = {Elsevier BV},
	title = {Internet of Things applications: A systematic review},
	url = {http://dx.doi.org/10.1016/j.comnet.2018.12.008},
	volume = {148},
	author = {Asghari, Parvaneh and Rahmani, Amir Masoud and Javadi, Hamid Haj Seyyed},
	pages = {241--261},
	date = {2019-01},
	year = {2019},
	month = {1},
}

@misc{IOTSTATS,
  title       = {30 Internet of Things (IoT) Stats in 2024 From Reputable Sources},
  author      = {AIMultiple},
  url         = {https://research.aimultiple.com/iot-stats/},
  note        = {Accessed: 2024-09-23}
}

@article{IOTRECENTSURVEY,
	journal = {International Journal of Sensors, Wireless Communications and Control},
	doi = {10.2174/2210327913666221216160435},
	issn = {2210-3279},
	number = {9},
	language = {en},
	publisher = {Bentham Science Publishers Ltd.},
	title = {A Concise Review on Internet of Things: Architecture, Enabling Technologies,
Challenges, and Applications},
	url = {http://dx.doi.org/10.2174/2210327913666221216160435},
	volume = {12},
	author = {Saqib, Manasha and Moon, Ayaz Hassan},
	pages = {629--650},
	date = {2022-11},
	year = {2022},
	month = {11},
}

@misc{ATNTSTRATEGICPRIVATE,
  title       = {5G Strategic Design Considerations for Private Cellular},
  author      = {{AT\&T}},
  url         = {https://www.business.att.com/content/dam/attbusiness/infographics/strategic-design-considerations-for-pcn-white-paper.pdf},
  note        = {Accessed: 2024-09-23}
}

@misc{GSMAPRIVATENET,
  title       = {Private 5G Industrial Networks: An Analysis of Use Cases and Deployment},
  author      = {{GSMA}},
  year        = {2023},
  url         = {https://www.gsma.com/iot/wp-content/uploads/2023/09/GSMA-Private-5G-Industrial-Networks-Report_June-23-.pdf},
  note        = {Accessed: 2024-09-23}
}

@techreport{ITUGOALSFOR5G,
  title       = {Setting the Scene for 5G: Opportunities \& Challenges},
  author      = {{International Telecommunication Union (ITU)}},
  year        = {2018},
  url         = {https://www.itu.int/en/ITU-D/Documents/ITU_5G_REPORT-2018.pdf},
  note        = {Accessed: 2024-09-23}
}

@INPROCEEDINGS{RLWIFILTE,
  author={Mishra, Garima and Rath, Hemant Kumar and Nadaf, Shameem and Mukhopadhyay, Raja and Menon, M.},
  booktitle={2024 IEEE 13th International Conference on Communication Systems and Network Technologies (CSNT)}, 
  title={A Reinforcement Learning Approach to Improve WiFi Network Performance Coexisting with LTE}, 
  year={2024},
  volume={},
  number={},
  pages={12-17},
  doi={10.1109/CSNT60213.2024.10546221}}

@article{MULTIARMED,
	journal = {Sensors},
	doi = {10.3390/s23156718},
	issn = {1424-8220},
	number = {15},
	language = {en},
	publisher = {MDPI AG},
	title = {A Multiarmed Bandit Approach for LTE-U/Wi-Fi Coexistence in a Multicell Scenario},
	url = {http://dx.doi.org/10.3390/s23156718},
	volume = {23},
	author = {Diógenes do Rego, Iago and de Castro Neto, José M. and Neto, Sildolfo F. G. and de Santana, Pedro M. and de Sousa, Jr., Vicente A. and Vieira, Dario and Venâncio Neto, Augusto},
	pages = {6718},
	date = {2023-07-27},
	year = {2023},
	month = {7},
	day = {27},
}

@INPROCEEDINGS{NRUDRL,
  author={Liu, Yan and Zhou, Hui and Deng, Yansha and Allanathan, Arumugam N},
  booktitle={GLOBECOM 2022 - 2022 IEEE Global Communications Conference}, 
  title={{DRL-based Channel Access in NR Unlicensed Spectrum for Downlink URLLC}}, 
  year={2022},
  volume={},
  number={},
  pages={591-596},
  doi={10.1109/GLOBECOM48099.2022.10000882}}

@ARTICLE{GAMETHEORY,
  author={Rahman, Aniq Ur and Kishk, Mustafa A. and Alouini, Mohamed-Slim},
  journal={IEEE Transactions on Cognitive Communications and Networking}, 
  title={A Game-Theoretic Framework for Coexistence of WiFi and Cellular Networks in the 6-GHz Unlicensed Spectrum}, 
  year={2023},
  volume={9},
  number={1},
  pages={239-251},
  doi={10.1109/TCCN.2022.3213732}}

\end{document}